\documentclass[preprint,12pt]{elsarticle}

\usepackage{amssymb}
 \usepackage{amsthm}
\usepackage{amsmath}
\usepackage{lineno}

\usepackage{subfigure}
\usepackage{enumerate}
\usepackage{here}
\usepackage{indentfirst}
\usepackage{bm}
\usepackage{url}
\usepackage{color}
\usepackage{here}

\def\Xint#1{\mathchoice
	{\XXint\displaystyle\textstyle{#1}}%
	{\XXint\textstyle\scriptstyle{#1}}%
	{\XXint\scriptstyle\scriptscriptstyle{#1}}%
	{\XXint\scriptscriptstyle\scriptscriptstyle{#1}}%
	\!\int}
\def\XXint#1#2#3{{\setbox0=\hbox{$#1{#2#3}{\int}$ }
		\vcenter{\hbox{$#2#3$ }}\kern-.6\wd0}}

\def\dashint{\Xint-}

\journal{Elsevier}

\begin{document}

\begin{frontmatter}



\title{Topology optimization of acoustic metasurfaces by using a two-scale homogenization method }

\author{Yuki~Noguchi\corref{cor1}} 
\ead{noguchi@mech.t.u-tokyo.ac.jp}
\cortext[cor1]{Corresponding author.
	Tel.: +81-3-5841-0294;
	Fax: +81-3-5841-0294.}
\author{Takayuki~Yamada \corref{}}

\address{Department of Strategic Studies, Institute of Engineering Innovation, School of Engineering, The University of Tokyo, Yayoi 2-11-16, Bunkyo--ku, Tokyo 113-8656, Japan.}

\begin{abstract}
In this paper, we propose a level set-based topology optimization method for the unit-cell design of acoustic metasurfaces by using a two-scale homogenization method. 
Based on previous works, we first propose a homogenization method for acoustic metasurfaces that can be combined with topology optimization. In this method, a nonlocal transmission condition depending on the unit cell of the metasurface appears in a macroscale  problem.
Next, we formulate an optimization problem within the framework of a level set-based topology optimization method, wherein an objective functional is expressed as the macroscopic responses obtained through the homogenization, and material distributions in the unit cell are set as design variables.
A sensitivity analysis is conducted based on the concept of the topological derivative.
To confirm the validity of the proposed method, two-dimensional numerical examples are provided. 
First, \textcolor{black}{we provide a numerical example that supports the validity of the homogenization method}, and we then
perform optimization calculations based on the waveguide settings of the acoustic metasurfaces. In addition, we discuss the mechanism of the obtained optimized structures.
\end{abstract}


\end{frontmatter}


\section{Introduction}
\label{intro}
Acoustic metamaterials are artificial composite materials that exhibit unusual acoustic performance that cannot be achieved by naturally existing acoustic media.
The concept of acoustic metamaterials is derived from that of the metamaterials of electromagnetic waves, which was first proposed by Veselago \cite{veselago1968electrodynamics}.
Several researchers have reported on electromagnetic metamaterials and their many unusual properties, represented by a negative refractive index.
The first extension of the concept of metamaterials to acoustic waves was a locally resonant sonic material proposed by Liu et al. \cite{liu2000locally}.
It is composed of a periodic array of unit cells filled with various elastic media, and the local resonance phenomenon is induced in the unit cell, producing a bandgap that prevents \textcolor{black}{the transmission of} acoustic waves.
After this pioneering work, various studies have proposed acoustic metamaterials exhibiting various characteristics, such as
negative bulk modulus \cite{fang2006ultrasonic}, negative mass density \cite{huang2009negative,huang2009wave}, and negative refractive index \cite{li2004double,ding2007metamaterial}.
These unusual acoustic properties can be used for the efficient control of acoustic waves and development of novel acoustic devices, such as an acoustic cloaking device \cite{zigoneanu2014three} and acoustic hyperlens \cite{li2009experimental}.

Although typical acoustic metamaterials have three-dimensional arrays of unit cells, current research interests are focused on \textcolor{black}{the} planar type of acoustic metamaterials\textcolor{black}{,} called acoustic metasurfaces.
Acoustic metasurfaces are based on the two-dimensional array of unit cells with finite thickness; they efficiently control acoustic waves in a smaller region compared to the bulk acoustic metamaterials.
Various types of metasurfaces that \textcolor{black}{exhibit} unusual acoustic properties have been proposed, for example, the sound-absorbing metasurface using the resonance of membrane structure \cite{ma2014acoustic}, a metasurface that manipulates the wavefront of acoustic waves by shifting their phase by using the complex structure of the unit cells \cite{xie2014wavefront}.

These extraordinary properties of metasurfaces strongly depend on their unit-cell structure. Therefore, the structural design of a unit cell is essential to obtain \textcolor{black}{the} desired acoustic behaviors. \textcolor{black}{As} metasurfaces control acoustic waves in a narrow region compared to bulk metamaterials, efficient structural design is required. Hence, we introduce a topology optimization method, with the highest degree of design freedom among structural optimization methods. Since the introduction of the method for linear elasticity problems \cite{bendsoe1988generating}, it has been applied in a wide range of fields including wave-propagation problems \cite{sigmund2003systematic}.
Regarding the applications of topology optimization to acoustic-wave-propagation problems, 
Wadbro and Berggren \cite{wadbro2006topology} proposed a topology optimization method for the design of an acoustic horn that transmits acoustic waves efficiently.
Du and Olhoff \cite{du2007minimization} optimized a bi-material structure to minimize sound radiation from the surface.
D\"uhring et al. \cite{duhring2008acoustic} proposed the SIMP method for acoustic problems to reduce indoor and outdoor noises.
Furthermore, acoustic-structural interaction problems can be considered in topology optimization, as suggested by \cite{dilgen2019topology}.

Topology optimization has also been applied for the design of metamaterials and metasurfaces.
Diaz and Sigmund \cite{Diaz2010} proposed a topology optimization method for electromagnetic metamaterials and showed that the obtained designs of the unit cell exhibited a negative permeability. 
Lu et al.\cite{lu2013topology} pointed out that acoustic metamaterials are optimized based on a level set-based topology optimization method and \textcolor{black}{exhibit} a negative bulk \textcolor{black}{modulus}.
The optimum design of an acoustic metasurface that converts longitudinal elastic waves into transverse elastic waves was derived in \cite{noguchi2015acoustic}.
\textcolor{black}{Christiansen and Sigmund \cite{christiansen2016designing} conducted topology optimization for a finite acoustic metamaterial slab, and they obtained its unit cell design inducing negative refraction.}
\textcolor{black}{Roca et al. \cite{roca2019computational} combined a multiscale homogenization approach based on a generalized Hill-Mandel principle with topology optimization; they obtained the unit cell design of locally resonant acoustic metamaterials.}

\textcolor{black}{When determining} the optimum design of acoustic \textcolor{black}{metasurfaces} using the topology optimization method, the computational cost must be considered because the optimization procedure comprises iterative analyses of the metasurface system, which is defined as the aggregation of unit cells with complex geometries.
The S-parameters-based retrieval method proposed by Smith et al. \cite{smith2005electromagnetic} is used to obtain the macroscopic properties of metamaterials and metasurfaces, and it was first applied to electromagnetic metamaterials\textcolor{black}{;} its application was later expanded to acoustic metamaterials \cite{fokin2007method}. 
Once the S-parameters representing the complex transmission and reflection coefficients are obtained, 
the effective material parameters of the metamaterials, such as \textcolor{black}{the} effective refractive index, can be estimated.
Although this method is simple and easy to implement, its application to general systems of a metasurface with complex incident-wave conditions is difficult\textcolor{black}{,} as it is based on the assumption that waves can be expressed as plane waves. 

Homogenization is another method used to estimate the macroscopic properties of metasurfaces.
The classical homogenization method \cite{sanchez127non,bakhvalov2012homogenisation,bensoussan1978asymptotic} is based on the asymptotic expansion of the solution using two \textcolor{black}{types} of characteristic scales: microscale and macroscale. By adopting this method \textcolor{black}{for} a system composed of a periodic array of unit cells, its complex structure is equivalently replaced with a homogeneous material, the properties of which are expressed through homogenized coefficients. 
This method holds for static or quasi-static problems, in which the wavelength of the waves traveling within the periodic structure is \textcolor{black}{significantly} longer than the size of the unit cell.
To address \textcolor{black}{the} problems involved with shorter wavelengths, where the quasi-static limit cannot be applied, a higher-order homogenization method \cite{santosa1991dispersive,smyshlyaev2000rigorous,abdulle2014finite,dohnal2014bloch,allaire2016comparison} was proposed. 
This is an extended version of the homogenization method and considers higher order terms, based on which the method allows for the modeling of the size effects of unit cells.
For a wavelength that is \textcolor{black}{considerably} shorter but \textcolor{black}{has} a similar order \textcolor{black}{as} the unit cell, a high-frequency homogenization method \cite{craster2010high} was proposed based on asymptotic expansions of the solution and frequency.
This method analyzes \textcolor{black}{the} perturbations of standing waves induced in the unit cell and can be applied to estimate the performance of metamaterials or photonic crystals \cite{antonakakis2013asymptotics}.
Furthermore, topology optimization was combined with this method for designing hyperbolic acoustic metamaterials \cite{noguchi2018topology}.

These homogenization methods target perfectly periodic infinite media composed of an array of unit cells, such as a square lattice in a two-dimensional problem.
\textcolor{black}{As} the metasurface has a finite thickness, special treatments are required in the homogenization method for dealing with such a metasurface.
Marigo and Maurel \cite{marigo2016homogenization,marigo2016two} proposed a homogenization method for metasurfaces in which higher-order approximation was introduced with inner and outer asymptotic expansions, which correspond to the region near the unit cells and the surrounding medium of the periodic array of unit cells, respectively.
Rohan and Luke\v{s} \cite{rohan2010homogenization} proposed a homogenization method for thin structures, the thickness of which was assumed to have the same order as the period of the unit cells.
In \cite{rohan2010homogenization}, the system of a rigid plate with periodic perforations was decomposed into a fictitious layer containing rigid obstacles and other regions filled with the background acoustic medium. The two-scale homogenization limit resulted in a homogeneous acoustic system with a nonlocal transmission condition imposed on the interface, which is a limited form of the fictitious layer.
This method was later extended to consider the oscillation of the elastic plate comprising the metasurface by introducing the Reissner--Mindlin plate model \cite{rohan2019homogenization}.
In the context of optimization, a shape sensitivity analysis was also conducted \cite{rohan2010sensitivity,rohan2013optimal} to be used in the shape optimization of the metasurface; however, no study has reported on topology optimization \textcolor{black}{thus far}.

In this research, we developed a topology optimization method for designing acoustic metasurfaces based on the homogenization method proposed by Rohan and Luke\v{s}.
Based on the proposed method, the material distribution in a single unit cell of the metasurface is optimized \textcolor{black}{such} that the metasurface composed of the optimized unit cells exhibits the desired macroscopic performance.
The remainder of this paper is organized as follows.
Section \ref{sec: Homogenization method} introduces the homogenization method for acoustic metasurfaces. 
We extend the previous method \cite{rohan2010homogenization} for rigid obstacles to the system composed of two media, in which waves are described by the Helmholtz equation, \textcolor{black}{to ensure} that topology optimization can be introduced.
Next, in section \ref{sec: Design problem for acoustic metasurfaces}, the design problem for acoustic metasurfaces is formulated based on the two-dimensional settings.
In section \ref{sec: Topology optimization for acoustic metasurfaces}, we explain the topology optimization for acoustic metasurfaces. The setting of the objective functional is provided within the framework of the homogenization method, and sensitivity analysis is conducted based on the concept of the topological derivative.
Then, section \ref{sec: Level set-based topology optimization} briefly explains the proposed level set-based topology optimization.
The numerical implementation is described in section \ref{sec:Numerical implementation}, which also presents the \textcolor{black}{optimization process} and discretization method using the finite element model (FEM) for \textcolor{black}{the} governing and adjoint equations.
In section \ref{sec: numerical examples}, several two-dimensional numerical examples are provided. 
Here, \textcolor{black}{an example that supports the validity of the proposed homogenization method is first provided}, and then the optimization results of waveguiding acoustic metasurfaces are \textcolor{black}{presented}. To confirm the obtained results, we conducted acoustic-wave-propagation analysis based on the FEM for the \textcolor{black}{entire} system of the metasurfaces without using the homogenization method. 
Finally, we provide the conclusions drawn from \textcolor{black}{this} study in section \ref{sec: Conclusion}.

\section{Homogenization method for acoustic metasurfaces}\label{sec: Homogenization method}
\begin{figure}[H]
	\centering
	\includegraphics[scale=0.3]{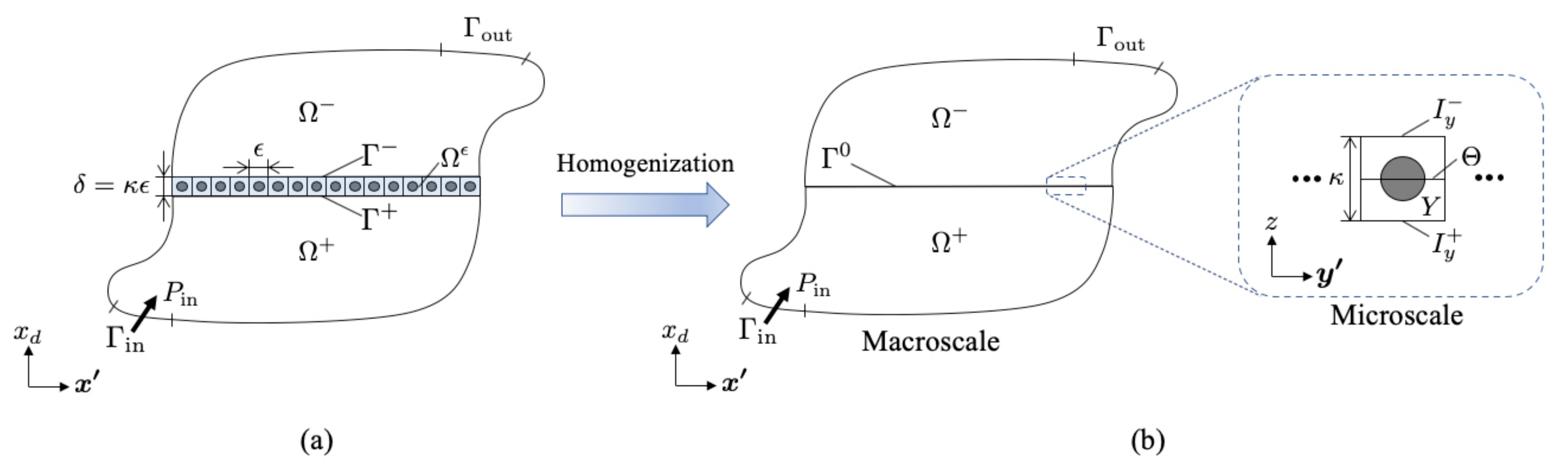}
	\caption{Geometrical settings: (a) Original system of an acoustic metasurface. (b) The homogenized  system of (a) with macroscale and microscale problems.}
\label{fig:Geom}       
\end{figure}

In this section, we introduce a homogenization method for acoustic metasurfaces based on a previous method \cite{rohan2010homogenization,rohan2019homogenization}.
Figure \ref{fig:Geom}(a) \textcolor{black}{presents} the system of an acoustic metasurface. 
Cartesian coordinate $\bm{x}=(\bm{x}',x_d)$ is used, and it comprises $\bm{x}' \in \mathbf{R}^{N-1}$ and $x_d \in \mathbf{R}$, where $N=2$ or $3$ represents the spatial dimension.
Unit cells with period $\epsilon$ of the metasurface are periodically arranged along $\bm{x}'$, and they have a finite thickness of $\delta$ along $x_d$.
For the homogenization procedure, thickness $\delta$ is assumed to be of the similar order as $\epsilon$. Then, $\delta$ can be expressed as $\delta = \kappa \epsilon$ with $\kappa>0$.
The array of unit cells forms a rectangular domain of $\Omega^\epsilon$, called as a transmission layer.
It is connected with the two outer regions of $\Omega^\pm$, where incident, reflected, and transmitted waves can propagate. 
$\Gamma^\pm$ represents the interfaces between $\Omega^\epsilon$ and $\Omega^\pm$, whereas
$\Gamma^0$ represents the mid-plane of $\Omega^\epsilon$.
We set the origin of coordinate $\bm{x}=(\bm{x'},x_d)$ \textcolor{black}{such} that $\Gamma^0$ can be expressed as $\Gamma^0=\{\bm{x}\in\Omega^G|~x_d = 0\}$, where $\Omega^G=\Omega^+\cup \Omega^- \cup \Omega^\epsilon$ represents the \textcolor{black}{entire} domain. 
Then, interfaces $\Gamma^\pm$ are expressed as $\Gamma^\pm=\{\bm{x}\in\Omega^G|~x_d = \mp \frac{\delta}{2} \}$.

By assuming a harmonic oscillation with angular frequency $\omega$, the boundary value problem corresponding to the system shown in Fig.~\ref{fig:Geom}(a) is described as follows:
\begin{align}
\nabla\cdot\left(\frac{1}{\rho(\bm{x})}\nabla p\right) + \frac{\omega^2}{K(\bm{x})}p &= 0~~\mathrm{in~}\Omega^G,
\\
\bm{n}\cdot\left(\frac{1}{\rho_0}\nabla p \right)+ \frac{i k_0}{\rho_0}p &=\frac{2i k_0}{\rho_0}P_\mathrm{in} ~~\mathrm{on~}\Gamma_\mathrm{in},\label{eq: incident wave}
\\
\bm{n}\cdot\left(\frac{1}{\rho_0}\nabla p \right)+ \frac{i k_0}{\rho_0}p &= 0~~\mathrm{on~}\Gamma_\mathrm{out},\label{eq: absorbing condition}
\\
\bm{n}\cdot \left(\frac{1}{\rho(\bm{x})} \nabla p\right)&=0~~\mathrm{on~}\partial_\mathrm{ext}{\Omega^G}\setminus(\Gamma_\mathrm{in}\cup \Gamma_\mathrm{out}),
\end{align}
where $p$ represents the complex amplitude of acoustic pressure\textcolor{black}{,} and 
$\partial_\mathrm{ext}\Omega^G$ represents the external boundary of the \textcolor{black}{entire} domain, $\Omega^G$.
\textcolor{black}{The mass} density $\rho$ and bulk modulus $K$ have piece-wise constant distributions in $\Omega^\epsilon$ corresponding to the structural configuration of the metasurface, \textcolor{black}{whereas} they are constant in $\Omega^\pm$, denoted by $\rho_0$ and $K_0$ \textcolor{black}{respectively}.
Eq.~(\ref{eq: incident wave}) represents an incident-wave boundary condition on $\Gamma_\mathrm{in}$, the amplitude and wavenumber of which are $P_\mathrm{in}$ and $k_0$, respectively.
Eq.~(\ref{eq: absorbing condition}) represents an absorbing boundary condition on $\Gamma_\mathrm{out}$ to reduce the reflected waves, and
sound-hard conditions are applied to the other external boundaries.

By introducing the homogenization method, we aim to 
replace the complex structure of the metasurface in $\Omega^\epsilon$ with an equivalent interface, $\Gamma^0$, by \textcolor{black}{considering} the limit $\epsilon \to 0$, on which there can be jumps in acoustic pressure and flux. When $\epsilon \to 0$, boundaries $\Gamma^\pm$ approach $\Gamma^0$ and transmission layer $\Omega^\epsilon$ degenerates to $\Gamma^0$.

As the metasurface has a finite thickness, the original boundary value problem must first be decomposed into the problems defined in \textcolor{black}{the} transmission layer $\Omega^\epsilon$ and outer regions $\Omega^\pm$.
Given the acoustic pressure on $\Gamma^\pm$ expressed by $p_g^\epsilon$, 
the boundary value problem for $\Omega^\pm$ is summarized as follows: 
\begin{align}
\frac{1}{\rho_0}\nabla^2 P^\pm + \frac{\omega^2}{K_0}P^\pm &= 0~~\mathrm{in~}\Omega^+ \cup \Omega^- 
\\
P^\pm &= p_g^\epsilon~~\mathrm{on~}\Gamma^\pm \label{eq: coupling1}
\\
\bm{n}\cdot\left(\frac{1}{\rho_0}\nabla P \right)+ \frac{i k_0}{\rho_0}P &= 0~~\mathrm{on~}\Gamma_\mathrm{out},
\\
\bm{n}\cdot\left(\frac{1}{\rho_0}\nabla P \right)+ \frac{i k_0}{\rho_0}P &=\frac{2i k_0}{\rho_0}P_\mathrm{in} ~~\mathrm{on~}\Gamma_\mathrm{in},
\\
\bm{n}\cdot\left(\frac{1}{\rho_0}\nabla P \right) &=0 ~~\mathrm{on~}\partial_\mathrm{ext}\Omega^\pm \setminus (\Gamma_\mathrm{in}\cup \Gamma_\mathrm{out}),
\end{align}
Similarly, given the fluxes on $\Gamma^\pm$ expressed by $g^{\epsilon \pm}$,
the boundary value problem for transmission layer $\Omega^\epsilon$ to \textcolor{black}{determine} the unknown $p^\epsilon$ is summarized as follows:
\begin{align}
\nabla\cdot\left(\frac{1}{\rho(\bm{x})}\nabla p^\epsilon\right) + \frac{\omega^2}{K(\bm{x})}p^\epsilon &= 0~~\mathrm{in~}\Omega^\epsilon,
\\
\bm{n}\cdot\left(\frac{1}{\rho_0}\nabla p^\epsilon \right) &= -g^{\epsilon \pm}~~\mathrm{on~}\Gamma^\pm,
\\
\bm{n}\cdot\left(\frac{1}{\rho_0}\nabla p^\epsilon \right) &= 0~~\mathrm{on~}\partial\Omega^\epsilon \setminus \Gamma^\pm.
\end{align}
The \textcolor{black}{abovementioned} decoupled problems are equivalent to the original boundary value problem if the following coupling conditions on $\Gamma^\pm$ hold:
\begin{align}
p^\epsilon &= p_g^\epsilon = P^\pm~~~\mathrm{on~}\Gamma^\pm,
\\
g^{\epsilon \pm} &= \bm{n}^\pm \cdot \left(\frac{1}{\rho_0}\nabla P^\pm\right)~~\mathrm{on~}\Gamma^\pm.
\label{eq: coupling in the outer regions}
\end{align}
Next, we establish a homogenized system for transmission layer $\Omega^\epsilon$.
First, the weak form of the system in $\Omega^\epsilon$ is expressed as follows:
\begin{align}
\int_{\Omega^\epsilon}\frac{1}{\rho(\bm{x})}\nabla p^\epsilon \cdot \nabla q^\epsilon d\Omega 
-\int_{\Omega^\epsilon}\frac{\omega^2}{K(\bm{x})}p^\epsilon q^\epsilon d\Omega
=-\int_{\Gamma^\pm}g^{\epsilon\pm}q^\epsilon d\Gamma~~~\forall q^\epsilon \in H^1(\Omega^\epsilon).
\end{align}
We introduce the scaled coordinate in the direction of $x_d$ as $z=\frac{1}{\epsilon}x_d$.
Then, transmission layer $\Omega^\epsilon$ can be expressed as $\hat{\Omega}=\Gamma^0\times ]-\frac{\kappa}{2},\frac{\kappa}{2}[$ by using scaled coordinate $(\bm{x'}, z)$.
For this coordinate, equation (15) is modified as
\begin{align}
\int_{\hat{\Omega}}\frac{1}{\rho(\bm{x})}\left(\overline{\nabla} p^\epsilon \cdot \overline{\nabla} q^\epsilon
+\frac{1}{\epsilon^2} \frac{\partial p^\epsilon}{\partial z} \frac{\partial q^\epsilon}{\partial z}  \right) d\Omega 
-\int_{\hat{\Omega}}\frac{\omega^2}{K(\bm{x})}p^\epsilon  q^\epsilon d\Omega
=-\frac{1}{\epsilon}\int_{\Gamma^\pm}g^{\epsilon\pm}q^\epsilon d\Gamma,
\label{eq: weak formulation before homogenization}
\end{align}
where $\overline{\nabla}$ is an in-plane gradient, the components of which are denoted by $\frac{\partial}{\partial x_\alpha}~(\alpha = 1,...,N-1)$.

In previous studies \cite{rohan2010homogenization,rohan2019homogenization}, the periodic unfolding method was used to homogenize the system \textcolor{black}{with the} unfolding operator $T_\epsilon$.
\textcolor{black}{The operator} $T_\epsilon$ associates solution $v\in L^p(\Omega)$ with $T_\epsilon(v)\in L^p(\Omega\times Y)$, where domain $\Omega$ contains a periodic structure characterized by the representative unit cell, $Y$.
One of the important properties of $T_\epsilon$ is the so-called integral conservation, which is represented as follows:
\begin{align}
\int_{\Omega}v(\bm{x})d\Omega_x = \frac{1}{|Y|}\int_{\Omega\times Y}T_{\epsilon}(v)(\bm{x},\bm{y})d\Omega_x d\Omega_y .
\end{align}
\textcolor{black}{Details regarding} the periodic unfolding method and the properties of $T_\epsilon$ can be referenced from \cite{cioranescu2008periodic}.

To introduce the periodic unfolding method into the system of the metasurface, 
a scaled coordinate is defined in the direction of $\bm{x'}$ as $\bm{y'}=\frac{\bm{x'}}{\epsilon}$. 
\textcolor{black}{The microscale} coordinate $\bm{y}=(\bm{y'},z)$ is utilized to express the representative unit cell, $Y$, as shown in Fig.~\ref{fig:Geom}(b).
\textcolor{black}{Thereafter, the} unfolding operator $T_\epsilon$ is defined \textcolor{black}{such} that it maps solution $p^\epsilon \in L^2(\Omega^\epsilon)$ to $T_\epsilon(p^\epsilon) \in L^2(\Gamma^0 \times Y)$.

We impose the following assumption on fluxes $g^{\epsilon \pm}$:
\begin{align}
g^{\epsilon+}(\bm{x'})=g^0(\bm{x'})+\epsilon g^{1+}(\bm{x'},\frac{\bm{x'}}{\epsilon}),\\
g^{\epsilon-}(\bm{x'})=-g^0(\bm{x'})-\epsilon g^{1-}(\bm{x'},\frac{\bm{x'}}{\epsilon}),
\end{align}
These assumptions assure the continuity of the lowest order of fluxes across $\Gamma^\pm$ and are expressed as $g^0(\bm{x'})$. The opposite signs in the definition of $g^{\epsilon\pm}$ are due to the outward normal vector on $\Gamma^\pm$.
Under this assumption and the weak form in Eq.~(\ref{eq: weak formulation before homogenization}),
a priori estimates to the solution (see \cite{rohan2010homogenization,rohan2019homogenization}) lead to the following convergence results for $\epsilon \to 0$:
\begin{align}
T_\epsilon(p^\epsilon) \rightharpoonup p^0~~~&\mathrm{weakly~in~}L^2(\Gamma_0 \times Y), \label{eq: convergence_first}\\
T_\epsilon(\overline{\nabla}p^\epsilon) \rightharpoonup  \overline{\nabla}_x  p^0 + \overline{\nabla}_y  p^1~~~&\mathrm{weakly~in~}L^2(\Gamma_0 \times Y),\\
\frac{1}{\epsilon}T_\epsilon(\frac{\partial p^\epsilon}{\partial z}) \rightharpoonup \frac{\partial p^1}{\partial z}~~~&\mathrm{weakly~in~}L^2(\Gamma_0 \times Y), \label{eq: convergence_last}
\end{align}
where $\overline{\nabla}_x$ and $\overline{\nabla}_y$ denote in-plane gradients with respect to $\bm{x'} \in \Gamma_0$ and $\bm{y'} \in Y$, respectively.
$p^0 \in H^1(\Gamma^0)$ and $p^1 \in L^2(\Gamma^0; H^1_{\underline{\sharp}})$ are asymptotic expanded pressures, where $H^1_{\underline{\sharp}}(Y)$ represents a subspace of $H^1(Y)$ that satisfies the periodic boundary conditions in the direction of $y_\alpha~(\alpha=1,..,N-1)$.

\textcolor{black}{By using these results, the following homogenized equation is obtained:}
\textcolor{black}{
\begin{align}
	&\sum_{\alpha=1}^{N-1}\sum_{\beta=1}^{N-1}\int_{\Gamma^0}A_{\alpha\beta}^\ast \frac{\partial p^0}{\partial x_\beta} \frac{\partial q^0}{\partial x_\alpha} d\Gamma_x
	-\omega^2\int_{\Gamma^0}K^{-1\ast} p^0 q^0 d\Gamma_x
	+\int_{\Gamma^0}g^0 \bm{B}^\ast \cdot \overline{\nabla}_x q^0 d\Gamma_x
	\nonumber\\
	&= -\int_{\Gamma^0}q^0\left(\dashint_{\Theta}\Delta g^1 d\Gamma_y\right) d\Gamma_x ,
	\label{eq: surface 1}
\end{align}
}
\textcolor{black}{where $\Delta g^1$ is used to express the difference between $g^{1\pm}$ as $\Delta g^1 = g^{1+}-g^{1-}$.	
$A_{\alpha \beta}^\ast$, $\bm{B}^\ast$, and $K^{-1\ast}$ are homogenized coefficients, and they are expressed as}
\textcolor{black}{
\begin{align}
	&A_{\alpha\beta}^\ast =\dashint_Y \frac{1}{\rho(\bm{y})}\left\{  \overline{\nabla}_y (\textcolor{black}{\eta}^\beta + y_\beta)\cdot\overline{\nabla}_y (\textcolor{black}{\eta}^\alpha + y_\alpha)
	+ \frac{\partial \textcolor{black}{\eta}^\beta}{\partial z} \frac{\partial \textcolor{black}{\eta}^\alpha}{\partial z}
	\right\}d\Omega_y,\label{eq: coeff A}
	\\
	&\bm{B}^\ast = \dashint_Y \frac{1}{\rho(\bm{y})}\overline{\nabla}_y \xi d\Omega_y,\label{eq: coeff B}
	\\
	&K^{-1\ast} =  \dashint_Y \frac{1}{K(\bm{y})}d\Omega_y.\label{eq: coeff K}
\end{align}
}
\textcolor{black}{To estimate these homogenized coefficients, functions, $\textcolor{black}{\eta}^\alpha$ and $\xi$, defined in the microscale $\bm{y}$, are introduced. They are the solutions of the following cell problems:
\begin{align}
	&\dashint_Y \frac{1}{\rho(\bm{y})}\left(
	\overline{\nabla}_y \textcolor{black}{\eta}^\alpha \cdot \overline{\nabla}_y \psi + \frac{\partial \textcolor{black}{\eta}^\alpha}{\partial z}\frac{\partial \psi}{\partial z}
	\right)d\Omega_y = -\dashint_Y \frac{1}{\rho(\bm{y})}\frac{\partial \psi}{\partial y_\alpha}d\Omega_y~~~\forall \psi\in H^1_{\underline{\sharp}}(Y) , \label{eq: local pi}
	\\
	&\dashint_Y \frac{1}{\rho(\bm{y})}\left(
	\overline{\nabla}_y \xi \cdot \overline{\nabla}_y \psi + \frac{\partial \xi}{\partial z}\frac{\partial \psi}{\partial z}
	\right)d\Omega_y =
	-\left(
	\dashint_{I_y^+} \psi d\Gamma_y - \dashint_{I_y^-} \psi d\Gamma_y
	\right)~~~\forall \psi\in H^1_{\underline{\sharp}}(Y),\label{eq: local xi}
\end{align}
where $I_y^+$ and $I_y^-$ represent the bottom and top surfaces, respectively, in unit cell $Y$, as shown in Fig.~\ref{fig:Geom}(b).
Details regarding the derivation of Eq.~(\ref{eq: surface 1}) are available in \ref{sec:Appendix homogenized equations}.
}

Next, we consider the coupling condition given in Eq.~(\ref{eq: coupling1}) in the weak sense.
By multiplying the equation with test function $\psi = \psi(\bm{x'})$ and by applying Green's formula, we \textcolor{black}{obtain}
\begin{align}
	\int_{\Gamma^-}P^- \psi d\Gamma - \int_{\Gamma^+}P^+ \psi d\Gamma 
	= \int_{\Gamma^0}\psi \int_{-\frac{\delta}{2}}^{\frac{\delta}{2}} \frac{\partial p^\epsilon}{\partial x_d} d\Gamma.
	\label{eq: Interface coupling}
\end{align}
\textcolor{black}{Considering limit $\epsilon\to \epsilon_0$ with a small positive number, $\epsilon_0$, and the procedure explained in \ref{sec:Appendix homogenized equations}, the following homogenized equation is obtained:
\begin{align}
	\int_{\Gamma^0}\left(
	\bm{B}^\ast \cdot \overline{\nabla}_x p^0 - F^\ast g^0
	\right)\psi d\Gamma
	=\frac{1}{\epsilon_0}\int_{\Gamma^0}({P}^+_m - {P}^-_m)\psi d\Gamma_x
	~~~\forall \psi \in L^2(\Gamma^0),
	\label{eq: surface 2}
\end{align} 
where ${P}^{\pm}_m(\bm{x}')$ are mapped acoustic pressures defined in the original coordinate, $(\bm{x'}, x_d)$, as follows:
\begin{align}
	{P}^+_m(\bm{x}') = P^+ (\bm{x}', -\frac{\delta}{2}), ~~~~~{P}^-_m(\bm{x}') = P^- (\bm{x}', \frac{\delta}{2}).
	\label{eq: mapped pressures}
\end{align}
A homogenized coefficient, $F^\ast$, in Eq.~(\ref{eq: surface 2}) is introduced, which is expressed as
\begin{align}
	F^\ast &= -\left(
	\dashint_{I_y^+}\xi d\Gamma_y - \dashint_{I_y^-}\xi d\Gamma_y 
	\right).\label{eq: coeff F}
\end{align}
}

To determine the relationship between the limit form of acoustic pressure $p^0$ and the external fields, $P^+$ and $P^-$, we consider the transmission layer, $\hat{\Omega}=\Gamma^0 \times ]-\frac{\kappa}{2},\frac{\kappa}{2}[$, which is expressed by scaled coordinate $(\bm{x'}, z)$ \textcolor{black}{with $z=\frac{x_d}{\epsilon}$}, and the following condition corresponding to the coupling condition in Eq.~(\ref{eq: coupling1}):
\begin{align}
	\int_{\hat{\Omega}}(p^\epsilon - P_b)\varphi d\Omega_x= 0 ~~~\forall \varphi \in L^2(\Gamma^0),
\end{align}
where $P_b$ is a blending function for $P^\pm$, and it is defined at coordinate $(\bm{x'}, z)$ as 
\begin{align}
	P_b(\bm{x'}, z) = \frac{1}{\kappa}\left\{ (z+\frac{\kappa}{2}){P}^-_m(\bm{x'}) -  (z-\frac{\kappa}{2}){P}^+_m(\bm{x'})      \right\}.
\end{align}
Then, by considering the case of $\varphi = \varphi(\bm{x'})$ with $\bm{x'} \in \Gamma_0$ and recalling the convergence result for $p^\epsilon$, the limit form of this integral results in the following condition:
\begin{align}
	\int_{\Gamma^0}\left(
	p^0-\frac{1}{2}({P}^+_m + {P}^-_m)
	\right)\varphi d\Gamma_x =0 ~~~\forall\varphi \in L^2(\Gamma^0).
	\label{eq: mean value}
\end{align} 

Finally, a weak form is considered in the outer regions. 
When scale parameter $\epsilon$ approaches a small number, $\epsilon_0$, 
the coupling condition for acoustic fluxes, as expressed in Eq.~(\ref{eq: coupling in the outer regions}), satisfies the following condition:
\begin{align}
	\bm{n}^\pm\cdot \left(\frac{1}{\rho_0}\nabla_x {P}^\pm_m\right) \to \dashint_{\Theta} T_{\epsilon_0}(g^{\epsilon_0 \pm}) d\Gamma_y
	~~~\mathrm{on~}\Gamma_0.
\end{align}
To further modify the right-hand side, we introduce the following variables with respect to integrated fluxes:
\begin{align}
	G_0^\pm(\bm{x}) &= \pm \dashint_{\Theta} T_{\epsilon_0}(g^{\epsilon_0 \pm})d\Gamma_y = \dashint_{\Theta}\left\{
	g^0(\bm{x}) + \epsilon_0 g^{1\pm}(\bm{x},\bm{y})
	\right\}d\Gamma_y\nonumber\\
	&=g^0(\bm{x}) + \epsilon_0 G^{1\pm}(\bm{x}),\\
	G^{1\pm} &= \dashint_{\Theta}g^{1\pm}(\bm{x},\bm{y})d\Gamma_y.
\end{align}
Therefore, the following relations are valid when $\epsilon \to \epsilon_0$,
\begin{align}
	g^0 &\approx \frac{1}{2}(G_0^+ + G_0^-),\nonumber\\
	\Delta G^1 &= G^{1+} - G^{1-} = \frac{1}{\epsilon_0}(G_0^+ - G_0^-).
	\label{eq: G0 and G1}
\end{align}
Then, the coupling condition, Eq.~(\ref{eq: coupling in the outer regions}), can be replaced with $G_0^\pm$ as follows:
\begin{align}
	&\bm{n}^+\cdot \left(\frac{1}{\rho_0}\nabla_x {P}^+\right) =G_0^+ ,~~~
	\bm{n}^-\cdot \left(\frac{1}{\rho_0}\nabla_x {P}^-\right) =-G_0^-.
	\label{eq: definition of G0}
\end{align}
By using $G_0^\pm$, the weak form in external regions $\Omega^\pm$ is given as
\begin{align}
	&\int_{\Omega^+ \cup \Omega^-} \frac{1}{\rho_0}\nabla_x P^\pm \cdot \nabla_x \tilde{P} d\Omega_x
	-\int_{\Omega^+ \cup \Omega^-}\frac{\omega^2}{K_0}P^\pm \tilde{P}d\Omega_x \nonumber\\
	&-\int_{\Gamma^0}G_0^+ \tilde{P} d\Gamma_x + \int_{\Gamma^0}G_0^- \tilde{P} d\Gamma_x\nonumber\\
	&+\int_{\Gamma_\mathrm{in}\cup \Gamma_\mathrm{out}}\frac{i k_0}{\rho_0}P^\pm \tilde{P} d\Gamma_x
	-\int_{\Gamma_\mathrm{in}}\frac{2 i k_0}{\rho_0}P_\mathrm{in} \tilde{P} d\Gamma_x = 0
	~~~\forall \tilde{P} \in H^1(\Omega^+ \cup \Omega^-).
\end{align}
Note that the acoustic pressure in the external regions, $P^\pm$, can be discontinuous across interface $\Gamma^0$ owing to the internal fluxes, $G_0^\pm$.

By using the relations expressed in Eq.~(\ref{eq: G0 and G1}), the homogenized acoustic system when $\epsilon\to 0$ is summarized as follows:
\begin{align}
	&\sum_{\alpha=1}^{N-1}\sum_{\beta=1}^{N-1}\int_{\Gamma^0}A_{\alpha\beta}^\ast \frac{\partial p^0}{\partial x_\beta} \frac{\partial q^0}{\partial x_\alpha} d\Gamma_x
	-\omega^2\int_{\Gamma^0}K^{-1\ast} p^0 q^0 d\Gamma_x\nonumber\\
	&+\frac{1}{2}\int_{\Gamma^0}(G_0^+ + G_0^-) \bm{B}^\ast \cdot \overline{\nabla}_x q^0 d\Gamma_x
	= -\frac{1}{\epsilon_0}\int_{\Gamma^0}q^0(G_0^+ - G_0^-) d\Gamma_x ,
	\label{eq:macro_first}\\
	&\int_{\Gamma^0}\left(
	\bm{B}^\ast \cdot \overline{\nabla}_x p^0 - \frac{1}{2} F^\ast (G_0^+ + G_0^-)
	\right)\psi d\Gamma_x
	=\frac{1}{\epsilon_0}\int_{\Gamma^0}({P}^+ - {P}^-)\psi d\Gamma_x, \\
	&\int_{\Gamma^0}\left(
	p^0-\frac{1}{2}({P}^+ + {P}^-)
	\right)\varphi d\Gamma_x =0,\\
	&\int_{\Omega^+ \cup \Omega^-} \frac{1}{\rho_0}\nabla_x P^\pm \cdot \nabla_x \tilde{P} d\Omega_x
	-\int_{\Omega^+ \cup \Omega^-}\frac{\omega^2}{K_0}P^\pm \tilde{P}d\Omega_x \nonumber\\
	&-\int_{\Gamma^0}G_0^+ \tilde{P} d\Gamma_x + \int_{\Gamma^0}G_0^- \tilde{P} d\Gamma_x\nonumber\\
	&+\int_{\Gamma_\mathrm{in}\cup \Gamma_\mathrm{out}}\frac{i k_0}{\rho_0}P^\pm \tilde{P} d\Gamma_x
	-\int_{\Gamma_\mathrm{in}}\frac{2 i k_0}{\rho_0}P_\mathrm{in} \tilde{P} d\Gamma_x = 0,
	\label{eq:macro_last}
\end{align}
where we replace notation $P_m^\pm$ with $P^\pm$ on $\Gamma^0$ 
because boundaries $\Gamma^\pm$ approach $\Gamma^0$ when $\epsilon \to 0$.
The \textcolor{black}{abovementioned} equations are solved at the macroscale, $\bm{x}=(\bm{x'},x_d)$, \textcolor{black}{using} the following procedure. 
Given the material distribution in unit cell $Y$, the cell problems, i.e., Eq.~(\ref{eq: local pi}) and (\ref{eq: local xi})\textcolor{black}{,} are solved first.
Then, homogenized coefficients $(\bm{A}^\ast, \bm{B}^\ast, K^{-1 \ast}, F^\ast)$ are evaluated based on Eqs.~(\ref{eq: coeff A}), (\ref{eq: coeff B}), (\ref{eq: coeff K}), and (\ref{eq: coeff F}).
By using these coefficients, we can solve the macroscale equations.

\textcolor{black}{
As we focused on metasurfaces that are composed of two types of media, the cell problems expressed by Eq.~(\ref{eq: local pi}) and (\ref{eq: local xi}) are defined over the unit cell $Y$. Therefore, the homogenized coefficients expressed by Eqs.~(\ref{eq: coeff A}), (\ref{eq: coeff B}), and (\ref{eq: coeff K}) are defined by the integrals over $Y$. This is  different from previous works \cite{rohan2010homogenization,rohan2019homogenization}, where the former work targeted acoustic transmission through rigid bodies and the latter tackled acoustic-elastic interaction problems.
}

\section{Design problem for acoustic metasurfaces}\label{sec: Design problem for acoustic metasurfaces}
\begin{figure}[H]
	\centering
	\includegraphics[scale=0.3]{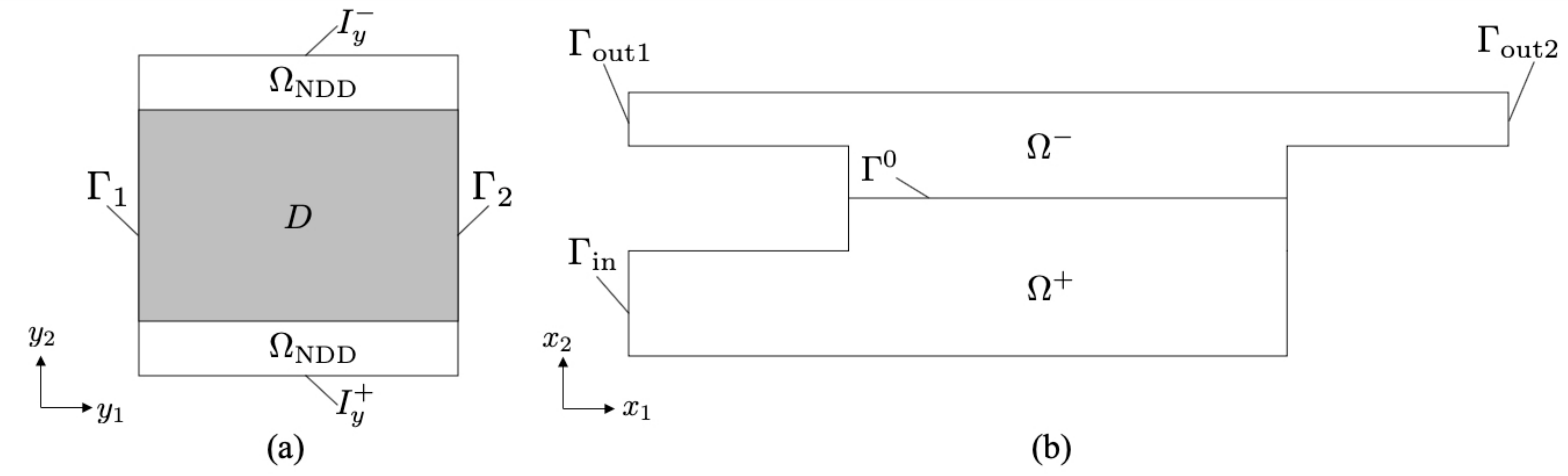}
	\caption{Setting of design domain and boundary conditions at the 
		(a) microscale and (b) macroscale.}
	\label{fig:Design_model}       
\end{figure}

Figure~\ref{fig:Design_model} \textcolor{black}{presents} the setup of the design domain and boundary conditions for the optimization of acoustic metasurfaces in this research.
We focused on the two-dimensional case of $N=2$, in which the metasurface composed of square unit cells is homogenized to be a line, $\Gamma^0$.
To simplify the notations for formulating the optimization problem, 
we used macroscale $\bm{x}=(x_1,x_2)$ and microscale $\bm{y}=(y_1, y_2)$.
Under these notations, components $x_2$ and $y_2$ correspond to $x_d$ and $z$ used in the previous section.

By exploiting the \textcolor{black}{benefits} of the \textcolor{black}{abovementioned} homogenization method, 
the structure of the metasurface at the microscale is optimized to achieve the desired responses at the macroscale.
Thus, we set design domain $D$ in unit cell $Y$ as shown in Fig.~\ref{fig:Design_model}(a).
Design domain $D$ is sandwiched between \textcolor{black}{non-design} domains $\Omega_\mathrm{NDD}$, with air as the medium. 
Periodic boundary conditions are applied to $\Gamma_1$ and $\Gamma_2$ but not to $I_y^+$ and $I_y^-$.
As both air and elastic media appear in $D$, this domain is represented by $D=\Omega_\mathrm{air}\cup \Omega_\mathrm{elastic}$, where $\Omega_\mathrm{air}$ and $\Omega_\mathrm{elastic}$ are the regions comprising air and elastic media, respectively.
We assume that wave propagation can be described by the corresponding Helmholtz equation, which is often used in topology optimization, as demonstrated in \cite{andkjaer2013topology}.  
Usually, acoustic--elastic  coupling effects, which are interactions between acoustic and elastic waves in acoustic and elastic media, respectively, should be considered. 
Therefore, the use of the Helmholtz equation is generally inappropriate to express the system.
However, the \textcolor{black}{abovementioned} assumption is justified if the corresponding media have a high-contrast ratio between their acoustic impedance.
In such a setting, almost all of the waves will be reflected on their interfaces, and the interactions between acoustic and elastic media can be \textcolor{black}{neglected}. 
In this study, we chose an elastic material to satisfy the \textcolor{black}{abovementioned} settings, the details of which are available in section \ref{sec: Validation of the HMM}.

Figure~\ref{fig:Design_model}(b) shows the settings of the geometric and boundary conditions at the macroscale. 
The incident plane wave impinges normally on boundary $\Gamma_\mathrm{in}$.
Two outlets $\Gamma_\mathrm{out1}$ and $\Gamma_\mathrm{out2}$ are set, on which the absorbing boundary conditions are applied. 
Interface $\Gamma^0$ represents the homogenized metasurface, which is characterized by the homogenized coefficients, and
sound-hard conditions are applied on the other boundaries.

Corresponding to these two-dimensional settings, the cell problems in $Y=D\cup \Omega_\mathrm{NDD}$ are defined as follows:
\begin{align}
&\int_Y \frac{1}{\rho(\bm{y})}
{\nabla}_y \textcolor{black}{\eta} \cdot {\nabla}_y \psi 
d\Omega_y = -\int_Y \frac{1}{\rho(\bm{y})}\frac{\partial \psi}{\partial y_1}d\Omega_y~~~\forall \psi\in H^1_{\underline{\sharp}}(Y) , \label{eq: local pi_2D}
\\
&\int_Y \frac{1}{\rho(\bm{y})}
{\nabla}_y \xi \cdot {\nabla}_y \psi d\Omega_y =
-\left(
\int_{I_y^+} \psi d\Gamma_y - \int_{I_y^-} \psi d\Gamma_y
\right)~~~\forall \psi\in H^1_{\underline{\sharp}}(Y).\label{eq: local xi_2D}
\end{align}
The average notations are omitted in Eqs. (\ref{eq: local pi_2D} and \ref{eq: local xi_2D}) as we focus on the metasurface composed of squares, which are described as $[0,1]\times [0,1]$ according to the microscale coordinate, $\bm{y}$.
Then, the homogenized coefficients are defined as
\begin{align}
&A_{11}^\ast =\int_Y \frac{1}{\rho(\bm{y})}\left\{  {\nabla}_y (\textcolor{black}{\eta} + y_1)\cdot{\nabla}_y (\textcolor{black}{\eta} + y_1)
\right\}d\Omega_y\nonumber\\
&=\int_Y \frac{1}{\rho(\bm{y})}\left( {\nabla}_y \textcolor{black}{\eta} \cdot{\nabla}_y \textcolor{black}{\eta} + \frac{\partial \textcolor{black}{\eta}}{\partial y_1}
\right)d\Omega_y
+\int_Y \frac{1}{\rho(\bm{y})}\left(\frac{\partial \textcolor{black}{\eta}}{\partial y_1}+1\right)d\Omega_y\nonumber\\
&=\int_Y \frac{1}{\rho(\bm{y})}\left(\frac{\partial \textcolor{black}{\eta}}{\partial y_1}+1\right)d\Omega_y,\label{eq: coeff A_2D}
\\
&{B}_1^\ast = \int_Y \frac{1}{\rho(\bm{y})}\frac{\partial \xi}{\partial y_1} d\Omega_y,
\label{eq: coeff B_2D}
\\
&F^\ast = -\left(
\int_{I_y^+}\xi d\Gamma_y - \dashint_{I_y^-}\xi d\Gamma_y 
\right),\label{eq: coeff F_2D}
\\
&K^{-1\ast} =  \int_Y \frac{1}{K(\bm{y})}d\Omega_y,\label{eq: coeff K_2D}
\end{align}
where we used Eq.~(\ref{eq: local pi_2D}) with test function $\psi = \textcolor{black}{\eta}$ to modify the form of $A_{11}^\ast$ in Eq.~(\ref{eq: coeff A_2D}).
Based on these coefficients, the macroscopic problem is defined as follows:
\begin{align}
&\int_{\Gamma^0}A_{11}^\ast \frac{\partial p^0}{\partial x_1} \frac{\partial q^0}{\partial x_1} d\Gamma_x
-\omega^2\int_{\Gamma^0}K^{-1\ast} p^0 q^0 d\Gamma_x\nonumber\\
&+\frac{1}{2}\int_{\Gamma^0}(G_0^+ + G_0^-) {B}_1^\ast \frac{\partial q^0}{\partial x_1}d\Gamma_x
= -\frac{1}{\epsilon_0}\int_{\Gamma^0}q^0(G_0^+ - G_0^-) d\Gamma_x ~~~\forall q^0\in H^1(\Gamma^0),
\label{eq:macro_first_2D}\\
&\int_{\Gamma^0}\left(
{B}_1^\ast \frac{\partial p^0}{\partial x_1}- \frac{1}{2} F^\ast (G_0^+ + G_0^-)
\right)\psi d\Gamma_x
=\frac{1}{\epsilon_0}\int_{\Gamma^0}({P}^+ - {P}^-)\psi d\Gamma_x ~~~\forall \psi \in L^2(\Gamma^0), \\
&\int_{\Gamma^0}\left(
p^0-\frac{1}{2}({P}^+ + {P}^-)
\right)\varphi d\Gamma_x =0~~~\forall \varphi\in L^2(\Gamma^0),\\
&\int_{\Omega^+} \frac{1}{\rho_0}\nabla_x P^+ \cdot \nabla_x \tilde{P} d\Omega_x
-\int_{\Omega^+}\frac{\omega^2}{K_0}P^+ \tilde{P}d\Omega_x \nonumber\\
&-\int_{\Gamma^0}G_0^+ \tilde{P} d\Gamma_x 
+\int_{\Gamma_\mathrm{in}}\frac{i k_0}{\rho_0}P^+ \tilde{P} d\Gamma_x
-\int_{\Gamma_\mathrm{in}}\frac{2 i k_0}{\rho_0}P_\mathrm{in} \tilde{P} d\Gamma_x = 0
~~~\forall \tilde{P}\in H^1(\Omega^+),\\
&\int_{\Omega^-} \frac{1}{\rho_0}\nabla_x P^- \cdot \nabla_x \tilde{P} d\Omega_x
-\int_{\Omega^-}\frac{\omega^2}{K_0}P^- \tilde{P}d\Omega_x \nonumber\\
&+ \int_{\Gamma^0}G_0^- \tilde{P} d\Gamma_x
+\int_{\Gamma_\mathrm{out 1}\cup\Gamma_\mathrm{out 2}}\frac{i k_0}{\rho_0}P^- \tilde{P} d\Gamma_x =0
~~~\forall \tilde{P}\in H^1(\Omega^-).\label{eq:macro_last_2D}
\end{align}

Based on these settings at the microscale and macroscale, 
we repeatedly conducted multiscale analysis using the homogenization method in the topology optimization procedure. 
First, the cell problems, i.e., Eqs.~(\ref{eq: local pi_2D}) and (\ref{eq: local xi_2D})\textcolor{black}{,} are solved in unit cell $Y$ to obtain the homogenized coefficients, as expressed through Eqs.~(\ref{eq: coeff A_2D})--(\ref{eq: coeff K_2D}) .
Then, these coefficients are used to solve the homogenized equations \textcolor{black}{in} Eqs.~(\ref{eq:macro_first_2D})--(\ref{eq:macro_last_2D}) defined in $\Omega^\pm$ and $\Gamma^0$.

\section{Topology optimization for acoustic metasurfaces}\label{sec: Topology optimization for acoustic metasurfaces}

\subsection{Formulation of the optimization problem}
Here, we formulate an optimization problem to obtain the structural design of the unit cell of the acoustic metasurfaces exhibiting the desired macroscopic performances.
As a typical example of the function of metasurfaces, we focused on waveguiding metasurfaces that efficiently control transmitted acoustic waves. 
We set an objective functional to minimize and maximize the amplitude of acoustic pressure on boundaries $\Gamma_\mathrm{min}$ and $\Gamma_\mathrm{max}$, respectively.
By introducing weighting factor $0 \le w \le 1$, this objective functional can be expressed as follows:
\begin{align}
J&=w J_1 - (1-w) J_2,\nonumber\\
J_1&= \frac{\int_{\Gamma_\mathrm{min}} |P^-|^2 d\Gamma_x}{\left[\int_{\Gamma_\mathrm{min}} |P^-|^2 d\Gamma_x\right]_\mathrm{init}},~~~~~
J_2=\frac{\int_{\Gamma_\mathrm{max}} |P^-|^2 d\Gamma_x}{\left[\int_{\Gamma_\mathrm{max}} |P^-|^2 d\Gamma_x\right]_\mathrm{init}},
\label{eq: objective functional}
\end{align}
where \textcolor{black}{the} subscript ``init" represents a quantity before optimization.
We assigned two outlets, $\Gamma_\mathrm{out1}$ and $\Gamma_\mathrm{out2}$, to $\Gamma_\mathrm{min}$ and $\Gamma_\mathrm{max}$ in the objective functional.
Within the framework of the homogenization method, objective functional $J$ is minimized by optimizing the material distribution in unit cell $Y$.
Then, the optimization problem is formulated as follows:
\begin{align}
\min_{\Omega}~&J\nonumber\\
\mathrm{subject~to~}&\mathrm{Governing~equations~in~}Y,\nonumber\\
&\mathrm{Governing~equations~in~}\Omega^\pm\mathrm{~and~on~}\Gamma^0,\nonumber\\
&\mathrm{Expressions~of~}({A}_{11}^\ast,{B}_1^\ast,K^{-1\ast},F^\ast).
\label{eq: optimization problem}
\end{align}
The constraint \textcolor{black}{on} the expressions of $({A}_{11}^\ast,{B}_1^\ast,K^{-1\ast},F^\ast)$ couples the microscale and macroscale problems.

\subsection{Sensitivity analysis}\label{sec:sensitivity analysis}
Sensitivity analysis was conducted based on the concept of the topological derivative, which measures the rate of change \textcolor{black}{in} objective functional $J$ when an infinitesimal circular inclusion, $\Omega_i$, characterized by its radius $\varepsilon > 0$, is inserted in the homogeneous material domain, $\Omega$.
The topological derivative is defined as follows:
\begin{align}
D_T J=\lim_{\varepsilon \to 0} \frac{J(\Omega \setminus \overline{\Omega_i})-J(\Omega)}{V(\varepsilon)},\label{eq:definition of TD_original_main}
\end{align}
where $V(\varepsilon)$ is a function of radius $\varepsilon$, 
and in \textcolor{black}{this} case, it was set to $V(\varepsilon)=-\pi \varepsilon^2$, as in \cite{carpio2008solving,carpio2008topological}.
To derive the topological derivative, we applied the topological-shape-sensitivity method proposed by Novotny et al. \cite{novotny2003topological} and Feij\'oo et al. \cite{feijoo2003topological}.
This method is based on the relationship between the topological derivative and limit form of the shape derivative.
Therefore, we first derive the shape derivative for objective functional $J$ and calculate its limit when $\varepsilon\to 0$ \textcolor{black}{in order} to derive the topological derivative.
\textcolor{black}{Details} of this procedure are summarized in \ref{sec:Appendix sensitivity analysis}.

The expression of the topological derivative to $J$ in Eq.~(\ref{eq: objective functional}) is derived as follows: 
\begin{align}
D_T J &= -\frac{1}{2\pi}\sum_{i=1}^4 v_{B i}(\bm{u}_{macro},\bm{v}_{macro}) I_i(\bm{u}_{micro}),
\end{align}
where $\bm{v}_B=(\lambda_{{A}_{11}^\ast},\lambda_{{B}_{1}^\ast},\lambda_{{F}^\ast},\lambda_{{K}^{-1 \ast}})$
are the Lagrange multipliers for the homogenized coefficients $({A}_{11}^\ast,{B}_1^\ast,K^{-1\ast},F^\ast)$ depending on the state variables in the macroscale $\bm{u}_{macro}=(p^0,P^\pm,G_0^\pm)$ and corresponding adjoint variables in $\bm{v}_{macro}=(q^0,Q^\pm,\Psi_0^\pm)$.
$I_i~(i=1,...,4)$ represents the functions of the state variables in the microscale $\bm{u}_{micro}=(\textcolor{black}{\eta},\xi)$.
The explicit formulas of $\bm{v_B}$ and $I_i$ and the adjoint equations for $\bm{v}_{macro}$ are summarized in \ref{sec:Appendix sensitivity analysis}.

\section{Level set-based topology optimization}\label{sec: Level set-based topology optimization}
To optimize the material distribution in unit cell $Y$, we used the level-set-based topology optimization method proposed by Yamada et al. \cite{yamada2010topology}.
In this method, the level set function representing the shape and topology of the optimizing structure is updated using a reaction--diffusion equation based on the topological derivative.

As explained earlier, the fixed design domain, $D$, comprises two regions: air-filled domain $\Omega_\mathrm{air}$ and elastic domain $\Omega_\mathrm{elastic}$.
These regions and their interfaces, $\Gamma$, are represented by the following level set function, $\phi$:
\begin{eqnarray}
\left\{
\begin{array}{ll}
0<\phi(\bm{y})\le 1 &\mathrm{if}~~\bm{y}\in \Omega_\mathrm{elastic}\\
\phi(\bm{y})= 0 &\mathrm{if}~~\bm{y}\in \Gamma\\
-1\le \phi(\bm{y})< 0 &\mathrm{if}~~\bm{y}\in \Omega_\mathrm{air}.\label{eq:profile of LSF}
\end{array}
\right. 
\end{eqnarray}
This level set function is different from a signed distance function that is usually used in a shape-optimization method \cite{allaire2004structural}.
The upper and lower limits of $\phi$, $1$, and $-1$ allow for the regularization of the optimization problem, as explained later.

The optimization problem to minimize objective functional $J$ by optimizing the material distribution in $D$ is formulated as
\begin{align}
\inf_{\chi_\phi}~~~&J,
\end{align}
where $\chi_\phi$ is the characteristic function in $D$ defined using the level-set function as
\begin{eqnarray}
\chi_\phi=
\left\{
\begin{array}{ll}
1 &\mathrm{if}~~\phi\ge 0 \\
0 &\mathrm{if}~~\phi< 0.
\end{array}
\right. 
\end{eqnarray}
\textcolor{black}{To elucidate the distribution of the level set function that minimizes the objective functional $J$, we introduce a fictitious time and replace the optimization problem with a time-evolution problem.}
Let $t$ denote \textcolor{black}{the} fictitious time used in the optimization.
\textcolor{black}{Let} a partial derivative of the level set function with respect to time $\frac{\partial \phi}{\partial t}$ \textcolor{black}{be} proportional to \textcolor{black}{the design sensitivity} $J'$, \textcolor{black}{which measures the rate of change in $J$ when the structural design of the metasurface is altered slightly. Thus, the time-evolution equation is expressed as}
\begin{align}
\frac{\partial \phi}{\partial t}=-K_\phi {J}',
\end{align}
where $K_\phi>0$ is a positive constant. To regularize the \textcolor{black}{abovementioned} optimization problem, the following regularization term is introduced:
\begin{align}
\frac{\partial \phi}{\partial t}=-K_\phi ({J}'-\tau \nabla_y^2 \phi),\label{eq:reaction-diffusion eq}
\end{align}
where $\tau >0$ controls the strength of the regularization. 
Eq.~(\ref{eq:reaction-diffusion eq}) is a reaction--diffusion equation with the diffusion and reaction terms. The reaction term corresponds to \textcolor{black}{the design sensitivity} $J'$, \textcolor{black}{while} the diffusion term ensures the smoothness of the level set function.
Smoother distributions of the level set function can be obtained with larger values of $\tau$, and the optimization problem can be regularized without disturbing the minimization of the objective functional by choosing an appropriate value of $\tau$.

\textcolor{black}{To obtain the optimized design of the metasurface, } this reaction--diffusion equation is solved in $D \subset Y$. 
As the metasurface is composed of a periodic array of the unit cells, we impose the periodic boundary conditions for $\phi$ on $\Gamma_{1,2}$.
By setting an appropriate initial condition, the system for $\phi$ can be summarized as follows:
\begin{eqnarray}
\left\{
\begin{array}{ll}
\cfrac{\partial \phi}{\partial t}=-K_\phi ({J}'-\tau \nabla_y^2 \phi) &\mathrm{in}~~D, \\
\bm{n}\cdot\nabla_y \phi =0 &\mathrm{on}~~\partial D \setminus (\Gamma_{1}\cup\Gamma_{2}),\\
\mathrm{Periodic~boundary~conditions} &\mathrm{on~} \Gamma_{1,2}, \label{eq:reaction-diffusion equation with B.C}\\
\phi(\bm{y},t=0)=\phi_0(\bm{y}).
\end{array}
\right. 
\end{eqnarray}
For simplicity, we imposed the Neumann boundary condition on the boundaries of $D$ except for $ \Gamma_{1,2}$; other boundary conditions can also be applied.
The fourth line shows the initial condition, at which the initial level-set function, $\phi_0(\bm{y})$, represents the initial configuration.

\textcolor{black}{The design sensitivity} $J'$ is \textcolor{black}{related to} the topological derivative, $D_T J$. 
According to the definition of $D_T J$ expressed in Eq.~(\ref{eq:definition of TD_original_main}) and the form of the reaction--diffusion equation, $J'$ can be written in terms of the topological derivative as follows:
\begin{eqnarray}
&{J}'&=
\left\{
\begin{array}{ll}
-D_T J^{\mathrm{air}\to\mathrm{elastic}} &\mathrm{if}~~\bm{y}\in \Omega_\mathrm{air}\\
D_T J^{\mathrm{elastic}\to\mathrm{air}} &\mathrm{if}~~\bm{y}\in \Omega_\mathrm{elastic}\,,
\end{array}
\right.
\end{eqnarray}
where $D_T J^{\mathrm{air}\to\mathrm{elastic}}$ is the topological derivative when an infinitesimal inclusion domain with the elastic medium appears in $\Omega_\mathrm{air}$, while $D_T J^{\mathrm{elastic}\to\mathrm{air}}$ represents the inverse case. \textcolor{black}{Details regarding} $D_T J^{\mathrm{air}\to\mathrm{elastic}}$ and $D_T J^{\mathrm{elastic}\to\mathrm{air}}$ are provided in \ref{sec:Appendix sensitivity analysis}.

\section{Numerical implementation}\label{sec:Numerical implementation}

\subsection{Optimization \textcolor{black}{process}}
This section provides a brief explanation of the \textcolor{black}{optimization process}.
First, the level set function is initialized, and the state problem is solved based on the homogenization method. As explained, the state problem is composed of the problems defined at the microscale and macroscale. At the microscale, the cell problems for the state variables, $\bm{u}_{micro}=(\textcolor{black}{\eta},\xi)$, are solved to obtain the homogenized coefficients of $({A}_{11}^\ast,{B}_1^\ast,K^{-1\ast},F^\ast)$. \textcolor{black}{During} this step, a remeshing process is applied 
to reduce numerical errors when solving the cell problems. This procedure is detailed in the next section.
Then, the macroscale state variables, $\bm{u}_{macro}=(p^0, P^\pm, G_0^\pm)$, are obtained by solving the homogenized equations. 
Next, objective function $J$ is evaluated using the macroscale solutions. 
If the objective function is converged, the process ends; otherwise,
 adjoint variables $\bm{v}_{macro}=(q^0,Q^\pm, \Psi_0^\pm)$ at the macroscale are computed, and then Lagrange multipliers $\bm{v}_B=(\lambda_{{A}_{11}^\ast},\lambda_{{B}_{1}^\ast}, \lambda_{{F}^\ast}, \lambda_{{K}^{-1 \ast}})$ are evaluated.
The state and adjoint variables are then used to compute topological derivative $D_T J$.
Based on the distribution of $D_T J$, the level-set function is updated using the reaction--diffusion equation, Eq.~(\ref{eq:reaction-diffusion equation with B.C}). The optimization routine then returns to the step of obtaining the state variables. These steps are repeated until the objective function is converged.

\textcolor{black}{As a convergence criterion, we introduce the 10-iteration moving average of the relative error between the values of $J$ for two adjacent iterations. If this value becomes sufficiently small after the optimization reaches a certain iteration, the optimization calculation is considered to have converged. Corresponding details are explained in Section~\ref{sec: Optimization results}.}

\subsection{FEM-based discretization of microscale and macroscale problems}
To obtain the state and adjoint variables at the microscale and macroscale, the governing and adjoint equations \textcolor{black}{need to} be discretized. In this research, we introduced a finite element program implemented by the open-source PDE solver, FreeFEM \cite{MR3043640}.

At the macroscale, we used the piecewise linear-continuous finite element for $(G_0^\pm,\Psi_0^\pm)$, whereas the piecewise quadratic-continuous finite element was used for $(p^0,P^\pm,q_0,Q^\pm)$.
These different choices of finite elements are inspired by the functional spaces, to which the state and adjoint variables belong. 

At the microscale, we used the piecewise quadratic-continuous finite element for $(\textcolor{black}{\eta},\xi)$.
As mentioned earlier, the design domain comprises two material domains\textcolor{black}{:} $\Omega_\mathrm{air}$ and $\Omega_\mathrm{elastic}$\textcolor{black}{. If} element division is not performed along these interfaces, numerical errors tend to occur in solution $(\textcolor{black}{\eta},\xi)$, \textcolor{black}{which} makes the optimization unstable. To avoid this issue, design domain $D$ was remeshed \textcolor{black}{such} that elements are fitted to their interfaces, $\Gamma$, which is represented by the level-set function, $\phi(\bm{y})=0$, when solving the cell problems in each iteration of the optimization. The implementation of this remeshing process is based on the open-source platform Mmg, whose algorithm is based on \cite{dapogny2014three}.

\section{Numerical examples}\label{sec: numerical examples}
\subsection{Validation of the homogenization method}\label{sec: Validation of the HMM}
\begin{figure}[H]
	\centering
	\includegraphics[scale=0.3]{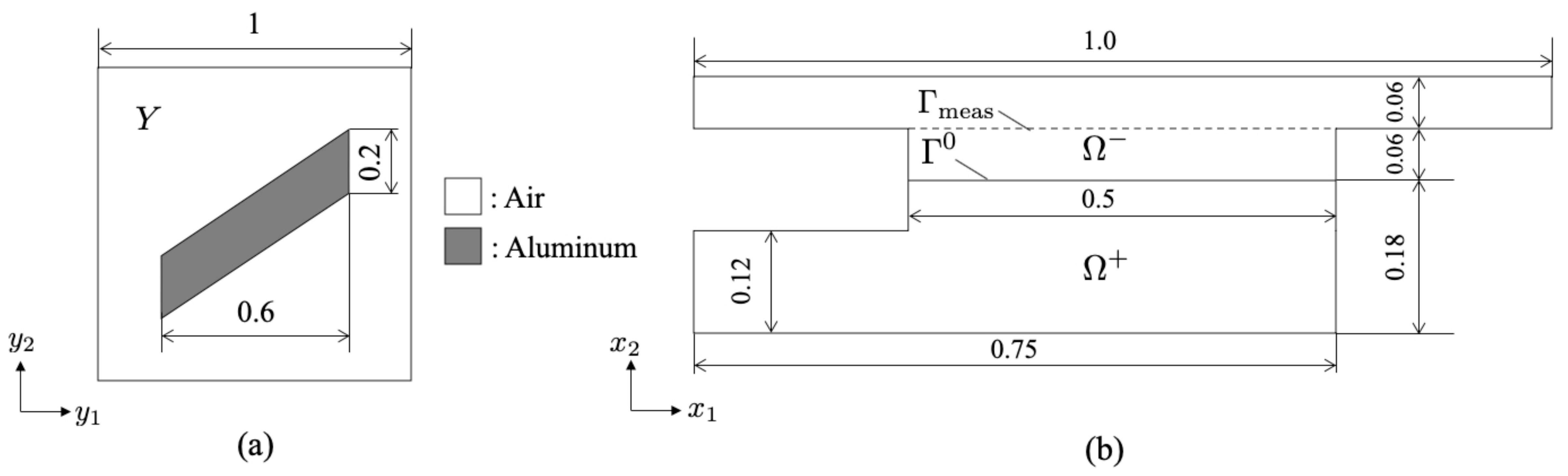}
	\caption{Settings of computational domains at the 
		(a) microscale and (b) macroscale. The \textcolor{black}{dimensions} in (b) are in [m].}
	\label{fig:dimensions}       
\end{figure}
\textcolor{black}{In this section, we provide a numerical example that supports the validity of the proposed homogenization method.}
Figure~\ref{fig:dimensions} shows the settings of the computational domains used in the multiscale analysis based on the homogenization method.
Unit cell $Y$ contains a parallelogram domain comprising aluminum surrounded by an air-filled region, as shown in Fig.~\ref{fig:dimensions}(a).
The mass density and bulk modulus of air are $1.2~\textcolor{black}{\mathrm{[kg~m^{-3}]}}$ and $\textcolor{black}{1.42}\times 10^{5}\mathrm{[Pa]}$, respectively, \textcolor{black}{whereas} those of aluminum are $2643~\textcolor{black}{\mathrm{[kg~m^{-3}]}}$ and $6.87\times 10^{10}\mathrm{[Pa]}$, respectively.
The finite size of unit cell $\epsilon_0$ used in the macroscopic equations was set to $0.01$\textcolor{black}{,} and its thickness $\delta$ was set to $0.01~\textcolor{black}{\mathrm{[m]}}$, i.e., $\kappa=1$.

\begin{figure}[H]
	\centering
	\includegraphics[scale=0.25]{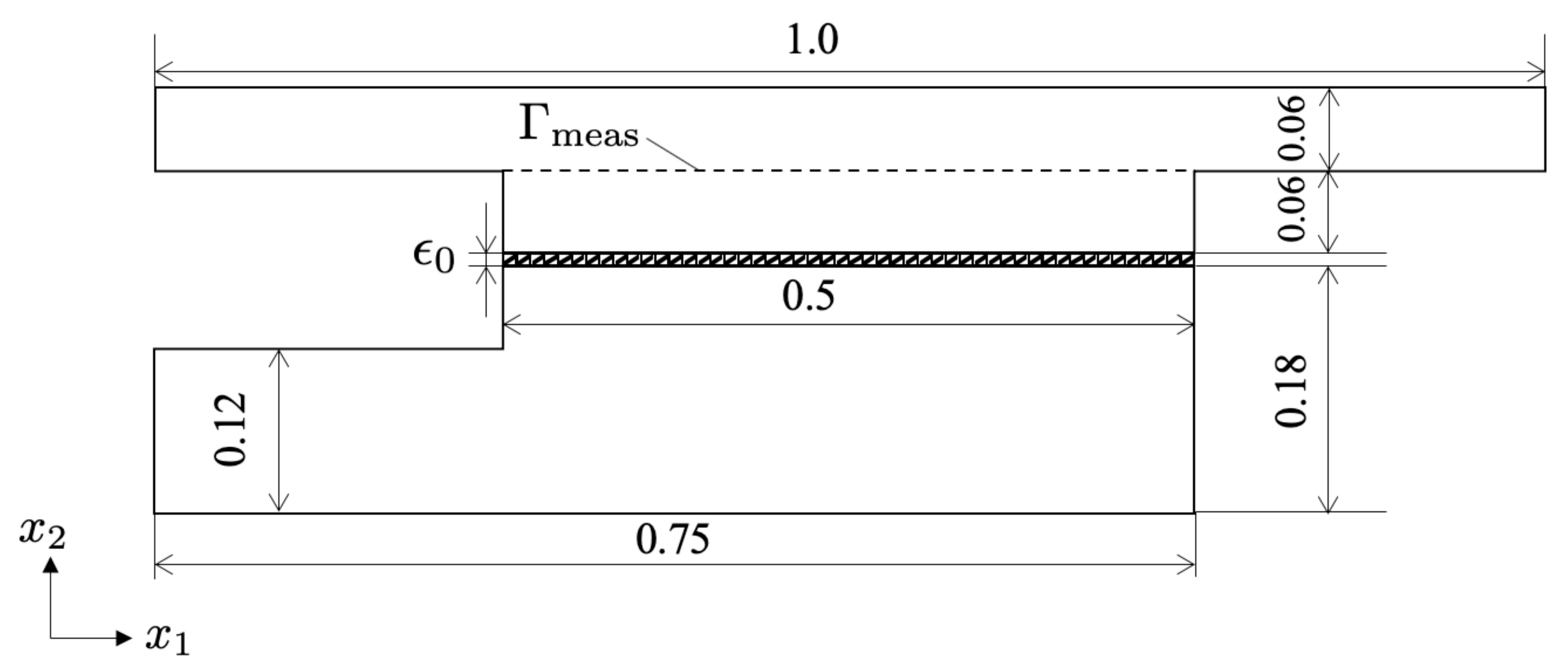}
	\caption{Settings of computational domains used in 
		the \textcolor{black}{conventional} FEM. \textcolor{black}{Dimensions} are in [m].}
	\label{fig:Geom_usualFEM}       
\end{figure}

\begin{figure}[H]
	\centering
	\includegraphics[scale=0.4]{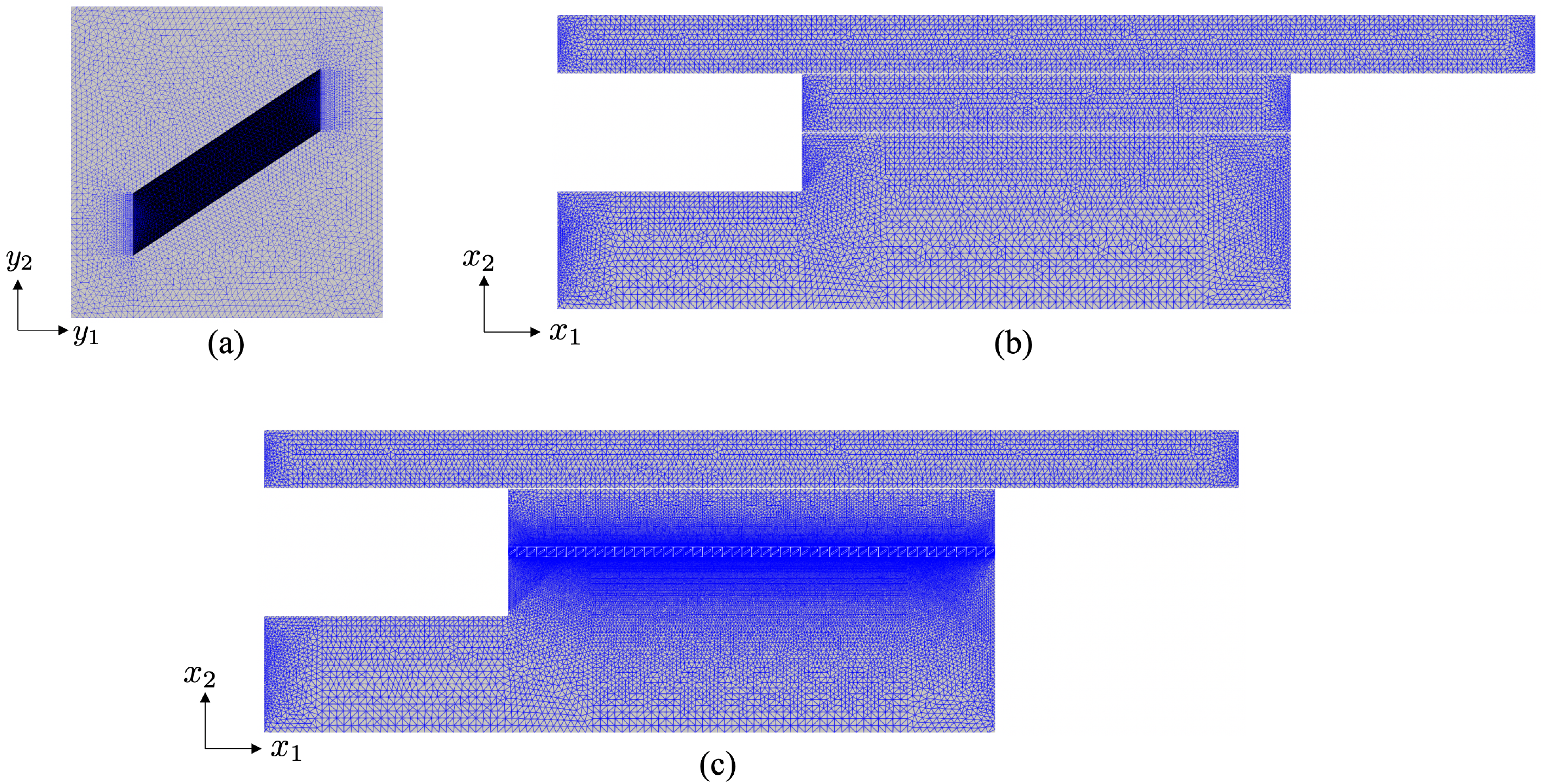}
	\caption{\textcolor{black}{Finite element discretization for (a) the microscale analysis, (b) macroscale analysis with homogenization, and (c) reference analysis.
	}}
	\label{fig:FE_discretization_section71}       
\end{figure}

For \textcolor{black}{a} comparison of the solution obtained by the homogenization method, we used the solution obtained when the \textcolor{black}{entire} system of the metasurface, including the array of unit cells, is solved using the FEM without the homogenization method. Hereafter, this solution is called the reference solution. 
Figure~\ref{fig:Geom_usualFEM} shows the settings of the computational domains for obtaining the reference solution, which corresponds to Fig.~\ref{fig:dimensions}. 
The shape and material distributions in $Y$ are the same as those described in Fig.~\ref{fig:dimensions}(a); however, they are embedded in the model with the finite value of the spacing, $\epsilon_0=0.01~\textcolor{black}{\mathrm{[m]}}$. 
We first compared the frequency responses of acoustic pressures obtained \textcolor{black}{via} the two aforementioned methods.
Let $p^{Ref}$ represent the reference solution of acoustic pressure. The following quantities were compared in a certain range of frequencies:
\begin{align}
h^{H}(\omega) = \int_{\Gamma_\mathrm{meas}}|P^-(\omega)|^2 d\Gamma_x,\nonumber\\
h^{Ref}(\omega) = \int_{\Gamma_\mathrm{meas}}|p^{Ref}(\omega)|^2 d\Gamma,\nonumber
\end{align}
where boundary $\Gamma_\mathrm{meas}$ is defined as shown in Figs.~\ref{fig:dimensions}(b) and ~\ref{fig:Geom_usualFEM}.
The wavenumber of incident wave $k_0$ was set to $5\le k_0 \le 60~\mathrm{[m^{-1}]}$, corresponding to the range of frequencies, $274 \le \frac{\omega}{2\pi} \le 3285~\mathrm{[Hz]}$.
The amplitude of the incident wave was set to $P_\mathrm{in}=1~\mathrm{[Pa]}$ on $\Gamma_\mathrm{in}$.
\textcolor{black}{Figure~\ref{fig:FE_discretization_section71} presents the finite element discretization for the homogenization method and the reference analysis.}
We used 10,212 triangular elements for discretizing the unit cell, as shown in \textcolor{black}{Fig.~\ref{fig:FE_discretization_section71}(a)}, and \textcolor{black}{16,705} triangular elements for discretizing the macroscopic model shown in \textcolor{black}{Fig.~\ref{fig:FE_discretization_section71}(b)}. Based on these settings, \textcolor{black}{an} analysis of the cell problems revealed that the unit cell in Fig.~\ref{fig:dimensions}(a) is characterized by the homogenized coefficients, $({A}_{11}^\ast,{B}_1^\ast,K^{-1\ast},F^\ast)$\\$=\textcolor{black}{(0.567~\textcolor{black}{\mathrm{[m^3~kg^{-1}]}}, 0.260, 6.20\times 10^{-6}~\textcolor{black}{\mathrm{[Pa^{-1}]}}, 1.88~\textcolor{black}{\mathrm{[kg~m^{-3}]}})}$.
For obtaining the reference solution, \textcolor{black}{50 unit cells with the discretization shown in Fig.~\ref{fig:FE_discretization_section71}(a) are periodically arrayed over the transmission layer, while 709,665} triangular elements were used to discretize the \textcolor{black}{entire} system, as depicted in \textcolor{black}{Fig.~\ref{fig:FE_discretization_section71}(c)}.

\begin{figure}[H]
	\centering
	\includegraphics[scale=0.6]{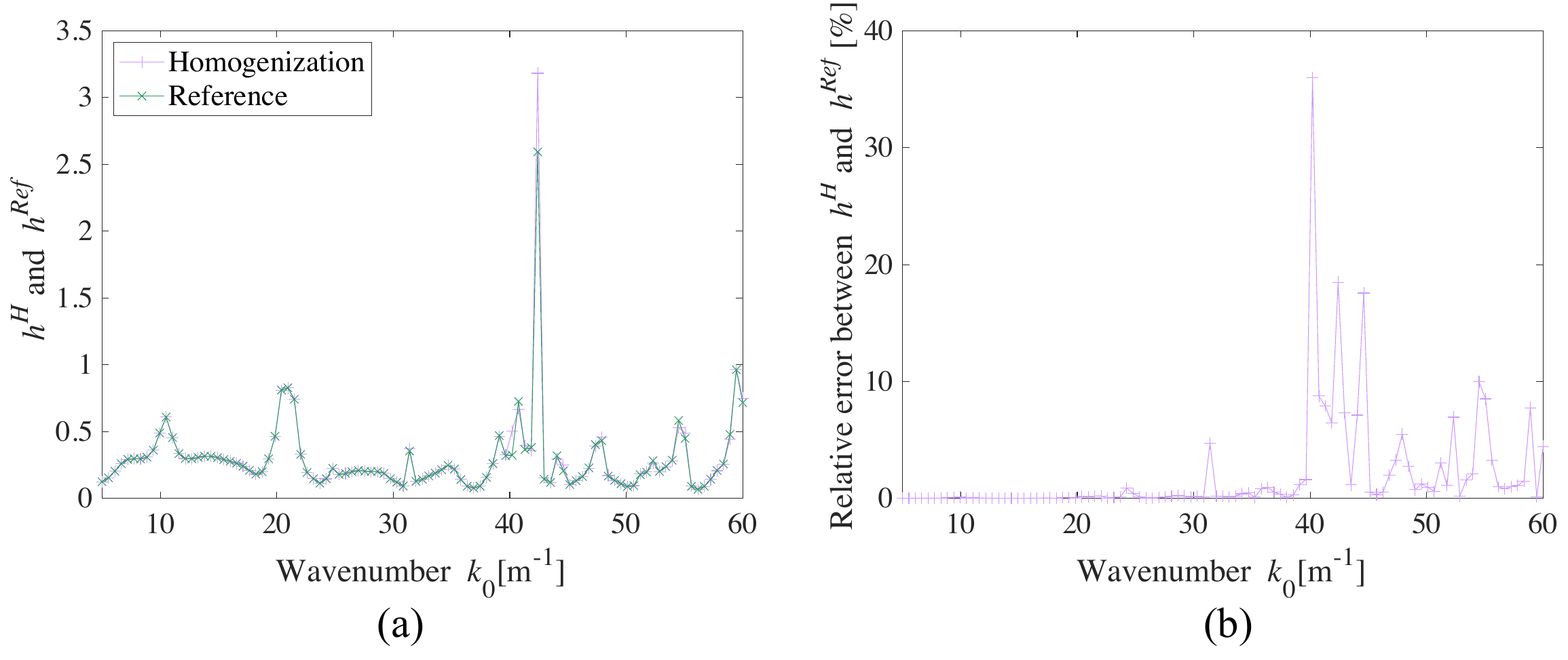}
	\caption{\textcolor{black}{(a) Frequency responses of $h^H$ and $h^{Ref}$. (b) Frequency responses of the relative error between $h^H$ and $h^{Ref}$.}}
	\label{fig:FreqResponse}       
\end{figure}

\textcolor{black}{Figure~\ref{fig:FreqResponse} displays the frequency responses of $h^H$ and $h^{Ref}$ and also those of the relative error between $h^H$ and $h^{Ref}$.}
\textcolor{black}{As shown in Fig.~\ref{fig:FreqResponse}(a), good congruence can be observed, except for the resonance frequency, especially around $k_0 = 42~\textcolor{black}{\mathrm{[m^{-1}]}}$. Figure~\ref{fig:FreqResponse}(b) also indicates that the proposed homogenization method can express the system of the metasurface with small errors at non-resonant frequencies. 
}

\begin{figure}[H]
	\centering
	\includegraphics[scale=0.5]{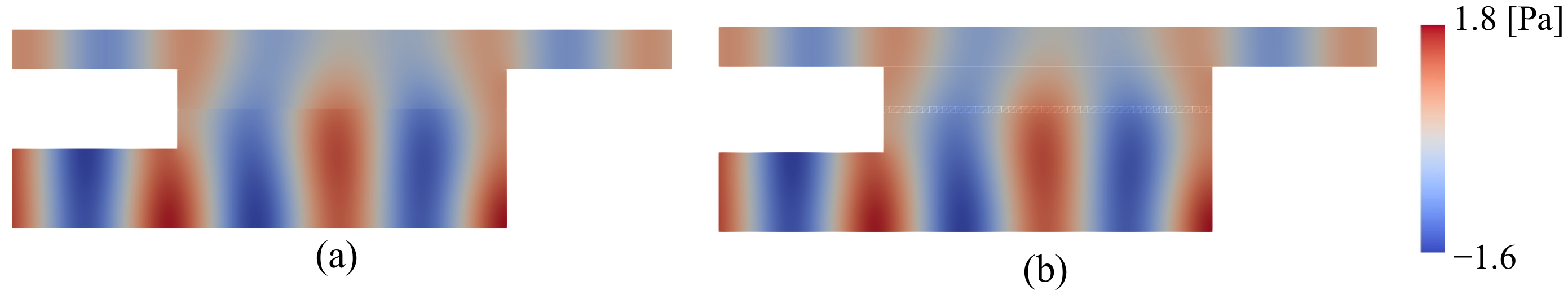}
	\caption{\textcolor{black}{Distribution of the real part of acoustic pressure at $k_0 = 25~\textcolor{black}{\mathrm{[m^{-1}]}}$ for (a) $\mathrm{Re}(P^\pm)$ and (b) $\mathrm{Re}(p^{Ref})$.} }
	\label{fig:Comparison}       
\end{figure}

\begin{figure}[H]
	\centering
	\includegraphics[scale=0.6]{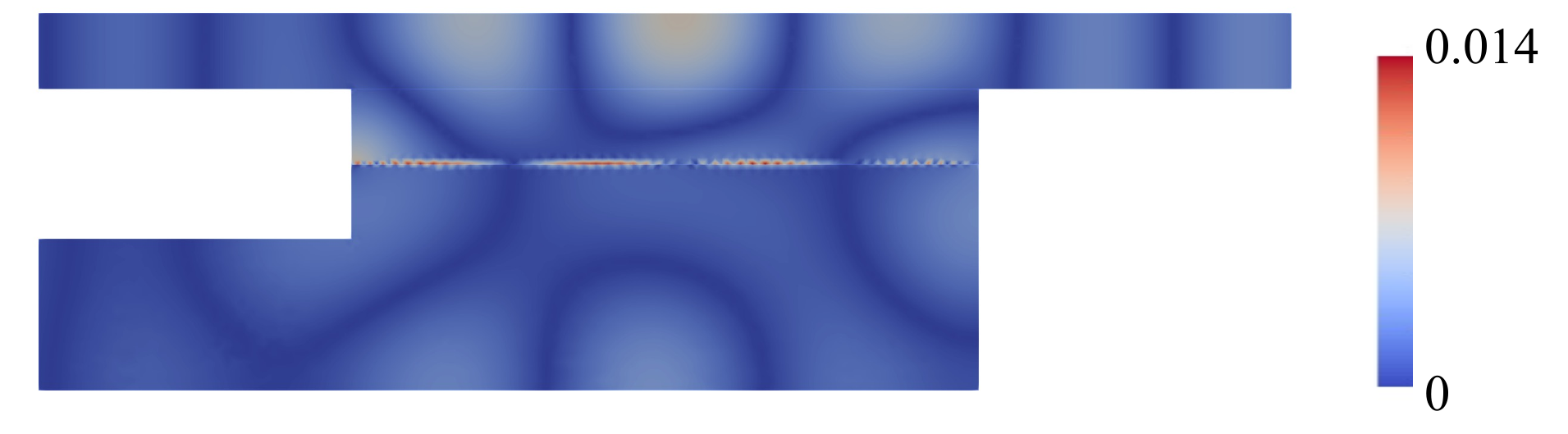}
	\caption{\textcolor{black}{Distribution of $e^\pm$ at $k_0 = 25~\textcolor{black}{\mathrm{[m^{-1}]}}$.}}
	\label{fig:Diff}       
\end{figure}

To show the validity at a \textcolor{black}{non-resonant} frequency, we compared the spatial distribution of acoustic pressures, $P^\pm$, obtained through the homogenization method and the reference solution, $p^{Ref}$.
In the analysis based on the homogenization method, we used 10,212 and \textcolor{black}{16,705} triangular elements for the microscale and macroscale problems, respectively.
In addition, \textcolor{black}{709,665} triangular elements were used for the reference solution.
Figure~\ref{fig:Comparison} shows the distributions of $P^\pm$ and $p^{Ref}$ at $k_0=25~\textcolor{black}{\mathrm{[m^{-1}]}}$.
As shown, the two solutions have similar distributions.
For \textcolor{black}{a} more precise verification, an error function is defined as follows:
\begin{align}
e^\pm(\bm{x}) = \frac{|\mathrm{Re}(P^\pm) - \mathrm{Re}(p^{Ref})|}{\mathrm{mean}(|\mathrm{Re}(P^\pm)|)},\nonumber
\end{align}
where the denominator represents the average values of $|\mathrm{Re}(P^\pm)|$ in domain $\Omega^\pm$ defined as
\begin{align}
\mathrm{mean}(|\mathrm{Re}(P^\pm)|)=\frac{\int_{\Omega^\pm}|\mathrm{Re}(P^\pm)| d\Omega_x }{|\Omega^\pm|}.\nonumber
\end{align}
Figure~\ref{fig:Diff} represents the distribution of $e^\pm$.
Although large values of $e^\pm$ can be found around $\Gamma^0$ and at the \textcolor{black}{corners} of the geometries,
they are less than \textcolor{black}{1.4\%}, and \textcolor{black}{this supports the validity of the proposed homogenization method.}

\textcolor{black}{The proposed homogenization method assumes periodicity in the microscale problems, but periodicity is not assumed in the macroscale problem. 
If this discrepancy was significant, errors would have been observed in the solution of the homogenization method, owing to the rigid side walls at both ends of the transmission layer. However, such errors were not confirmed, as indicated by Fig.~\ref{fig:Diff}.
This is because the unit cell size and the widths of the region where the unit cell is in contact with the outer boundaries are considerably smaller than the wavelength of the acoustic waves, and the structure of the metasurface can be replaced by a homogeneous material. 
In \cite{semin2018homogenization}, this type of discrepancy caused by the finite length of a periodic structure was examined by using the method of matched asymptotic expansions. As this error appears to be small for the practical use of the proposed model, we neglect this point.
}


\subsection{Optimization results}\label{sec: Optimization results}
Two numerical cases are solved to demonstrate the validity of the proposed optimization method. 
In case 1, the amplitude of acoustic pressure on $\Gamma_\mathrm{out 2}$ is minimized, \textcolor{black}{whereas} that on $\Gamma_\mathrm{out 1}$ is maximized. Here, we assign $\Gamma_\mathrm{out 2}$ to $\Gamma_\mathrm{min}$ and $\Gamma_\mathrm{out 1}$ to $\Gamma_\mathrm{max}$ in the objective functional expressed in Eq.~(\ref{eq: objective functional}).
Case 2 is the inverse of case 1, that is,
we assign $\Gamma_\mathrm{out 1}$ to $\Gamma_\mathrm{min}$ and $\Gamma_\mathrm{out 2}$ to $\Gamma_\mathrm{max}$.
\textcolor{black}{The weighting} factor $w$ in the objective functional was fixed at $w=0.5$ \textcolor{black}{for} both cases.
The computational domains in the macroscale and material properties in $D$ are set at the same values as those in section \ref{sec: Validation of the HMM}. 
\textcolor{black}{To represent the macroscale system, 17,251 triangular elements are used, similar to the example shown in Fig.~\ref{fig:FE_discretization_section71}(b). However, approximately 20,000 elements are used for the microscale system. Details regarding the finite element discretization in the microscale are explained in Section~\ref{sec: Mesh dependency}.
}
The wavenumber of the incident wave was set to $k_0=25~\textcolor{black}{\mathrm{[m^{-1}]}}$ corresponding to $1369~\mathrm{[Hz]}$ and the corresponding wave amplitude was set to $P_\mathrm{in}=1~\mathrm{[Pa]}$.

\begin{figure}[H]
	\centering
	\includegraphics[scale=0.5]{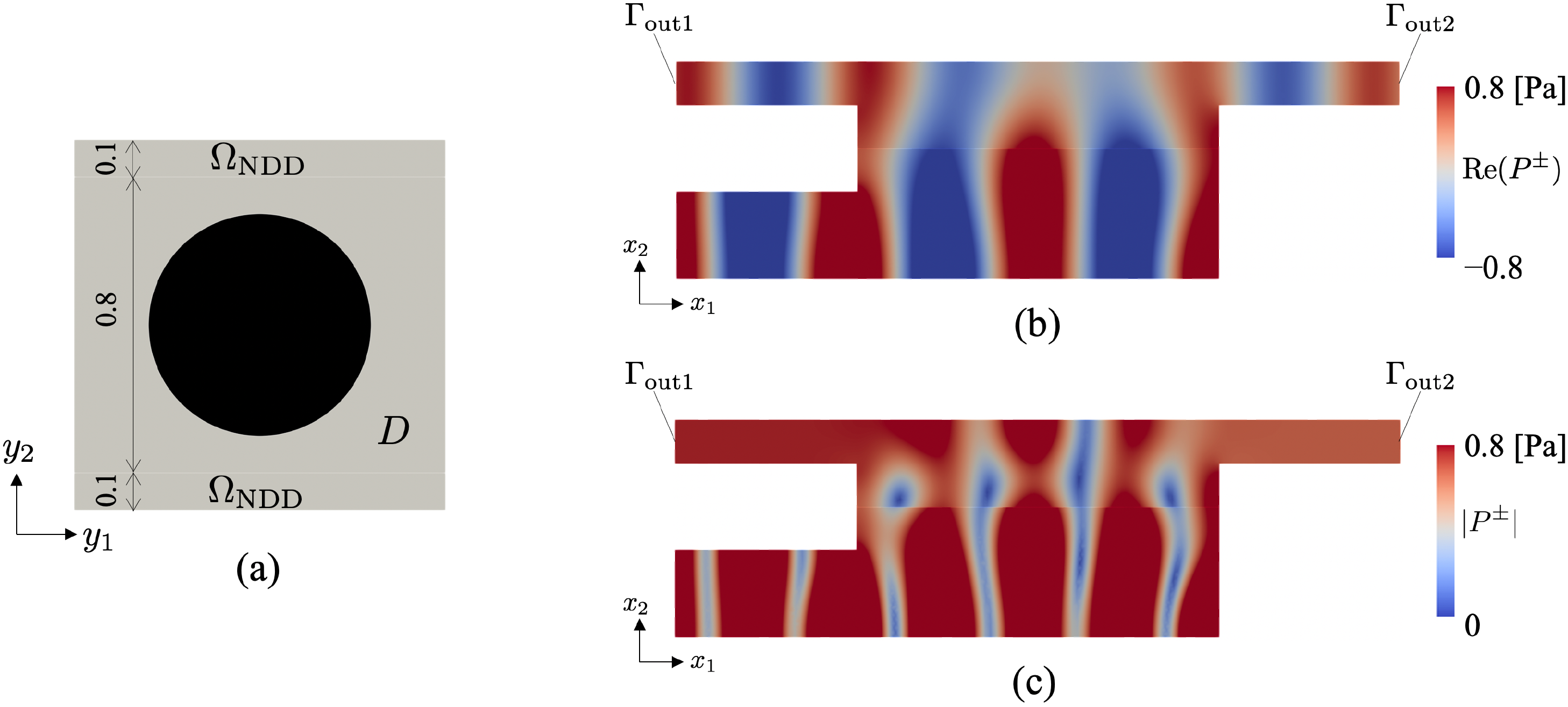}
	\caption{\textcolor{black}{(a) Computational domains and the initial configuration at the microscale. 
		(b) Distribution of the real part of acoustic pressure, $\mathrm{Re}(P^\pm)$. 
		(c) Distribution of the absolute value of acoustic pressure, $|P^\pm|$.}
	}
	\label{fig:Init_config}       
\end{figure}

Figure~\ref{fig:Init_config}(a) shows the settings of the computational domains at the microscale and the initial configuration in $D$ for both optimization cases.
The black-colored domain represents $\Omega_\mathrm{elastic}$ comprising aluminum, \textcolor{black}{whereas} the gray-colored domain represents $\Omega_\mathrm{air}$. 
A circular elastic domain was selected as an initial configuration, with a radius of $\textcolor{black}{0.3}$ in the $\bm{y}$ coordinate. The other dimensions are listed in the figure.
By solving the cell problems for the initial configuration at the microscale, the homogenized coefficients are evaluated as \textcolor{black}{$({A}_{11}^\ast,{B}_1^\ast,K^{-1\ast},F^\ast)=(0.466~\textcolor{black}{\mathrm{[m^3~kg^{-1}]}}, 3.20\times 10^{-9}, 5.05\times 10^{-6}~\textcolor{black}{\mathrm{[Pa^{-1}]}}, 2.18~\textcolor{black}{\mathrm{[kg~m^{-3}]}})$}.
The acoustic-wave propagation behavior in the macroscale is obtained using these values in the homogenized equations.
Figure~\ref{fig:Init_config}(b) shows the distribution of the real part of acoustic pressure at the initial configuration.
To \textcolor{black}{present} the optimization results clearly, the upper and lower limits of the color bar of the contour diagram are fixed to $0.8$ and $-0.8\mathrm{[Pa]}$, respectively\textcolor{black}{, in order} to emphasize the transmitted waves in the upper region of $\Omega^-$. 
\textcolor{black}{Similarly, Fig.~\ref{fig:Init_config}(c) shows the distribution of the absolute value of acoustic pressure. It is observed that $|P^-|$ around $\Gamma_{\mathrm{out1}}$ exceeds that around $\Gamma_{\mathrm{out2}}$.}
\textcolor{black}{Moreover,} the squared norm of the acoustic pressure on $\Gamma_{\mathrm{out1}}$ and $\Gamma_{\mathrm{out2}}$ is \textcolor{black}{$3.59\times 10^{-2}~[\mathrm{Pa}^2 ~ \mathrm{m}]$ and $2.70\times 10^{-2}~[\mathrm{Pa}^2~  \mathrm{m}]$}, respectively.
As in the definition of the objective functional, i.e., Eq.~(\ref{eq: objective functional}), these values were used for the normalization of $J$. According to this definition, the initial value of $J$ is $0$ with $J_1 = J_2 =1$.

\begin{figure}[H]
	\centering
	\includegraphics[scale=0.55]{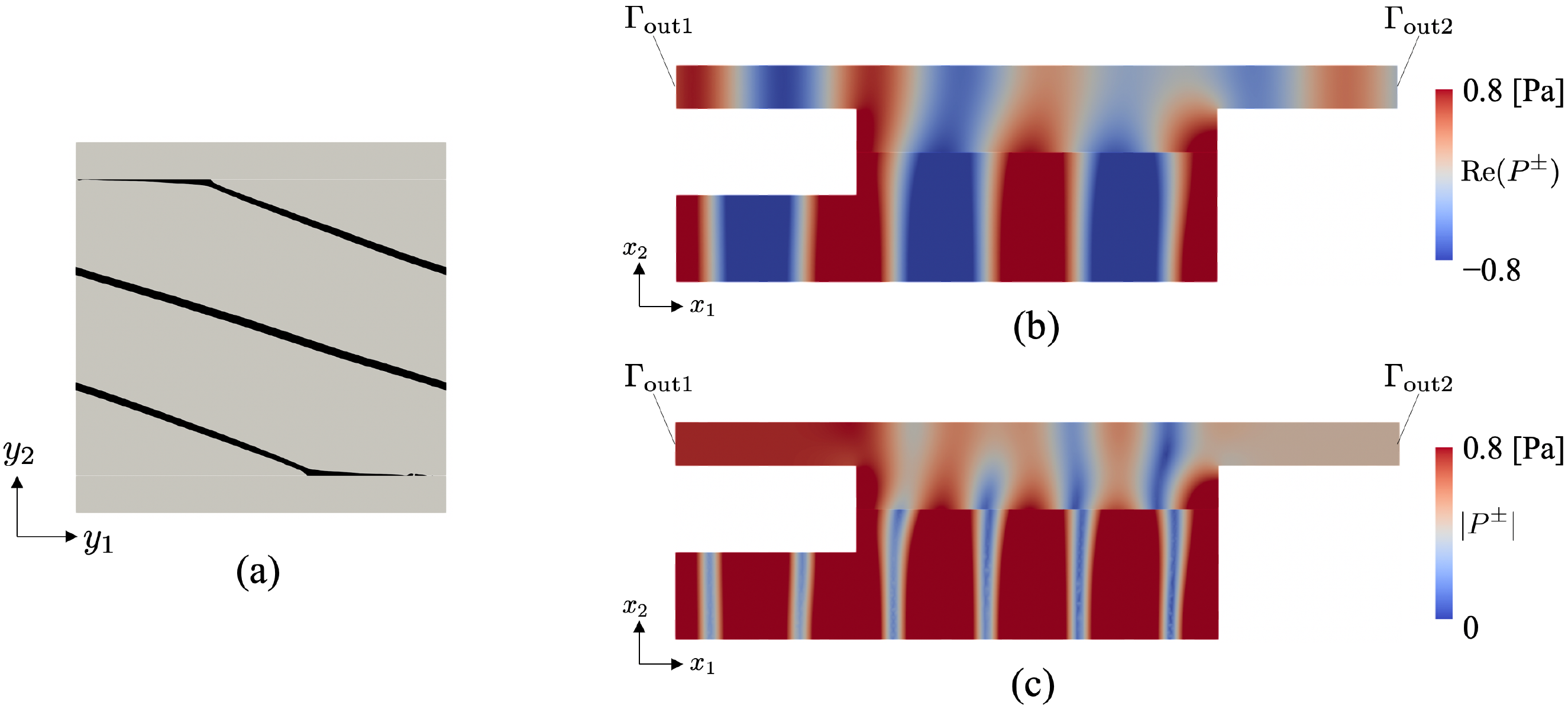}
	\caption{\textcolor{black}{
	(a) Optimized configuration for case 1. 
	(b) Distribution of the real part of acoustic pressure, $\mathrm{Re}(P^\pm)$.
	(c) Distribution of the absolute value of acoustic pressure, $|P^\pm|$.	}
}
	\label{fig:Opt_left}       
\end{figure}

\begin{figure}[H]
	\centering
	\includegraphics[scale=0.5]{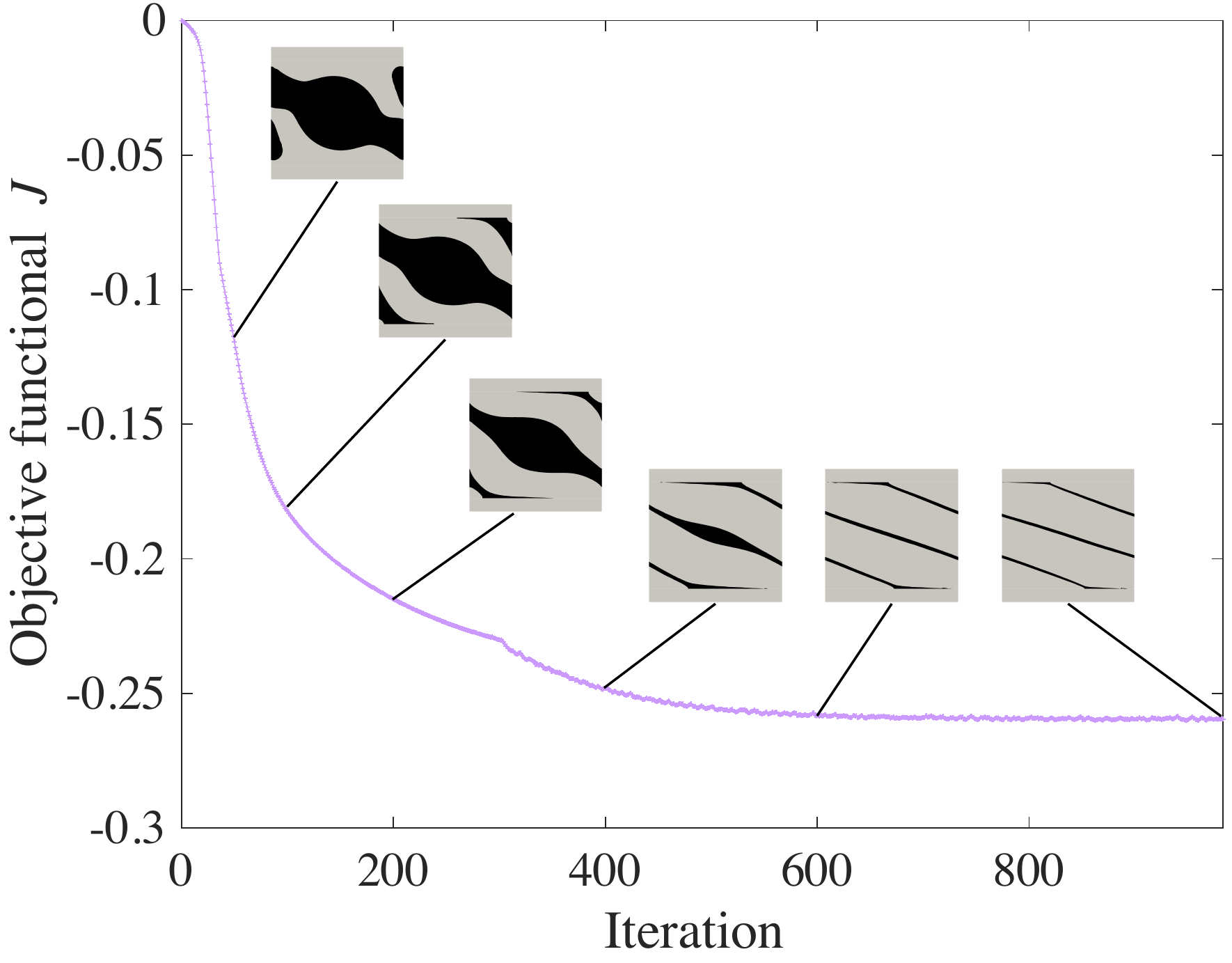}
	\caption{\textcolor{black}{History of objective functional $J$ with the intermediate and optimized designs for case 1.}}
	\label{fig:history_case1}       
\end{figure}

Figure~\ref{fig:Opt_left} represents the optimization results for case 1.
The \textcolor{black}{optimized} configuration is shown in Fig.~\ref{fig:Opt_left}(a), and it is characterized by the striped structures tilted \textcolor{black}{toward} the left. 
\textcolor{black}{The finite element discretization of the obtained design involves 26,264 elements, the details of which are explained in Section~\ref{sec: Mesh dependency}.}
This structure exhibits homogenized coefficients of $({A}_{11}^\ast,{B}_1^\ast,K^{-1\ast},F^\ast)=$\\$\textcolor{black}{(0.273~\textcolor{black}{\mathrm{[m^3~kg^{-1}]}}, -2.10, 6.66\times 10^{-6}~\textcolor{black}{\mathrm{[Pa^{-1}]}}, 10.14~\textcolor{black}{\mathrm{[kg~m^{-3}]}})}$.
By using these values, the macroscopic acoustic-pressure distribution is obtained as shown in Fig.~\ref{fig:Opt_left}(b).
The objective functional is calculated as $J = \textcolor{black}{-0.260}$ with $J_1 = \textcolor{black}{0.456}$ and $J_2 = \textcolor{black}{0.975}$.
\textcolor{black}{In other words}, the optimized result was obtained \textcolor{black}{such} that the amplitude of \textcolor{black}{the} acoustic pressure on $\Gamma_{\mathrm{out2}}$ was reduced\textcolor{black}{, whereas} that \textcolor{black}{of the pressure} on $\Gamma_{\mathrm{out1}}$ was retained.
\textcolor{black}{Figure~\ref{fig:Opt_left}(b) and (c) show this trend in the distribution of $\mathrm{Re}(P^-)$ and $|P^-|$.} Compared to the case of the initial configuration shown in \textcolor{black}{Fig.~\ref{fig:Init_config}(b) and (c)}, the amplitude \textcolor{black}{appears} small around the outlet, $\Gamma_{\mathrm{out2}}$.
\textcolor{black}{Figure~\ref{fig:history_case1} presents a history of the objective functional $J$ with the intermediate and optimized designs.
It was observed that a circular structure at the initial iteration was stretched to form the striped structures during optimization.
The optimization calculation was halted at the 983rd iteration, where the 10-iteration moving average of the relative error between the values of $J$ for two consecutive iterations was less than $3\times 10^{-4}$.
It is noted that the obtained design is dependent on the initial configuration.
At the microscale, the periodic boundary condition is applied in the $y_1$ direction; thus, the same performance as that of the optimized design is obtained for the shape translated along the $y_1$ direction.
If the position of the initial configuration is shifted, the optimized design will also be translated. 
Therefore, the optimized structure will change depending on the initial configuration.
}

\begin{figure}[H]
	\centering
	\includegraphics[scale=0.5]{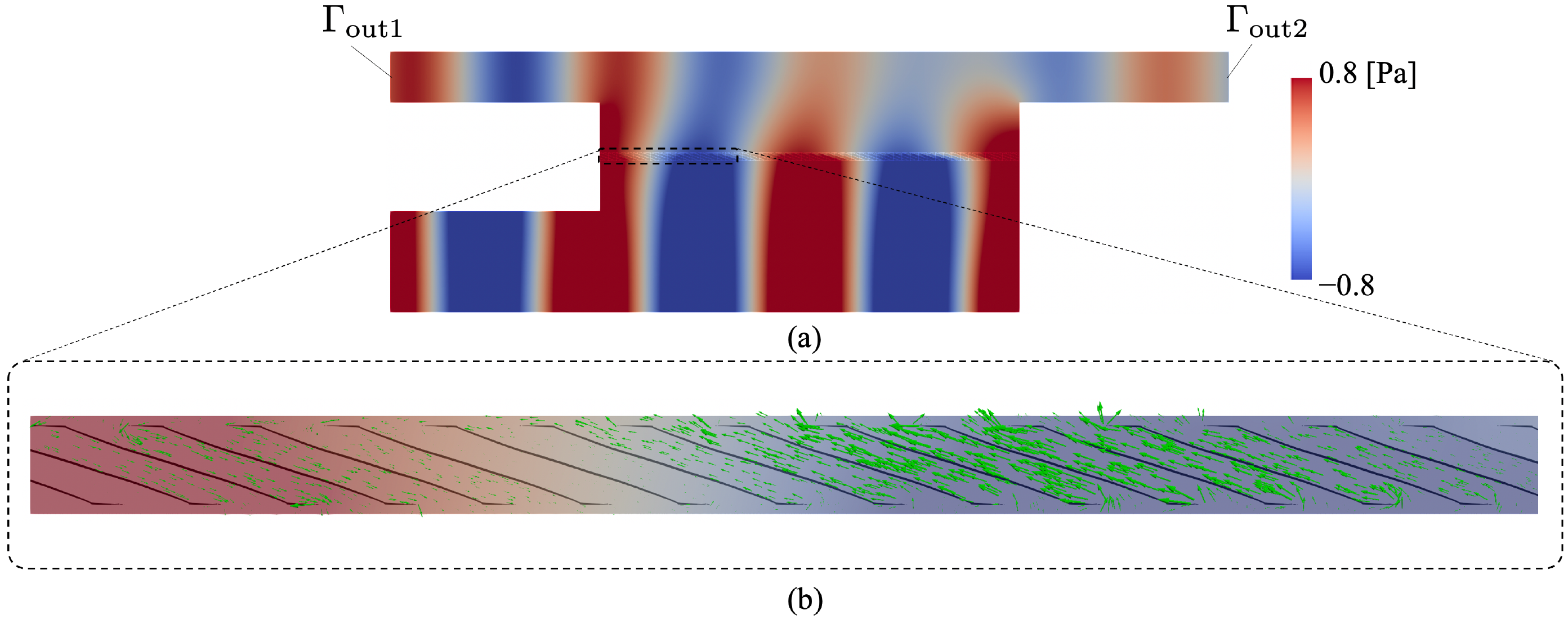}
	\caption{\textcolor{black}{(a) Distribution of the real part of acoustic pressure, $\mathrm{Re}(p^{Ref})$, corresponding to the optimized configuration in case 1. (b) The magnified view of (a). The green-colored arrows represent the sound intensity vector, $\bm{I}$, in air.}}
	\label{fig:Opt_left_usualFEM}       
\end{figure}

By considering interface $\Gamma^0$ in \textcolor{black}{Fig.~\ref{fig:Opt_left}(b) and (c)}, a strong discontinuity \textcolor{black}{in the} acoustic pressure can be observed.
To examine the details of this behavior around $\Gamma^0$, the reference solution is obtained for the \textcolor{black}{entire} system with an array of unit cells containing the \textcolor{black}{optimized} configuration, similar to that in section \ref{sec: Validation of the HMM}. 
Figure~\ref{fig:Opt_left_usualFEM}(a) \textcolor{black}{presents} the acoustic-pressure distribution for this reference analysis. The distribution is similar in the outer regions, $\Omega^\pm$, and demonstrates the validity of the proposed homogenization method.
Around the unit cells, the pressure contour is distorted owing to the \textcolor{black}{optimized} configuration. Figure~\ref{fig:Opt_left_usualFEM}(b) provides \textcolor{black}{additional} details on the acoustic-wave-propagation behavior through the magnified view of the dotted box in Fig.~\ref{fig:Opt_left_usualFEM}(a).
The green-colored arrows represent \textcolor{black}{the sound intensity vector} in air, as expressed by
\begin{align}
\bm{I}=\frac{1}{2}\mathrm{Re}\left(p^{Ref}\overline{\bm{u}^{Ref}}\right),\nonumber
\end{align}
where $\bm{u}^{Ref}=-\frac{1}{i \omega \rho_0}\nabla p^{Ref}$ denotes the particle velocity\textcolor{black}{,} and $\overline{\bm{u}^{Ref}}$ represents the complex conjugate of $\bm{u}^{Ref}$.
The sound intensity vector, $\bm{I}$, \textcolor{black}{indicates} the direction of energy flow. Within the unit cells, the direction of $\bm{I}$ is almost along the surface of the \textcolor{black}{optimized} configuration, and this results in \textcolor{black}{a} reduction of the transmission of acoustic waves toward outlet $\Gamma_{\mathrm{out2}}$.

\begin{figure}[H]
	\centering
	\includegraphics[scale=0.55]{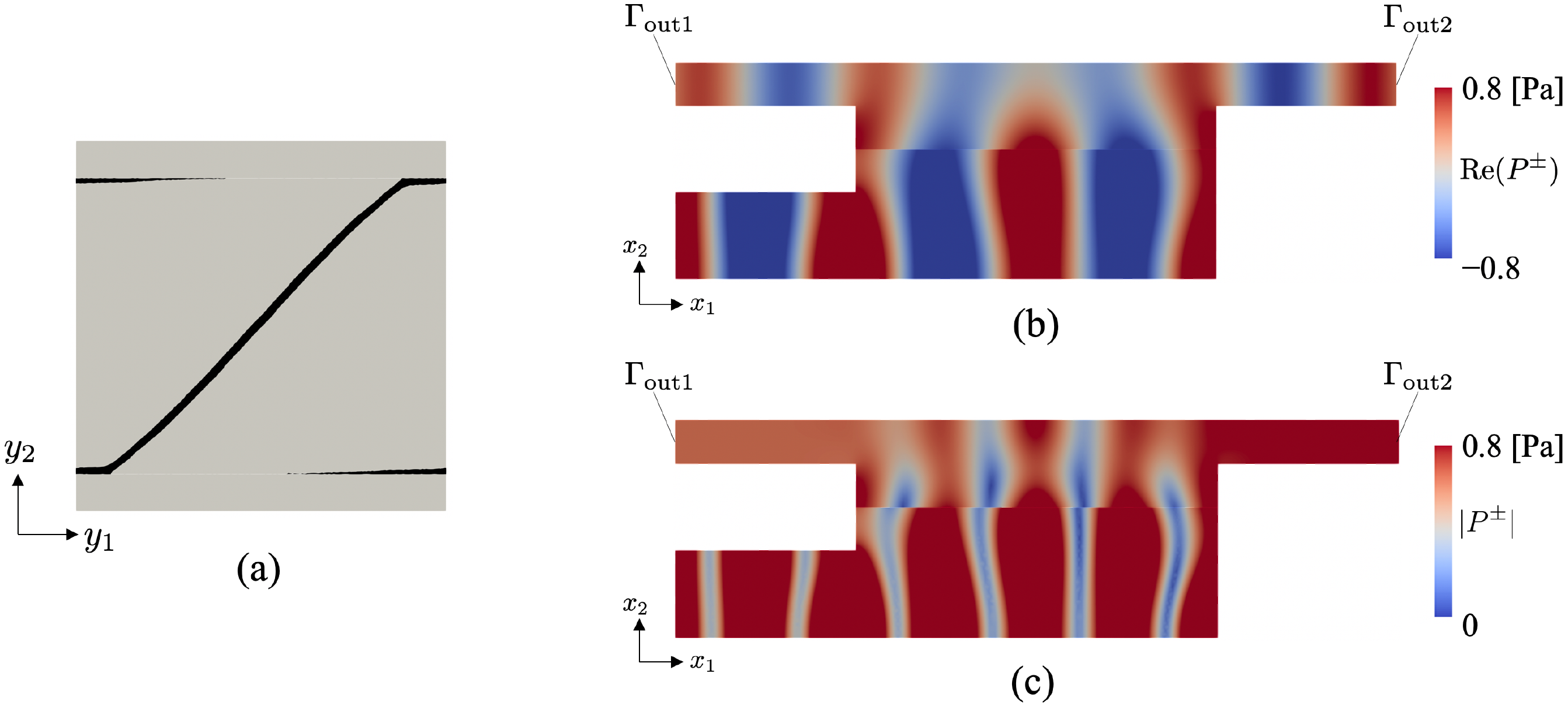}
	\caption{\textcolor{black}{(a) Optimized configuration for case 2. 
		(b) Distribution of the real part of acoustic pressure, $\mathrm{Re}(P^\pm)$. 
		(c) Distribution of the absolute value of acoustic pressure, $|P^\pm|$.}}
	\label{fig:Opt_right}       
\end{figure}

Figure~\ref{fig:Opt_right} represents the optimization results for case 2.
Similar to case 1, the \textcolor{black}{optimized} configuration shown in Fig.~\ref{fig:Opt_right}(a) is characterized by the striped structures\textcolor{black}{,} but these are tilted \textcolor{black}{toward} the right.
\textcolor{black}{The finite element discretization of the obtained design involves 21,182 elements, the details of which are explained in Section~\ref{sec: Mesh dependency}.}
The homogenized coefficients are evaluated as $({A}_{11}^\ast,{B}_1^\ast,K^{-1\ast},F^\ast)=\textcolor{black}{(0.234~\textcolor{black}{\mathrm{[m^3~kg^{-1}]}}, 1.10, 6.81\times 10^{-6}~\textcolor{black}{\mathrm{[Pa^{-1}]}}, 4.30~\textcolor{black}{\mathrm{[kg~m^{-3}]}})}$, 
and the acoustic-pressure distribution at the macroscale is obtained, as shown in Fig.~\ref{fig:Opt_right}\textcolor{black}{(b) and (c)}.
Compared to \textcolor{black}{the} case of \textcolor{black}{the} initial configuration, as shown in Fig.~\ref{fig:Init_config}(b), the amplitude of acoustic pressure around $\Gamma_{\mathrm{out1}}$ was reduced, \textcolor{black}{whereas} that \textcolor{black}{of the pressure} around $\Gamma_{\mathrm{out1}}$ increased with the use of the optimization formulation. \textcolor{black}{Moreover}, the value of objective functional $J$ was $\textcolor{black}{-0.366}$ with $J_1 = \textcolor{black}{0.718}$ and $J_2 = \textcolor{black}{1.45}$\textcolor{black}{;} this implies that both the properties of the metasurface for maximizing and minimizing the amplitude of acoustic pressure were improved.
\textcolor{black}{Although the value of $J$ was improved, a significant change in pressure distribution, as compared to case 1, could not be obtained. This was likely due to the settings of the computational domain in the macroscale and the objective functional. The value of the squared norm of the acoustic pressure on $\Gamma_{\mathrm{out1}}$ exceeds that of the pressure on $\Gamma_{\mathrm{out2}}$ during the initial configuration. Therefore, additional efforts are required to realize an opposite trend in pressure distribution than that in case 1. As the weighting factors in Eq.~(\ref{eq: objective functional}) were fixed during optimization and set to the same values as those in case 1, the optimization calculation did not proceed to realize such a significant change in pressure distribution.
}

\begin{figure}[H]
	\centering
	\includegraphics[scale=0.5]{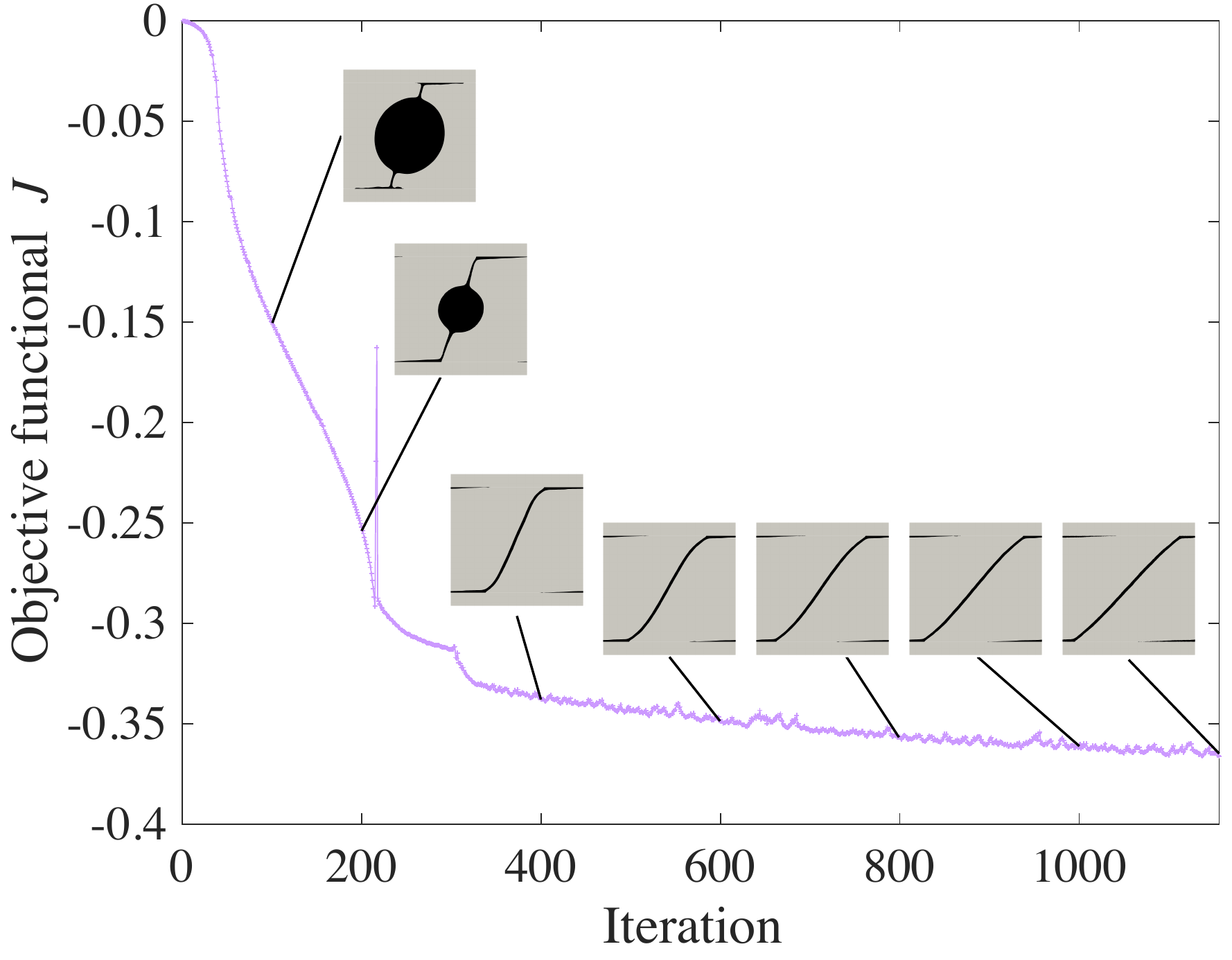}
	\caption{\textcolor{black}{History of objective functional $J$ with the intermediate and optimized designs for case 2.}}
	\label{fig:history_case2}       
\end{figure}

\textcolor{black}{Figure~\ref{fig:history_case2} represents a history of the objective functional $J$ with the intermediate and optimized designs.
An evolution behavior similar to that in case 1 was observed; however, the width of the striped structure during the initial iterations was less than that in case 1, which resulted in the unstable history of $J$, as compared to case 1. 
Consequently, we loosened the convergence criterion based on the moving average, and it was applied after the 900th iteration. 
The optimization calculation was halted at the 1156th iteration, where the 10-iteration moving average of the relative error between the values of $J$ for two consecutive iterations was less than $8\times 10^{-4}$.
}

\begin{figure}[H]
	\centering
	\includegraphics[scale=0.5]{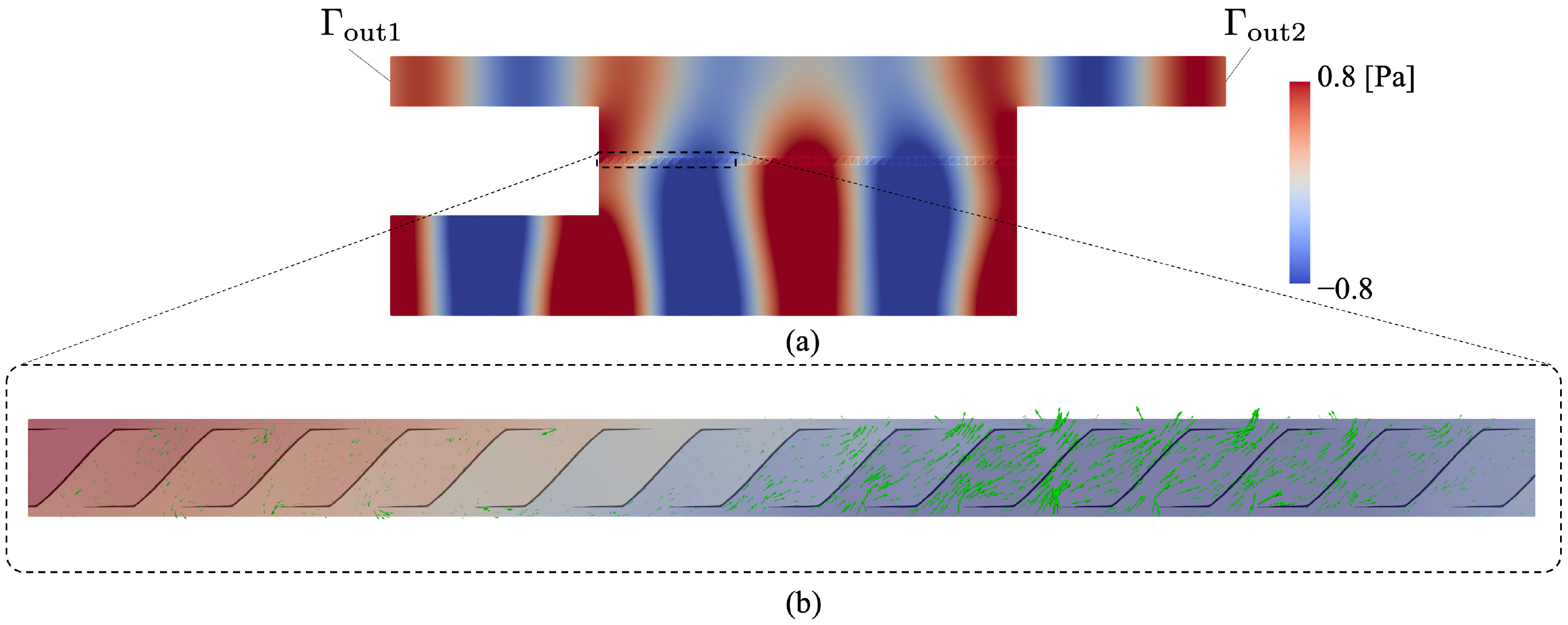}
	\caption{\textcolor{black}{(a) Distribution of the real part of acoustic pressure, $\mathrm{Re}(p^{Ref})$, corresponding to the optimized configuration in case 2. (b) The magnified view of (a). The green-colored arrows represent the sound intensity vector, $\bm{I}$, in air.}}
	\label{fig:Opt_right_usualFEM}       
\end{figure}

\textcolor{black}{For clear observations} around interface $\Gamma^0$, we conducted the reference analysis for the \textcolor{black}{entire} system containing the \textcolor{black}{optimized} configuration.
Figure~\ref{fig:Opt_right_usualFEM}(a) \textcolor{black}{presents} the acoustic-pressure distribution. 
As shown, the pressure contour is distorted owing to \textcolor{black}{the} \textcolor{black}{optimized} configuration, similar \textcolor{black}{to} that in case 1.
Figure~\ref{fig:Opt_right_usualFEM}(b) shows the magnified view around the dotted box shown in Fig.~\ref{fig:Opt_right_usualFEM}(a).
The direction of \textcolor{black}{the} sound intensity vector $\bm{I}$ is almost along the surface of the \textcolor{black}{optimized} configuration, and this reduces the transmission of acoustic waves toward outlet $\Gamma_{\mathrm{out1}}$ and \textcolor{black}{enables} transmission toward $\Gamma_{\mathrm{out2}}$.
Therefore, the mechanism of \textcolor{black}{controlling} the direction of wave propagation \textcolor{black}{appears identical to} that in case 1.

These optimization results are summarized with the values of the homogenized coefficients and objective functional in Table~\ref{tab: Homogenized coeffs_sum}.
Among these coefficients, the most significant change can be observed in the value of coefficient ${B}_1^\ast$ \textcolor{black}{for} both cases 1 and 2. Here, we discuss this change \textcolor{black}{during} the optimization.
One of the homogenized equations in the macroscale containing coefficient ${B}_1^\ast$ is described as follows:
\begin{align}
&\int_{\Gamma^0}\left(
{B}_1^\ast \frac{\partial p^0}{\partial x_1}- \frac{1}{2} F^\ast (G_0^+ + G_0^-)
\right)\psi d\Gamma_x
=\frac{1}{\epsilon_0}\int_{\Gamma^0}({P}^+ - {P}^-)\psi d\Gamma_x ~~~\forall \psi \in L^2(\Gamma^0). 
\label{eq: homogenized eq in numerical example}
\end{align}
This equation  implies that the jump between $P^\pm$ on interface $\Gamma^0$ depends on the values of ${B}_1^\ast$ and ${F}^\ast$.
The left-hand side of Eq.~(\ref{eq: homogenized eq in numerical example}) comprises two quantities concerning the gradient of acoustic pressure.
The first quantity is the tangential derivative, $\frac{\partial p^0}{\partial x_1}$, along interface $\Gamma_0$, \textcolor{black}{while} the second is $G_0^\pm$, which has a relationship with the normal derivative to $\Gamma^0$, as \textcolor{black}{indicated by} the definition in Eq.~(\ref{eq: definition of G0}).
These gradient quantities are reflected in the jump denoted as $(P^+ - P^-)$ on $\Gamma^0$, depending on the values of ${B}_1^\ast$ and ${F}^\ast$.
In our setting of the optimization problem, the phase shift of acoustic pressure along $\Gamma^0$ \textcolor{black}{appears} essential for minimizing the objective functional\textcolor{black}{, based on} the results shown in Figs.~\ref{fig:Opt_left_usualFEM}(b) and ~\ref{fig:Opt_right_usualFEM}(b).
The first term in Eq.~(\ref{eq: homogenized eq in numerical example}), i.e., ${B}_1^\ast\frac{\partial p^0}{\partial x_1}$, plays an important role as the tangential gradient represents the phase shift along $\Gamma^0$. The sign of ${B}_1^\ast$ determines the direction of the phase shift along $\Gamma^0$; this can be observed in our optimization results. In case 1, the phase shift to the left can be confirmed across $\Gamma^0$ from $\Omega^+$ to $\Omega^-$, as shown in Fig.~\ref{fig:Opt_left_usualFEM}(b), where the sign of ${B}_1^\ast$ is negative. \textcolor{black}{By} contrast, \textcolor{black}{in} the case 2, the phase shifts to the right across $\Gamma^0$, as shown in Fig.~\ref{fig:Opt_right_usualFEM}(b), where the sign of ${B}_1^\ast$ is positive.
Therefore, we consider that the optimization calculation proceeded such that the desired phase shift that minimizes the objective functional can be obtained by adjusting the value of ${B}_1^\ast$.

\begin{table}[htb]
	\begin{center}\caption{\textcolor{black}{Values of the homogenized coefficients and objective functional.}}
		\textcolor{black}{
		\begin{tabular}{|c|l|l|l|} \hline
			&\begin{tabular}{c}
				Initial configuration
			\end{tabular}&
			\begin{tabular}{c}
				Case 1
			\end{tabular}&
			\begin{tabular}{c}
				Case 2
			\end{tabular} \\ \hline 
			${A}_{11}^\ast ~\textcolor{black}{\mathrm{[m^3~kg^{-1}]}}$& $0.466$ &$0.273$ &$0.234$\\
			${B}_1^\ast$&$3.20\times 10^{-9}$ &$-2.10$&$1.10$\\
			$K^{-1\ast}~\textcolor{black}{\mathrm{[Pa^{-1}]}}$&$5.05\times 10^{-6}$&$6.66\times 10^{-6}$&$6.81\times10^{-6}$\\
			$F^\ast~\textcolor{black}{\mathrm{[kg~m^{-3}]}}$&$2.18$&$10.14$&$4.30$\\
			$J$&$0$&$-0.260$&$-0.366$\\ 
			$J_1$&$1$&$0.456$&$0.718$\\ 
			$J_2$&$1$&$0.975$&$1.45$\\ \hline
		\end{tabular}
	}
		\label{tab: Homogenized coeffs_sum}
	\end{center}
\end{table}

\subsection{\textcolor{black}{Mesh dependency of optimized designs}}\label{sec: Mesh dependency}
\begin{figure}[H]
	\centering
	\includegraphics[scale=0.5]{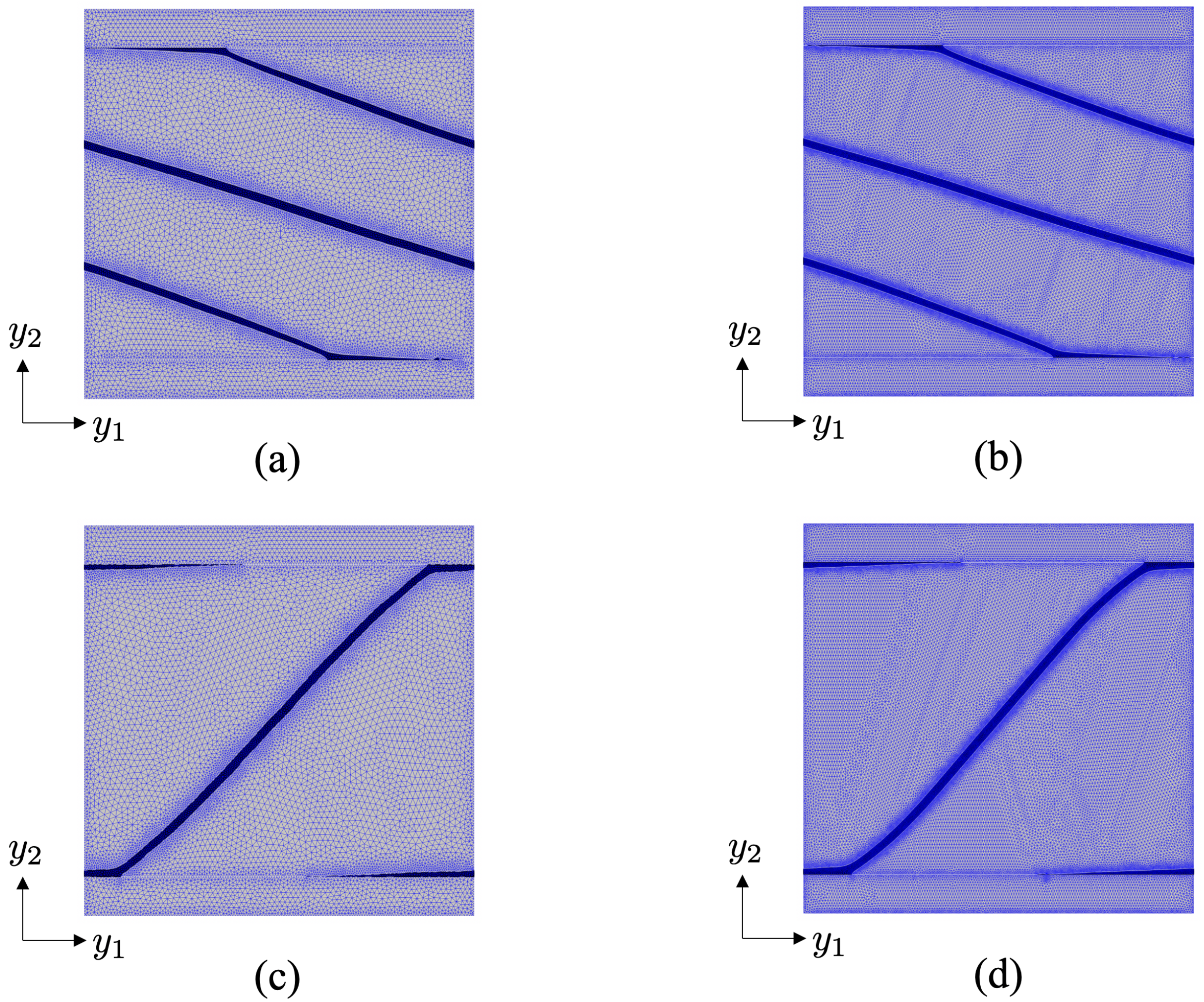}
	\caption{\textcolor{black}{Optimized results with different finite element discretization: 
		(a) Case 1 with 26,264 elements, (b) Case 1 with 76,406 elements, (c) Case 2 with 21,182 elements, and (d) Case 2 with 63,100 elements.
	}}
	\label{fig:Mesh_dependency}       
\end{figure}

\textcolor{black}{To examine the mesh dependency of the optimized designs, optimization calculations are conducted for case 1 and 2 using finer meshes in the microscale systems than those used for the results shown in Fig.~\ref{fig:Opt_left} and Fig.~\ref{fig:Opt_right}.
Figure~\ref{fig:Mesh_dependency} shows a comparison of the obtained configurations with finite element discretization.
The previous results for case 1 and 2 are shown in Fig.~\ref{fig:Mesh_dependency}(a) and (c), respectively.
Figure~\ref{fig:Mesh_dependency}(b) and (d) present the optimized results for case 1 and 2 when using the finer meshes, respectively.
The total number of elements in (b) is 76,406, whereas that in (d) is  63,100.
These are roughly three times more than those used for the previous results.
All the designs are characterized by striped structures, and there appears to be no difference between the results.
As the objective functional in Eq.~(\ref{eq: objective functional}) aims to simultaneously maximize and minimize the amplitude of the transmitted wave at different outlets, it is deduced that the striped structures with a certain finite width are required for each case.
Therefore, low mesh dependency is observed for the optimized designs. 
}

\subsection{\textcolor{black}{Discussion about computational cost}}
\begin{figure}[H]
	\centering
	\includegraphics[scale=0.3]{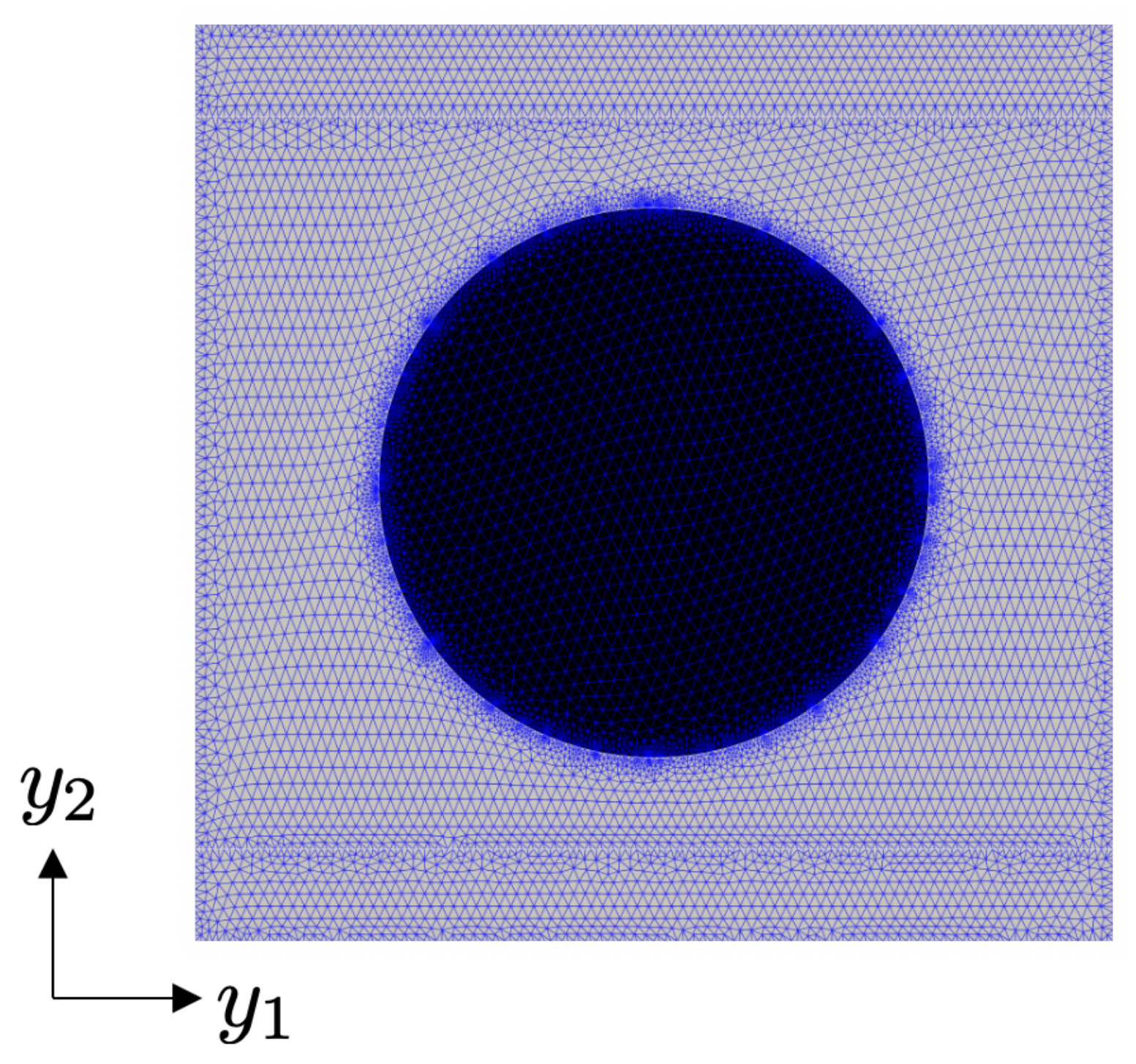}
	\caption{\textcolor{black}{
			Finite element discretization for the microscale system with 21,360 elements.}
	}
	\label{fig:Init_mesh}       
\end{figure}

\textcolor{black}{
Here, the computational effort required in the proposed method is presented.
As the proposed homogenization method decomposes the entire system of the metasurface into the macroscale and microscale, the total degree of freedom (DOF) required to analyze the system can be reduced.
Let $\mathrm{DOF}(u)$ denote the number of DOF for a variable $u$ in the FEM.
The total number of the DOF per single optimization loop, $\mathrm{DOF}_{hom}$, can be estimated as follows:
\begin{align}
		\mathrm{DOF}_{hom}= \{\mathrm{DOF}(\eta)+\mathrm{DOF}(\xi)+\mathrm{DOF}(\phi)\}
		 + \sum_{i=1}^5 \mathrm{DOF}(u_{macro}^i) + \sum_{i=1}^5\mathrm{DOF}(v_{macro}^i),
		\nonumber
\end{align}
where $\bm{u}_{macro}=(p^0, P^+,P^-, G_0^+,G_0^-)$ represents the state variables in the macroscale, whereas
$\bm{v}_{macro}=(q^0,Q^+,Q^-, \Psi_0^+,\Psi_0^-)$ represents the adjoint variables in the macroscale.
To determine the efficiency of the proposed homogenization model, we calculate the DOF of the system of the metasurface, whose unit cell structure is the same as that in the initial configuration, shown in Fig.~\ref{fig:Init_config}(a). 
Figure~\ref{fig:Init_mesh} presents the finite element discretization in the microscale, where 21,360 triangular elements are used.
For the macroscale, we use the same discretization as discussed in Section~\ref{sec: Optimization results}.
The DOF for each variable in this system is calculated and summarized in Table~\ref{tab: DOF_HOM}.
Based on Table~\ref{tab: DOF_HOM} and the abovementioned equation, $\mathrm{DOF}_{hom}$ is estimated as 167,512.
}

\begin{table}[htb]
	\begin{center}\caption{\textcolor{black}{
				Degree of freedom (DOF) required in the FEM for the initial iteration of the proposed optimization method.}}
		\begin{tabular}{|l|l|l|} \hline
			&\begin{tabular}{c}
				Variables
			\end{tabular}&
			\begin{tabular}{c}
				DOF
			\end{tabular}\\ \hline 
			Micro-& $({\eta},{\xi})$ &$(43361,43361)$\\
			scale& $\phi$ &$9226$\\
			Macro-&$(p^0, P^+,P^-, G_0^+,G_0^-)$
			&$(201,17474,17905,101,101)$\\
			scale&$(q^0,Q^+,Q^-, \Psi_0^+,\Psi_0^-)$
			&$(201,17474,17905,101,101)$\\
			\hline
		\end{tabular}
		\label{tab: DOF_HOM}
	\end{center}
\end{table}

\textcolor{black}{
In the case without homogenization, the DOF is estimated as a sum of the DOF of the state variable, adjoint variable corresponding to ${\bm{v}}_{macro}$, and level set function for representing the material distribution in the layer of the metasurface.
To estimate this, a reference analysis is introduced in Section \ref{sec: Validation of the HMM}.
In other words, we arrayed 50 unit cells over the transmission layer; the finite element discretization in each unit cell is depicted in Fig.~\ref{fig:Init_mesh}. To represent the entire system, 1,847,847 triangular elements were required.
Under this setting, the total DOF of the reference analysis, $\mathrm{DOF}_{ref}$, is estimated as 7,854,809,
which is approximately 47 times larger than $\mathrm{DOF}_{hom}$.
The DOF reduction afforded by the proposed method is attributable to the small number of finite elements required for solving the macroscale system. 
}

\textcolor{black}{
This reduction in the DOF resulted in a shorter computational time than that required for the standard FEM.
The computational time required for obtaining the state variables at the initial iteration by using the proposed method is compared with that for the conventional FEM without homogenization. 
The discretization conditions are the same as those explained above.
We used a desktop computer (Intel Core i9 CPU 3.6 GHz, 10 cores, 128 GB memory) for both analyses, and their FEM implementation is based on FreeFEM.
The computational time for $(\eta,\xi)$ in a unit cell was 0.59 [s], whereas that for $(p^0, P^+,P^-, G_0^+,G_0^-)$ in the macroscale was 0.58 [s].
Subsequently, the total computational time for the state variables per optimization loop could be estimated as 1.17 [s].
By contrast, the reference analysis without homogenization required 108.88 [s].
Based on this comparison, we concluded that the proposed method can analyze the system of the metasurface efficiently, with less computational cost than that of the standard FEM, which is beneficial for optimization.
}

\textcolor{black}{
Although we targeted two-dimensional metasurfaces that function at a single frequency, it is expected that the proposed method can be extended to three-dimensional or multi-frequency problems.
The efficiency of the proposed method is apparent in the case of multi-frequency optimizations.
As the microscale problem is independent of frequency, we only need to solve the microscale system once.
Although the macroscale analysis requires iterations with various input frequencies, it needs considerably less computational time than the conventional FEM, as evidenced by the abovementioned example.
The proposed homogenization method is also suitable for three-dimensional problems, as explained in Section \ref{sec: Homogenization method}. The number of cell problems increases to three, whereas the two-dimensional case involves two. Furthermore, the homogenized equations are defined in the three-dimensional external regions $\Omega^\pm$ and the two-dimensional surface $\Gamma^0$, whereas those for the abovementioned results are defined in the two-dimensional external regions $\Omega^\pm$ and one-dimensional boundary $\Gamma^0$. This will increase the computational costs; however, the standard FEM also requires a larger number of finite elements to analyze such a system. 
If the metasurface is composed of a periodic array of unit cells with complex structures, the DOF without homogenization will be significantly higher than that in the case with homogenization. Thus, the proposed method will require less computational time.
}


\section{Conclusion}\label{sec: Conclusion}
In this paper, we proposed a topology optimization method for the design of acoustic metasurfaces based on the homogenization method. We summarize the results of this study as follows\textcolor{black}{:}
	
	\begin{enumerate}
		\item This study introduces a homogenization method for acoustic metasurfaces based on the method proposed by Rohan and Luke\v{s} \cite{rohan2010homogenization,rohan2019homogenization}. We extend their approach to a metasurface system comprising both acoustic and elastic media.
		The proposed method can decompose the \textcolor{black}{entire} metasurface system, including the complex structures of unit cells, into problems defined at the microscale and macroscale.
		The microscale problem involves the so-called cell problem defined in the unit cell with appropriate periodic boundary conditions, and the homogenized coefficients expressing the feature of the unit cell can be estimated by solving the cell problems.
		The macroscale problem is defined in all regions, except the domain formed by the array of unit cells. 
		The complex structure of the metasurface is replaced with a boundary comprising the homogenized coefficients, and this reduces \textcolor{black}{the} computational \textcolor{black}{costs}.
			
		\item An optimization problem was formulated within the framework of the proposed homogenization method\textcolor{black}{,} and \textcolor{black}{it} includes a level set-based topology optimization. Acoustic responses at the macroscale were set to the objective functional, and the material distribution at the microscale was optimized to minimize the objective functional.
		As a typical macroscopic response, we chose the amplitude of transmitted acoustic waves at a certain target frequency and set them to the objective functional.

		\item A sensitivity analysis was conducted based on the concept of the topological derivative. We used the topological-shape sensitivity method to derive the topological derivative, which contains contributions at the macroscale and microscale of the objective functional. 
		The macroscale contribution can be estimated by solving the state and adjoint equations at the macroscale, \textcolor{black}{whereas} the microscale contribution can be obtained by solving the cell problems.

		\item An optimization algorithm that incorporates the homogenization method and the level set-based topology optimization method \textcolor{black}{was proposed}. In addition, we noted some numerical treatments to implement the algorithm using an FEM, especially for the selected shape functions and mesh refinement using the level set function.
		 		
		\item Numerical examples were provided to confirm the validity of the proposed method.
		First, \textcolor{black}{we provided an example that supports the validity of the proposed homogenization method} and compared the solutions with those obtained using the standard FEM (without homogenization); good congruence was observed between both, except for the \textcolor{black}{resonance} frequency.
		Then, we optimized both results, which were examined with respect to the settings of the objective functional. In both cases, each \textcolor{black}{optimized} configuration was characterized by striped structures of the elastic medium. In addition, phase shifts were observed around the array of unit cells, which play a key role in minimizing the objective functional. 
		
	\end{enumerate}

Although our optimization results target a single frequency, the method can be extended to include a range of frequencies by considering such a range in the settings of the objective functional. 
A three-dimensional optimization problem can also be addressed\textcolor{black}{,} as the homogenization method is valid three-dimensionally, as discussed in \cite{rohan2013optimal,rohan2019homogenization}.
Furthermore, our method could help in the optimum design of graded metasurfaces, which correspond to the spatial distribution of the homogenized coefficients at the macroscale.
In this research, the design variable was restricted to the material distribution at the microscale;
however, if the distribution of the homogenized coefficients at the macroscale is also considered as the design variable, the design space could \textcolor{black}{be increased}, and more effective control \textcolor{black}{over} acoustic waves could be realized. 
We plan to \textcolor{black}{achieve} these extensions in our future research.

\section*{Acknowledgment}
Funding: This work was supported in part by JSPS KAKENHI [grant number 20K14636] and Ono Charitable Trust for Acoustics.

We would like to thank Editage (www.editage.com) for English language editing.

\appendix
\section{Details for deriving the homogenized equations}\label{sec:Appendix homogenized equations}
\textcolor{black}{In this section, additional details on obtaining the homogenized equations in Section \ref{sec: Homogenization method} are explained.}

\textcolor{black}{First, the procedure to obtain Eq.~(\ref{eq: surface 1}) is provided.}
By substituting the convergence results for acoustic pressure, as expressed in Eqs.~(\ref{eq: convergence_first})--(\ref{eq: convergence_last}), and corresponding test function into the weak form [Eq.~(\ref{eq: weak formulation before homogenization})], we \textcolor{black}{obtain}
\begin{align}
&\int_{\Gamma^0} \dashint_Y \frac{1}{\rho(\bm{y})}(\overline{\nabla}_x p^0 + \overline{\nabla}_y p^1)\cdot
(\overline{\nabla}_x q^0 + \overline{\nabla}_y q^1) d\Omega_y d\Gamma_x
+\int_{\Gamma^0} \dashint_Y\frac{1}{\rho(\bm{y})}\frac{\partial p^1}{\partial z}\frac{\partial q^1}{\partial z} d\Omega_y d\Gamma_x
\nonumber\\
&-\int_{\Gamma^0} \dashint_Y\frac{\omega^2}{K(\bm{y})}p^0 q^0 d\Omega_y d\Gamma_x
= - \int_{\Gamma^0}\left\{ q^0 \dashint_{\Theta}\Delta g^1 + g^0 (\dashint_{I_y^+}q^1 - \dashint_{I_y^-}q^1)
\right\}d\Gamma_y d\Gamma_x,
\end{align}
where $\Theta$ represents the mid-plane of the unit cell, as shown in Fig.~\ref{fig:Geom}(b).
$\dashint= \frac{1}{|\Theta|}\int$ \textcolor{black}{is defined as an operator that averages the integrand over the cross-sectional area of the unit cell, $|\Theta|$.}
$\Delta g^1$ is used to express the difference between $g^{1\pm}$ as $\Delta g^1 = g^{1+}-g^{1-}$.

Next, we derive the so-called cell problems defined in $Y$. By setting the test functions as $q^0=0$ and $q^1\neq 0$, the following equation is obtained:
\begin{align}
&\int_{\Gamma^0}\dashint_Y \frac{1}{\rho(\bm{y})}
\left(\overline{\nabla}_y p^1 \cdot \overline{\nabla}_y q^1 
+\frac{\partial p^1}{\partial z}\frac{\partial q^1}{\partial z}
\right)d\Omega_y d\Gamma_x  \nonumber\\
&=-\int_{\Gamma^0}\overline{\nabla}_x p^0\cdot \left(\dashint_Y\frac{1}{\rho(\bm{y})}\overline{\nabla}_y q^1\right) d\Omega_y d\Gamma_x
-\int_{\Gamma^0} g^0\left(\dashint_{I_y^+}q^1 - \dashint_{I_y^-}q^1 
\right)d\Gamma_y d\Gamma_x.
\end{align}
This equation can be regarded as a weak form of the unknown, $p^1$.
Owing to linearity, $p^1$ is expressed as
\begin{align}
p^1(\bm{x'},\bm{y} ) = \sum_{\alpha=1}^{N-1} \left(\textcolor{black}{\eta}^\alpha (\bm{y})\frac{\partial p^0}{\partial x_\alpha} (\bm{x'})\right)+ \xi(\bm{y})g^0(\bm{x'}),
\end{align}
\textcolor{black}{where the functions $\textcolor{black}{\eta}^\alpha$ and $\xi$ are the solutions of the cell problems in Eq.~(\ref{eq: local pi}) and (\ref{eq: local xi}).}

Next, the macroscale problem defined on $\Gamma_0$ is derived by substituting $q^0\neq 0$ and $q^1= 0$ into the weak form [Eq.~(\ref{eq: weak formulation before homogenization})] as follows:
\begin{align}
&\int_{\Gamma^0}\dashint_Y \frac{1}{\rho(\bm{y})}(\overline{\nabla}_x p^0 + \overline{\nabla}_y p^1)\cdot \overline{\nabla}_x q^0 d\Omega_y d\Gamma_x 
-\int_{\Gamma^0}\dashint_Y \frac{\omega^2}{K(\bm{y})} p^0 q^0 d\Omega_y d\Gamma_x 
\nonumber\\
&=-\int_{\Gamma^0}q^0 \left(\dashint_\Theta \Delta q^1\right)d\Gamma_y d\Gamma_x 
\end{align}
By using the expression of $p^1$, this equation can be modified as
\begin{align}
&\sum_{\alpha=1}^{N-1}\sum_{\beta=1}^{N-1}\int_{\Gamma^0}\frac{\partial p^0}{\partial x_\beta} \frac{\partial q^0}{\partial x_\alpha}\left[
\dashint_Y \frac{1}{\rho(\bm{y})}\left\{  \overline{\nabla}_y (\textcolor{black}{\eta}^\beta + y_\beta)\cdot\overline{\nabla}_y (\textcolor{black}{\eta}^\alpha + y_\alpha)
+ \frac{\partial \textcolor{black}{\eta}^\beta}{\partial z} \frac{\partial \textcolor{black}{\eta}^\alpha}{\partial z}
\right\}d\Omega_y
\right]d\Gamma_x \nonumber\\
&+\int_{\Gamma^0} g^0 \overline{\nabla}_x q^0 \cdot \left(
\dashint_Y \frac{1}{\rho(\bm{y})} \overline{\nabla}_y \xi
d\Omega_y\right)d\Gamma_x
-\omega^2\int_{\Gamma^0}p^0 q^0\left(\dashint_Y\frac{1}{K(\bm{y})}d\Omega_y\right) d\Gamma_x
\nonumber\\
&=-\int_{\Gamma^0}q^0\left(\dashint_{\Theta}\Delta g^1 d\Gamma_y\right) d\Gamma_x.
\label{eq:macro_before_coef}
\end{align} 
\textcolor{black}{By using the homogenized coefficients expressed with Eqs.~(\ref{eq: coeff A})--(\ref{eq: coeff K}),}
 Eq.~(\ref{eq:macro_before_coef}) can be rewritten as \textcolor{black}{Eq.~(\ref{eq: surface 1}).}

\textcolor{black}{Next, the procedure to obtain Eq.~(\ref{eq: surface 2}) is explained.}
Using the mapped acoustic pressures defined in Eq.~(\ref{eq: mapped pressures}), the left-hand side of Eq.~(\ref{eq: Interface coupling}) can be considered \textcolor{black}{for} the mid-plane of $\Gamma_0$. 
The introduction of the scaled coordinate, $z=\frac{x_d}{\epsilon}$, and the multiplication of Eq.~(\ref{eq: Interface coupling}) with  $\frac{1}{\epsilon}$ \textcolor{black}{yield}
\begin{align}
\frac{1}{\epsilon}\int_{\Gamma^0}({P}^-_m - {P}^+_m)\psi d\Gamma = 
\int_{\Gamma^0}\psi \int_{-\frac{\kappa}{2}}^{\frac{\kappa}{2}}\frac{1}{\epsilon}\frac{\partial p^\epsilon}{\partial z} d\Gamma.
\label{eq: surface}
\end{align}
Considering limit $\epsilon \to 0$, the right-hand side of Eq.~(\ref{eq: surface}) takes the following form:
\begin{align}
\int_{\Gamma^0}\psi \int_{-\frac{\kappa}{2}}^{\frac{\kappa}{2}}\frac{1}{\epsilon}\frac{\partial p^\epsilon}{\partial z} d\Gamma
&\to
\int_{\Gamma^0}\psi \int_{-\frac{\kappa}{2}}^{\frac{\kappa}{2}}\frac{1}{\epsilon}
\left(\dashint_{\Theta}\frac{\partial T_\epsilon(p^\epsilon)}{\partial z } d\Gamma_y  \right)dz d\Gamma_x
\nonumber\\
&=\int_{\Gamma^0}\psi \int_{-\frac{\kappa}{2}}^{\frac{\kappa}{2}}\dashint_{\Theta}\frac{\partial p^1}{\partial z } d\Gamma_y  dz d\Gamma_x
\nonumber\\
&=\int_{\Gamma^0}\psi \dashint_{Y}\frac{\partial p^1}{\partial z } d\Omega_y d\Gamma_x
\nonumber\\
&=\int_{\Gamma^0}\psi \left(
\dashint_{I_y^-}p^1 d\Gamma_y - \dashint_{I_y^+}p^1 d\Gamma_y  \right)d\Gamma_x\nonumber\\
&=\int_{\Gamma^0}\psi \sum_{\alpha=1}^{N-1} \frac{\partial p^0}{\partial x_\alpha}\left(
\dashint_{I_y^-}\textcolor{black}{\eta}^\alpha d\Gamma_y - \dashint_{I_y^+}\textcolor{black}{\eta}^\alpha d\Gamma_y
\right)d\Gamma_x\nonumber\\
&+\int_{\Gamma^0}\psi g^0\left(
\dashint_{I_y^-}\xi d\Gamma_y - \dashint_{I_y^+}\xi d\Gamma_y
\right)d\Gamma_x.
\end{align}
To obtain the last line, the expression of $p^1$ is used.
Furthermore, \textcolor{black}{the} homogenized coefficient $F^\ast$\textcolor{black}{,} \textcolor{black}{expressed by Eq.~(\ref{eq: coeff F})}\textcolor{black}{,}
was introduced\textcolor{black}{,} and $\bm{B}^\ast$ was modified using the weak forms of the cell problems as follows:
\begin{align}
B^\ast_\alpha &= \dashint_{Y} \frac{1}{\rho(\bm{y})}\frac{\partial \xi}{\partial y_\alpha}d\Omega_y\nonumber\\
&=-\dashint_{Y}\frac{1}{\rho(\bm{y})}\left(
\overline{\nabla}_y\xi \cdot \overline{\nabla}_y\textcolor{black}{\eta}^\alpha +\frac{\partial \xi}{\partial z}\frac{\partial \textcolor{black}{\eta}^\alpha}{\partial z}
\right)d\Omega_y\nonumber\\
&=\dashint_{I_y^+}\textcolor{black}{\eta}^\alpha d\Gamma_y- \dashint_{I_y^-}\textcolor{black}{\eta}^\alpha d\Gamma_y.
\end{align}
Then, the right-hand side of Eq.~(\ref{eq: surface}) \textcolor{black}{under} the limit of $\epsilon \to 0$ can be modified as follows:
\begin{align}
\int_{\Gamma^0}\psi \int_{-\frac{\kappa}{2}}^{\frac{\kappa}{2}}\frac{1}{\epsilon}\frac{\partial p^\epsilon}{\partial z} d\Gamma
&\to
-\int_{\Gamma^0}\psi \overline{\nabla}_x p^0 \cdot \bm{B}^\ast d\Gamma_x
+\int_{\Gamma^0}\psi g^0 F^\ast d\Gamma_x
\end{align}
Considering limit $\epsilon\to \epsilon_0$ with a small positive number, $\epsilon_0$, at the left-hand side of Eq.~(\ref{eq: surface}),
\textcolor{black}{we finally obtain the homogenized equations in Eq.~(\ref{eq: surface 2}).}

\section{Sensitivity analysis}\label{sec:Appendix sensitivity analysis}
This section details the sensitivity analysis based on the concept of the topological derivative, which can be defined as

\begin{align}
	D_T J=\lim_{\varepsilon \to 0} \frac{J(\Omega \setminus \overline{\Omega_i})-J(\Omega)}{V(\varepsilon)},\label{eq:definition of TD_original}
\end{align}
where $V(\varepsilon)$ is a function depending on radius $\varepsilon$ of the inclusion domain $\Omega_i$.
The form of $V(\varepsilon)$ is chosen to include the limit value of the right-hand side of Eq. (\ref{eq:definition of TD_original}); we set $V(\varepsilon)=-\pi \varepsilon^2$ at the same value as in \cite{carpio2008solving,carpio2008topological}.
Novotny et al. \cite{novotny2003topological} and Feij\'oo et al. \cite{feijoo2003topological} proposed the topological-shape-sensitivity method for deriving the topological derivative by considering the relationship between the topological and shape derivatives.
The shape derivative of $J$ for the deformation of the inclusion domain $\Omega_i$ is defined as
\begin{align}
	DJ(\Omega)\cdot \bm{\theta}=\left.\frac{d}{d\beta}J(\phi_\beta(\Omega_i))\right|_{\beta=0\,,}\label{eq:def of shape derivative}
\end{align}
where $\phi_\beta$ represents the deformation mapping of $\Omega_i$ and is defined as
\begin{eqnarray}
	\phi_\beta(\bm{x}) = \bm{x}+\beta \bm{\theta}(\bm{x}).\label{eq:deforamation_mapping}\nonumber
\end{eqnarray}  
The shape derivative can be linked with the topological derivative via vector field $\bm{\theta}$, which is assumed to point \textcolor{black}{toward} the direction of the outward-pointing normal unit vector, $\bm{n}^{(i)}$, on the boundary of \textcolor{black}{the} inclusion domain $\Omega_i$.
In this case, vector $\bm{\theta}$ is expressed as $\bm{\theta}=-\theta_n \bm{n}^{(i)}$
with a negative constant, $\theta_n$.
Then, the topological derivative can be estimated as the limit value of the shape derivative when $\varepsilon \to 0$, as follows:
\begin{align}
	D_T J=\lim_{\varepsilon \to 0}\frac{1}{V'(\varepsilon)\left|\theta_n\right|}DJ(\Omega)\cdot \bm{\theta},\label{eq:definition of TD}
\end{align}
where $V'(\varepsilon)$ is the derivative of $V(\varepsilon)$ with respect to $\varepsilon$.
The topological derivative is derived using Eq.~(\ref{eq:definition of TD}) based on the following procedure.
First, the shape derivative is derived using the adjoint variable method.
Next, the asymptotic behaviors of the state and adjoint variables are examined according to radius $\varepsilon \to 0$ to estimate the limit form of the shape derivative.
Then, by using Eq.~($\ref{eq:definition of TD}$) with the shape derivative in the limit form, the explicit form of the topological derivative is obtained.

\begin{figure}[H]
	\centering
	\includegraphics[scale=0.4]{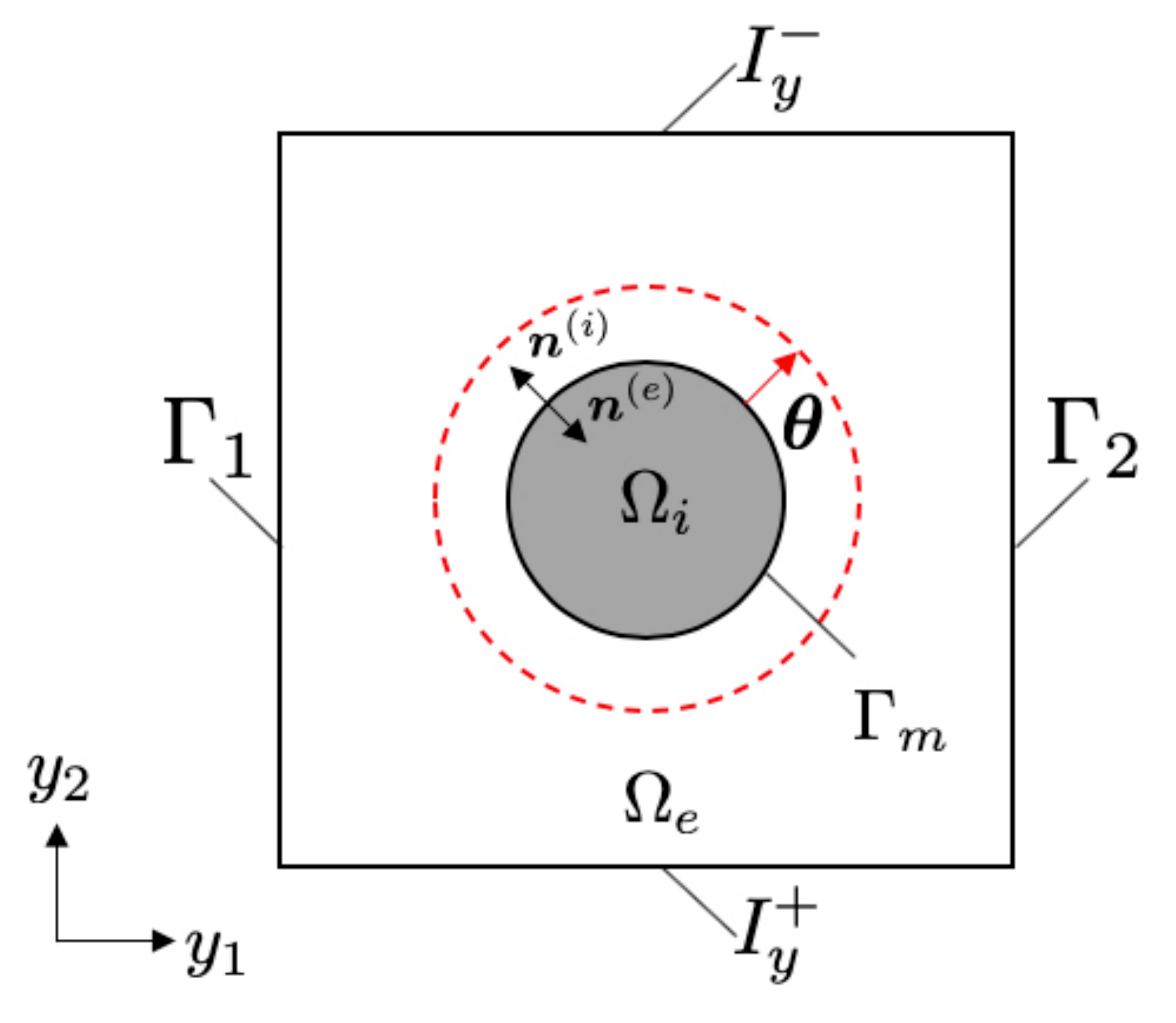}
	\caption{Geometrical settings at the microscale for \textcolor{black}{deriving} the shape derivative. }
	\label{fig:Geom_shape_derivative}       
\end{figure}
~\\

\textbf{Step1:} Derivation of the shape derivative
~\\

Here, we define the shape derivative for the optimization problem as expressed in Eq.~(\ref{eq: optimization problem}).
Corresponding to Eq.~(\ref{eq: optimization problem}), we assume that objective functional $J$ has the following form:
\begin{align}
J(P^-)=\int_{\Gamma_\mathrm{out1}}g_1(P^-,\overline{P^-})d\Gamma_x + \int_{\Gamma_\mathrm{out2}}g_2(P^-,\overline{P^-})d\Gamma_x,
\end{align}
where $\overline{P^-}$ represents the complex conjugate of $P^-$. Integrands $g_1$ and $g_2$ are assumed to satisfy $\overline{g_1}=g_1$ and $\overline{g_2}=g_2$, respectively.
Figure~\ref{fig:Geom_shape_derivative} shows the geometrical setting at the microscale for the derivation of the shape derivative. We consider a case in which the inclusion domain, $\Omega_i$, with radius $\varepsilon > 0$ is placed in unit cell $Y$. Then, unit cell $Y$ can be expressed as $Y=\Omega_i \cup \Omega_e$, where $\Omega_e$ represents the external domain in the unit cell. The interface of $\Omega_i$ and $\Omega_e$ is denoted by $\Gamma_m$, on which the outward-pointing normal vectors $\bm{n}^{(m)}~(m=i,e)$ are defined, respectively.
\textcolor{black}{A} change \textcolor{black}{in the shape} of domain $\Omega_i$ at the microscale results in \textcolor{black}{variations} in the homogenized coefficients, and this finally results in \textcolor{black}{a} change \textcolor{black}{in} the objective functional defined by the acoustic pressure at the macroscale.
Based on C\'ea's method \cite{Cea1986}, Lagrangian $L$ is defined by considering the microscale and macroscale as follows:
\begin{align}
L(\Omega,\hat{\bm{U}},\hat{\bm{V}}) &= J(\hat{P}^-) +2\mathrm{Re}\left[\sum_{i=1}^{5}C_i(\hat{\bm{u}}_{macro},\hat{\bm{u}}_{B},\hat{\bm{v}}_{macro})\right]\nonumber\\
&+\hat{\lambda}_{A_{11}^\ast}\left(
A_{11}^\ast(\Omega,\hat{\textcolor{black}{\eta}}) - \hat{A}_{11}^\ast
\right)
+\hat{\lambda}_{B_{1}^\ast}\left(
{B}_{1}^\ast(\Omega,\hat{\xi}) - \hat{B}_{1}^\ast
\right)\nonumber\\
&+\hat{\lambda}_{F^\ast}\left(
{F}^\ast(\hat{\xi}) - \hat{F}^\ast
\right)
+\hat{\lambda}_{K^{-1\ast}}\left(
K^{-1 \ast}(\Omega) - \hat{K}^{-1\ast}
\right)\nonumber\\
&-\hat{\lambda}_{A_{11}^\ast}a_{\textcolor{black}{\eta}}(\Omega,\hat{\textcolor{black}{\eta}},\hat{v}_{\textcolor{black}{\eta}})
-\hat{\lambda}_{B_{1}^\ast}a_{\xi 1}(\Omega,\hat{\xi},\hat{v}_{\xi_1})
+\hat{\lambda}_{F^\ast}a_{\xi 2}(\Omega,\hat{\xi},\hat{v}_{\xi_2}),
\label{eq: Lagrangian}
\end{align}
where the variables used in the Lagrangian are summarized in Table~(\ref{tab: Lagrangian variables}).
\begin{table}[htb]
	\begin{center}\caption{Variables used in the Lagrangian, Eq.~(\ref{eq: Lagrangian}).}
	\begin{tabular}{|c|l|l|} \hline
		  &\begin{tabular}{c}
		  	Variables $\hat{\bm{U}}$ corresponding \\to state variables ${\bm{U}}$
		  \end{tabular}&
	 	  \begin{tabular}{c}
	  	    Lagrange multipliers $\hat{\bm{V}}$ \\corresponding  to adjoint variables ${\bm{V}}$
	      \end{tabular}\\ \hline 
		Microscale& $\hat{\bm{u}}_{micro}=(\hat{\textcolor{black}{\eta}},\hat{\xi})$ &$\hat{\bm{v}}_{micro}=(\hat{v}_{\textcolor{black}{\eta}},\hat{v}_{\xi 1},\hat{v}_{\xi 2})$\\
		     &$\hat{\bm{u}}_{B}=(\hat{A}_{11}^\ast, \hat{B}_{1}^\ast, \hat{F}^\ast, \hat{K}^{-1 \ast})$
		&$\hat{\bm{v}}_{B}=(\hat{\lambda}_{{A}_{11}^\ast}, \hat{\lambda}_{{B}_{1}^\ast},\hat{\lambda}_{{F}^\ast}, \hat{\lambda}_{{K}^{-1\ast}} )$\\
		Macroscale&$\hat{\bm{u}}_{macro}=(\hat{p}^0, \hat{P}^\pm,\hat{G}_0^\pm)$
		&$\hat{\bm{v}}_{macro}=(\hat{q}^0, \hat{Q}^\pm, \hat{\Psi}_0^\pm)$\\ \hline
	\end{tabular}
	\label{tab: Lagrangian variables}
	\end{center}
\end{table}
The first line in Eq.~(\ref{eq: Lagrangian}) represents the macroscale contribution to Lagrangian $L$ with the objective functional $J$ and constraints for the global equations, $C_i(\hat{\bm{u}}_{macro},\hat{\bm{u}}_{B},\hat{\bm{v}}_{macro})$ $(i=1,...,5)$, and this is expressed as
\begin{align}
C_1(\hat{\bm{u}}_{macro},\hat{\bm{u}}_{B},\hat{\bm{v}}_{macro})&=
\int_{\Omega^+}\frac{1}{\rho_0}\nabla_x \hat{P}^+\cdot \nabla_x \hat{Q}^+ d\Omega_x -\int_{\Omega^+}\frac{\omega^2}{K_0}\hat{P}^+\hat{Q}^+ d\Omega_x\nonumber\\
&-\int_{\Gamma^0}\hat{G}_0^+ \hat{Q}^+ d\Gamma_x
+\int_{\Gamma_\mathrm{in}}\frac{i k_0}{\rho_0}\hat{P}^+\hat{Q}^+ d\Gamma_x
-\int_{\Gamma_\mathrm{in}}\frac{2 i k_0}{\rho_0}P_\mathrm{in}\hat{Q}^+ d\Gamma_x,\\
C_2(\hat{\bm{u}}_{macro},\hat{\bm{u}}_{B},\hat{\bm{v}}_{macro})&=
\int_{\Omega^-}\frac{1}{\rho_0}\nabla_x \hat{P}^-\cdot \nabla_x \hat{Q}^- d\Omega_x -\int_{\Omega^-}\frac{\omega^2}{K_0}\hat{P}^-\hat{Q}^- d\Omega_x\nonumber\\
&+\int_{\Gamma^0}\hat{G}_0^- \hat{Q}^- d\Gamma_x
+\int_{\Gamma_\mathrm{out1}\cup \Gamma_\mathrm{out2}}\frac{i k_0}{\rho_0}\hat{P}^-\hat{Q}^- d\Gamma_x,\nonumber\\
C_3(\hat{\bm{u}}_{macro},\hat{\bm{u}}_{B},\hat{\bm{v}}_{macro})&=
\hat{A}_{11}^\ast \int_{\Gamma^0} \frac{\partial \hat{p}_0}{\partial x_1}
\frac{\partial \hat{q}_0}{\partial x_1} d\Gamma_x
-\hat{K}^{-1\ast}\omega^2 \int_{\Gamma^0}\hat{p}_0\hat{q}_0 d\Gamma_x\nonumber\\
&+\frac{1}{2}\hat{B}_1^\ast \int_{\Gamma^0}(\hat{G}_0^+ + \hat{G}_0^-)\frac{\partial \hat{q}_0}{\partial x_1} d\Gamma_x
+\frac{1}{\epsilon_0} \int_{\Gamma^0}(\hat{G}_0^+ - \hat{G}_0^-)\hat{q}_0 d\Gamma_x,\nonumber\\
C_4(\hat{\bm{u}}_{macro},\hat{\bm{u}}_{B},\hat{\bm{v}}_{macro})&=\hat{B}_1^\ast\int_{\Gamma^0}\frac{\partial \hat{p}_0}{\partial x_1}\hat{\Psi}_0^+ d\Gamma_x -\frac{1}{2}\hat{F}^{\ast}\int_{\Gamma^0}(\hat{G}_0^+ + \hat{G}_0^-)\hat{\Psi}_0^+ d\Gamma_x\nonumber\\
&-\frac{1}{\epsilon_0}\int_{\Gamma^0}(\hat{P}^+ - \hat{P}^-)\hat{\Psi}_0^+ d\Gamma_x,\\
C_5(\hat{\bm{u}}_{macro},\hat{\bm{u}}_{B},\hat{\bm{v}}_{macro})&=\int_{\Gamma^0}\left\{
\hat{p}_0 - \frac{1}{2}(\hat{P}^+ + \hat{P}^-)
\right\}\hat{\Psi}_0^- d\Gamma_x.
\end{align}
The second and third lines in Eq.~(\ref{eq: Lagrangian}) show the contribution of the expression of homogenized coefficients, and are defined as 
\begin{align}
{A}_{11}^\ast(\Omega,\hat{\textcolor{black}{\eta}}) &=\sum_{m=i,e}\int_{\Omega_m}\frac{1}{\rho^{(m)}}\left(\frac{\partial \hat{\textcolor{black}{\eta}}}{\partial y_1} + 1\right)d\Omega_y, \\
{B}_{1}^\ast(\Omega,\hat{\xi}) &= \sum_{m=i,e}\int_{\Omega_m}\frac{1}{\rho^{(m)}}\frac{\partial \hat{\xi}}{\partial y_1}d\Omega_y, \\
{F}^\ast(\hat{\xi}) &= -\left(\int_{I_y^+}\hat{\xi}d\Gamma_y - \int_{I_y^-}\hat{\xi}d\Gamma_y  \right),\\
{K}^{-1 \ast}(\Omega)&=\sum_{m=i,e}\int_{\Omega_m}\frac{1}{K^{(m)}}d\Omega_y,
\end{align}
where \textcolor{black}{the} superscript $(m)$ represents a quantity in domain $\Omega_m~(m=i,e)$
and summation $\sum_{m=i,e}$ represents an integral over unit cell $Y$.
The last line in Eq.~(\ref{eq: Lagrangian}) shows the contribution \textcolor{black}{at} the microscale, \textcolor{black}{where} $a_{\textcolor{black}{\eta}}$, $a_{\xi 1}$, and $a_{\xi 2}$ are constraints for the cell problems: 
\begin{align}
a_{\textcolor{black}{\eta}}(\hat{\textcolor{black}{\eta}},\hat{v}_{\textcolor{black}{\eta}}) &=\sum_{m=i,e}\int_{\Omega_m}\left\{
\nabla_y \cdot \left(\frac{1}{\rho^{(m)}}\nabla_y \hat{\textcolor{black}{\eta}}^{(m)}\right)
+\nabla_y\cdot\left(\frac{1}{\rho^{(m)}}\bm{e_1}\right)
\right\}\hat{v}_{\textcolor{black}{\eta}}^{(m)} d\Omega_y\nonumber\\
&-\int_{I_y^\pm}\left\{\bm{n}\cdot\left(\frac{1}{\rho^{(e)}}\nabla_y \hat{{\textcolor{black}{\eta}}}^{(e)}\right)
+\bm{n}\cdot \left(\frac{1}{\rho^{(e)}}\bm{e_1}\right)\right\}\hat{v}_{\textcolor{black}{\eta}}^{(e)}d\Gamma_y\nonumber\\
&+\frac{1}{2}\int_{\Gamma_m}(\hat{{\textcolor{black}{\eta}}}^{(i)} - \hat{{\textcolor{black}{\eta}}}^{(e)})\left[
\left\{
\bm{n}^{(i)}\cdot\left(\frac{1}{\rho^{(i)}}\nabla_y \hat{v}_{\textcolor{black}{\eta}}^{(i)}\right) + \bm{n}^{(i)}\cdot\left(\frac{1}{\rho^{(i)}}\bm{e_1}\right)
\right\}\right.\nonumber\\
&\left.-\left\{
\bm{n}^{(e)}\cdot\left(\frac{1}{\rho^{(e)}}\nabla_y \hat{v}_{\textcolor{black}{\eta}}^{(e)}\right) + \bm{n}^{(e)}\cdot\left(\frac{1}{\rho^{(e)}}\bm{e_1}\right)
\right\}
\right]d\Gamma_y\nonumber\\
&-\frac{1}{2}\int_{\Gamma_m}(\hat{v}_{\textcolor{black}{\eta}}^{(i)}+\hat{v}_{\textcolor{black}{\eta}}^{(e)})\left[
\left\{\bm{n}^{(i)}\cdot \left(\frac{1}{\rho^{(i)}}\nabla_y \hat{{\textcolor{black}{\eta}}}^{(i)}\right)+\bm{n}^{(i)}\cdot \left(
\frac{1}{\rho^{(i)}}\bm{e_1}\right)
\right\}\right.\nonumber\\
&\left.+\left\{\bm{n}^{(e)}\cdot \left(\frac{1}{\rho^{(e)}}\nabla_y \hat{{\textcolor{black}{\eta}}}^{(e)}\right)+\bm{n}^{(e)}\cdot \left(
\frac{1}{\rho^{(e)}}\bm{e_1}\right)
\right\}
\right]d\Gamma_y,
\end{align}
\begin{align}
a_{\xi 1}(\hat{\xi},\hat{v}_{\xi 1}) &=\sum_{m=i,e}\int_{\Omega_m}\left\{
\nabla_y \cdot \left(\frac{1}{\rho^{(m)}}\nabla_y \hat{\xi}^{(m)}\right)
\right\}\hat{v}_{\xi 1}^{(m)} d\Omega_y\nonumber\\
&-\int_{I_y^+}\left\{\bm{n}\cdot\left(\frac{1}{\rho^{(e)}}\nabla_y \hat{\xi}^{(e)}\right) +1 \right\}\hat{v}_{\xi 1}^{(e)}d\Gamma_y
-\int_{I_y^-}\left\{\bm{n}\cdot\left(\frac{1}{\rho^{(e)}}\nabla_y \hat{\xi}^{(e)}\right) -1 \right\}\hat{v}_{\xi 1}^{(e)}d\Gamma_y
\nonumber\\
&+\frac{1}{2}\int_{\Gamma_m}(\hat{\xi}^{(i)} - \hat{\xi}^{(e)})\left[
\left\{
\bm{n}^{(i)}\cdot\left(\frac{1}{\rho^{(i)}}\nabla_y \hat{v}_{\xi 1}^{(i)}\right) + \bm{n}^{(i)}\cdot\left(\frac{1}{\rho^{(i)}}\bm{e_1}\right)
\right\}\right.\nonumber\\
&\left.-\left\{
\bm{n}^{(e)}\cdot\left(\frac{1}{\rho^{(e)}}\nabla_y \hat{v}_{\xi 1}^{(e)}\right) + \bm{n}^{(e)}\cdot\left(\frac{1}{\rho^{(e)}}\bm{e_1}\right)
\right\}
\right]d\Gamma_y\nonumber\\
&-\frac{1}{2}\int_{\Gamma_m}(\hat{v}_{\xi 1}^{(i)}+\hat{v}_{\xi 1}^{(e)})
\left\{\bm{n}^{(i)}\cdot \left(\frac{1}{\rho^{(i)}}\nabla_y \hat{\xi}^{(i)}\right)
+\bm{n}^{(e)}\cdot \left(\frac{1}{\rho^{(e)}}\nabla_y \hat{\xi}^{(e)}\right)
\right\}
d\Gamma_y,
\end{align}
\begin{align}
a_{\xi 2}(\hat{\xi},\hat{v}_{\xi 2}) &=\sum_{m=i,e}\int_{\Omega_m}\left\{
\nabla_y \cdot \left(\frac{1}{\rho^{(m)}}\nabla_y \hat{\xi}^{(m)}\right)
\right\}\hat{v}_{\xi 2}^{(m)} d\Omega_y\nonumber\\
&-\int_{I_y^+}\left\{\bm{n}\cdot\left(\frac{1}{\rho^{(e)}}\nabla_y \hat{\xi}^{(e)}\right) +1 \right\}\hat{v}_{\xi 2}^{(e)}d\Gamma_y
-\int_{I_y^-}\left\{\bm{n}\cdot\left(\frac{1}{\rho^{(e)}}\nabla_y \hat{\xi}^{(e)}\right) -1 \right\}\hat{v}_{\xi 2}^{(e)}d\Gamma_y
\nonumber\\
&+\frac{1}{2}\int_{\Gamma_m}(\hat{\xi}^{(i)} - \hat{\xi}^{(e)})
\left\{
\bm{n}^{(i)}\cdot\left(\frac{1}{\rho^{(i)}}\nabla_y \hat{v}_{\xi 2}^{(i)}\right) 
-\bm{n}^{(e)}\cdot\left(\frac{1}{\rho^{(e)}}\nabla_y \hat{v}_{\xi 2}^{(e)}\right) 
\right\}
d\Gamma_y\nonumber\\
&-\frac{1}{2}\int_{\Gamma_m}(\hat{v}_{\xi 2}^{(i)}+\hat{v}_{\xi 2}^{(e)})
\left\{\bm{n}^{(i)}\cdot \left(\frac{1}{\rho^{(i)}}\nabla_y \hat{\xi}^{(i)}\right)
+\bm{n}^{(e)}\cdot \left(\frac{1}{\rho^{(e)}}\nabla_y \hat{\xi}^{(e)}\right)
\right\}
d\Gamma_y.
\end{align}
At the stationary point of the Lagrangian, the following optimality conditions hold:
\begin{align}
\left.\left<\frac{\partial L}{\partial \bm{\hat{u}_{micro}}},\delta \bm{\hat{u}_{micro}}\right>\right|_\mathrm{opt} &= 0,\label{eq: opt_u_micro}\\
\left.\left<\frac{\partial L}{\partial \bm{\hat{u}_{B}}},\delta \bm{\hat{u}_{B}}\right>\right|_\mathrm{opt} &= 0,\label{eq: opt_u_B}\\
\left.\left<\frac{\partial L}{\partial \bm{\hat{u}_{macro}}},\delta \bm{\hat{u}_{macro}}\right>\right|_\mathrm{opt} &= 0,\label{eq: opt_u_macro}\\
\left.\left<\frac{\partial L}{\partial \bm{\hat{v}_{micro}}},\delta \bm{\hat{v}_{micro}}\right>\right|_\mathrm{opt} &= 0,\label{eq: opt_v_micro}\\
\left.\left<\frac{\partial L}{\partial \bm{\hat{v}_{B}}},\delta \bm{\hat{v}_{B}}\right>\right|_\mathrm{opt} &= 0,\label{eq: opt_v_B}\\
\left.\left<\frac{\partial L}{\partial \bm{\hat{v}_{macro}}},\delta \bm{\hat{v}_{macro}}\right>\right|_\mathrm{opt} &= 0,\label{eq: opt_v_macro}
\end{align}
where the expressions within the brackets represent the directional derivatives of the functional.
The optimality conditions, defined by Eq.~(\ref{eq: opt_v_micro})--(\ref{eq: opt_v_macro}), reveal that variable $\hat{\bm{U}}$ coincides with state variable $\bm{U}$.

The optimality conditions defined by Eq.~(\ref{eq: opt_u_micro})--(\ref{eq: opt_u_macro}) are considered in the following equations. 
First, Eq.~(\ref{eq: opt_u_macro}) is examined as follows:
\begin{align}
0=\left.\left<\frac{\partial L}{\partial \bm{\hat{u}_{macro}}},\delta \bm{\hat{u}_{macro}}\right>\right|_\mathrm{opt} &=2\mathrm{Re}\left[
\int_{\Gamma_\mathrm{out1}}\frac{\partial g_1(P^-)}{\partial {P^-}}\delta \hat{P}^- d\Gamma_x
+\int_{\Gamma_\mathrm{out2}}\frac{\partial g_2(P^-)}{\partial {P^-}}\delta \hat{P}^- d\Gamma_x
\right]
\nonumber\\
&+2\mathrm{Re}\left[
\sum_{i=1}^{5}\left.\left<\frac{\partial C_i}{\partial \hat{\bm{u}}_{macro} },\delta \hat{\bm{u}}_{macro}\right>\right|_\mathrm{opt}
\right],
\end{align}
where $\left.\left<\frac{\partial C_i}{\partial \hat{\bm{u}}_{macro} },\delta \hat{\bm{u}}_{macro}\right>\right|_\mathrm{opt} ~(i=1,...,5)$
are obtained as follows:
\begin{align}
\left.\left<\frac{\partial C_1}{\partial \hat{\bm{u}}_{macro} },\delta \hat{\bm{u}}_{macro}\right>\right|_\mathrm{opt}&=
\int_{\Omega^+}\frac{1}{\rho_0}\nabla_x \delta \hat{P}^+\cdot \nabla_x {Q}^+ d\Omega_x -\int_{\Omega^+}\frac{\omega^2}{K_0}\delta\hat{P}^+{Q}^+ d\Omega_x\nonumber\\
&-\int_{\Gamma^0}\delta\hat{G}_0^+ {Q}^+ d\Gamma_x
+\int_{\Gamma_\mathrm{in}}\frac{i k_0}{\rho_0}\delta\hat{P}^+{Q}^+ d\Gamma_x\nonumber\\
\left.\left<\frac{\partial C_2}{\partial \hat{\bm{u}}_{macro} },\delta \hat{\bm{u}}_{macro}\right>\right|_\mathrm{opt}&=
\int_{\Omega^-}\frac{1}{\rho_0}\nabla_x \delta\hat{P}^-\cdot \nabla_x {Q}^- d\Omega_x -\int_{\Omega^-}\frac{\omega^2}{K_0}\delta\hat{P}^-{Q}^- d\Omega_x\nonumber\\
&+\int_{\Gamma^0}\delta\hat{G}_0^- {Q}^- d\Gamma_x
+\int_{\Gamma_\mathrm{out1}\cup \Gamma_\mathrm{out2}}\frac{i k_0}{\rho_0}\delta\hat{P}^-{Q}^- d\Gamma_x,\nonumber\\
\left.\left<\frac{\partial C_3}{\partial \hat{\bm{u}}_{macro} },\delta \hat{\bm{u}}_{macro}\right>\right|_\mathrm{opt}&=
{A}_{11}^\ast \int_{\Gamma^0} \frac{\partial \delta\hat{p}_0}{\partial x_1}
\frac{\partial {q}_0}{\partial x_1} d\Gamma_x
-{K}^{-1\ast}\omega^2 \int_{\Gamma^0}\delta\hat{p}_0{q}_0 d\Gamma_x\nonumber\\
&+\frac{1}{2}{B}_1^\ast \int_{\Gamma^0}(\delta\hat{G}_0^+ + \delta\hat{G}_0^-)\frac{\partial {q}_0}{\partial x_1} d\Gamma_x
+\frac{1}{\epsilon_0} \int_{\Gamma^0}(\delta\hat{G}_0^+ - \delta\hat{G}_0^-){q}_0 d\Gamma_x,\nonumber\\
\left.\left<\frac{\partial C_4}{\partial \hat{\bm{u}}_{macro} },\delta \hat{\bm{u}}_{macro}\right>\right|_\mathrm{opt}&=
{B}_1^\ast\int_{\Gamma^0}\frac{\partial \delta\hat{p}_0}{\partial x_1}{\Psi}_0^+ d\Gamma_x -\frac{1}{2}{F}^{\ast}\int_{\Gamma^0}(\delta\hat{G}_0^+ + \delta\hat{G}_0^-){\Psi}_0^+ d\Gamma_x\nonumber\\
&-\frac{1}{\epsilon_0}\int_{\Gamma^0}(\delta\hat{P}^+ - \delta\hat{P}^-){\Psi}_0^+ d\Gamma_x,\\
\left.\left<\frac{\partial C_5}{\partial \hat{\bm{u}}_{macro} },\delta \hat{\bm{u}}_{macro}\right>\right|_\mathrm{opt}&=
\int_{\Gamma^0}\left\{
\delta\hat{p}_0 - \frac{1}{2}(\delta\hat{P}^+ + \delta\hat{P}^-)
\right\}{\Psi}_0^- d\Gamma_x.
\end{align}
To satisfy the optimality condition [Eq. (\ref{eq: opt_u_macro})], $\bm{v}_{macro}=({q}^0, {Q}^\pm, {\Psi}_0^\pm)$ should satisfy the following adjoint equation defined at the macroscale:
\begin{align}
&\int_{\Omega^+}\frac{1}{\rho_0}\nabla_x \tilde {Q}\cdot \nabla_x {Q}^+ d\Omega_x -\int_{\Omega^+}\frac{\omega^2}{K_0}\tilde{Q}{Q}^+ d\Omega_x+\int_{\Gamma_\mathrm{in}}\frac{i k_0}{\rho_0}\tilde{Q}{Q}^+ d\Gamma_x\nonumber\\
&-\frac{1}{\epsilon_0}\int_{\Gamma^0}\tilde{Q} {\Psi}_0^+ d\Gamma_x
- \frac{1}{2}\int_{\Gamma^0}\tilde{Q} {\Psi}_0^- d\Gamma_x=0~~~\forall \tilde{Q}\in H^1(\Omega^+),\nonumber\\
&\int_{\Omega^-}\frac{1}{\rho_0}\nabla_x \tilde {Q}\cdot \nabla_x {Q}^- d\Omega_x -\int_{\Omega^-}\frac{\omega^2}{K_0}\tilde{Q}{Q}^- d\Omega_x+\int_{\Gamma_\mathrm{out1}\cup \Gamma_\mathrm{out2}}\frac{i k_0}{\rho_0}\tilde{Q}{Q}^- d\Gamma_x\nonumber\\
&+\frac{1}{\epsilon_0}\int_{\Gamma^0}\tilde{Q} {\Psi}_0^+ d\Gamma_x
- \frac{1}{2}\int_{\Gamma^0}\tilde{Q} {\Psi}_0^- d\Gamma_x\nonumber\\
&=-\left(
\int_{\Gamma_\mathrm{out1}}\frac{\partial g_1(P^-)}{\partial {P^-}} \tilde{Q} d\Gamma_x
+\int_{\Gamma_\mathrm{out2}}\frac{\partial g_2(P^-)}{\partial {P^-}} \tilde{Q} d\Gamma_x
\right)
~~~\forall \tilde{Q}\in H^1(\Omega^-),\nonumber\\
&{A}_{11}^\ast \int_{\Gamma^0} \frac{\partial \tilde{q}}{\partial x_1}
\frac{\partial {q}_0}{\partial x_1} d\Gamma_x
-{K}^{-1\ast}\omega^2 \int_{\Gamma^0}\tilde{q}{q}_0 d\Gamma_x\nonumber\\
&+{B}_1^\ast\int_{\Gamma^0}\frac{\partial \tilde{q}}{\partial x_1}{\Psi}_0^+ d\Gamma_x
+\int_{\Gamma^0}\tilde{q} {\Psi}_0^- d\Gamma_x=0~~~\forall \tilde{q}\in H^1(\Gamma^0),\nonumber\\
&-\int_{\Gamma^0}\tilde{\Psi} {Q}^+ d\Gamma_x
+\frac{1}{2}{B}_1^\ast \int_{\Gamma^0}\tilde{\Psi}\frac{\partial {q}_0}{\partial x_1} d\Gamma_x\nonumber\\
&+\frac{1}{\epsilon_0} \int_{\Gamma^0}\tilde{\Psi} {q}_0 d\Gamma_x
-\frac{1}{2}{F}^{\ast}\int_{\Gamma^0}\tilde{\Psi}{\Psi}_0^+ d\Gamma_x=0~~~\forall \tilde{\Psi}\in L^2(\Gamma_0),\nonumber\\
&\int_{\Gamma^0}\tilde{\Psi} {Q}^- d\Gamma_x
+\frac{1}{2}{B}_1^\ast \int_{\Gamma^0}  \tilde{\Psi}\frac{\partial {q}_0}{\partial x_1} d\Gamma_x\nonumber\\
&-\frac{1}{\epsilon_0} \int_{\Gamma^0}  \tilde{\Psi}{q}_0 d\Gamma_x
-\frac{1}{2}{F}^{\ast}\int_{\Gamma^0} \tilde{\Psi}{\Psi}_0^+ d\Gamma_x=0~~~\forall \tilde{\Psi}\in L^2(\Gamma_0).
\end{align}

Next, Eq.~(\ref{eq: opt_u_B}) is considered.
\begin{align}
0=\left.\left<\frac{\partial L}{\partial \hat{\bm{u}}_{B} },\delta \hat{\bm{u}}_{B}\right>\right|_\mathrm{opt}
&=2\mathrm{Re}\left[
\sum_{i=1}^{5}\left.\left<\frac{\partial C_i}{\partial \hat{\bm{u}}_{B} },\delta \hat{\bm{u}}_{B}\right>\right|_\mathrm{opt}
\right]\nonumber\\
&-\lambda_{{A}_{11}^\ast}\delta \hat{A}_{11}^\ast
-\lambda_{{B}_{1}^\ast}\delta \hat{B}_{1}^\ast\nonumber\\
&-\lambda_{{F}^\ast}\delta \hat{F}^\ast
-\lambda_{{K}^{-1 \ast}}\delta \hat{K}^{-1 \ast},
\end{align}
where $\left.\left<\frac{\partial C_i}{\partial \hat{\bm{u}}_{B} },\delta \hat{\bm{u}}_{B}\right>\right|_\mathrm{opt}$ are calculated as follows:
\begin{align}
\left.\left<\frac{\partial C_1}{\partial \hat{\bm{u}}_{B} },\delta \hat{\bm{u}}_{B}\right>\right|_\mathrm{opt}
&=0,~~~
\left.\left<\frac{\partial C_2}{\partial \hat{\bm{u}}_{B} },\delta \hat{\bm{u}}_{B}\right>\right|_\mathrm{opt}
=0,\nonumber\\
\left.\left<\frac{\partial C_3}{\partial \hat{\bm{u}}_{B} },\delta \hat{\bm{u}}_{B}\right>\right|_\mathrm{opt}
&=\delta \hat{A}_{11}^\ast \int_{\Gamma^0} \frac{\partial {p}_0}{\partial x_1}
\frac{\partial {q}_0}{\partial x_1} d\Gamma_x
-\delta \hat{K}^{-1\ast}\omega^2 \int_{\Gamma^0}{p}_0 {q}_0 d\Gamma_x\nonumber\\
&+\frac{1}{2}\delta\hat{B}_1^\ast \int_{\Gamma^0}({G}_0^+ + {G}_0^-)\frac{\partial {q}_0}{\partial x_1} d\Gamma_x,\nonumber\\
\left.\left<\frac{\partial C_4}{\partial \hat{\bm{u}}_{B} },\delta \hat{\bm{u}}_{B}\right>\right|_\mathrm{opt}
&=\delta\hat{B}_1^\ast\int_{\Gamma^0}\frac{\partial {p}_0}{\partial x_1}{\Psi}_0^+ d\Gamma_x 
-\frac{1}{2}\delta \hat{F}^{\ast}\int_{\Gamma^0}({G}_0^+ + {G}_0^-){\Psi}_0^+ d\Gamma_x\nonumber\\
\left.\left<\frac{\partial C_5}{\partial \hat{\bm{u}}_{B} },\delta \hat{\bm{u}}_{B}\right>\right|_\mathrm{opt}
&=0,
\end{align}
where the adjoint variables at the macroscale of $\bm{v}_{macro}=(q^0, Q^\pm, \Psi_0^\pm)$ are used.
To satisfy the optimality condition [Eq. (\ref{eq: opt_u_B})], the optimal Lagrange multipliers of $\bm{v}_B =({\lambda}_{{A}_{11}^\ast}, {\lambda}_{{B}_{1}^\ast},{\lambda}_{{F}^\ast}, {\lambda}_{{K}^{-1\ast}} )$ are defined as follows:
\begin{align}
{\lambda}_{{A}_{11}^\ast} &=2\mathrm{Re}\left[
\int_{\Gamma^0}\frac{\partial p^0}{\partial x_1}\frac{\partial q^0}{\partial x_1}d\Gamma_x\right],\nonumber\\
{\lambda}_{{B}_{1}^\ast} &=2\mathrm{Re}\left[
\frac{1}{2} \int_{\Gamma^0}({G}_0^+ + {G}_0^-)\frac{\partial {q}_0}{\partial x_1} d\Gamma_x
+\int_{\Gamma^0}\frac{\partial {p}_0}{\partial x_1}{\Psi}_0^+ d\Gamma_x 
\right],\nonumber\\
{\lambda}_{{F}^\ast}&=2\mathrm{Re}\left[
-\frac{1}{2}\int_{\Gamma^0}({G}_0^+ + {G}_0^-){\Psi}_0^+ d\Gamma_x
\right],\nonumber\\
{\lambda}_{{K}^{-1\ast}}&=2\mathrm{Re}\left[
-\omega^2 \int_{\Gamma^0}{p}_0 {q}_0 d\Gamma_x
\right].
\end{align}

Finally, Eq.~(\ref{eq: opt_u_micro}) is considered.
\begin{align}
0=\left.\left<\frac{\partial L}{\partial \bm{\hat{u}_{micro}}},\delta \bm{\hat{u}_{micro}}\right>\right|_\mathrm{opt} &=
\lambda_{{A}_{11}^\ast}\left(\left.\left<\frac{\partial {A}_{11}^\ast}{\partial \hat{{\textcolor{black}{\eta}}}},\delta \hat{{\textcolor{black}{\eta}}}\right>\right|_\mathrm{opt}
-\left.\left<\frac{\partial a_{\textcolor{black}{\eta}}}{\partial \hat{{\textcolor{black}{\eta}}}},\delta \hat{{\textcolor{black}{\eta}}}\right>\right|_\mathrm{opt}\right)\nonumber\\
&+\lambda_{{B}_{1}^\ast}\left(
\left.\left<\frac{\partial B_1^\ast}{\partial \hat{\xi}},\delta \hat{\xi}\right>\right|_\mathrm{opt}
-\left.\left<\frac{\partial a_{\xi 1}}{\partial \hat{\xi}},\delta \hat{\xi}\right>\right|_\mathrm{opt}\right)\nonumber\\
&+\lambda_{{F}^\ast}\left(
\left.\left<\frac{\partial F^\ast}{\partial \hat{\xi}},\delta \hat{\xi}\right>\right|_\mathrm{opt}
+\left.\left<\frac{\partial a_{\xi 2}}{\partial \hat{\xi}},\delta \hat{\xi}\right>\right|_\mathrm{opt}\right),
\label{eq: dLdu_micro}
\end{align}
where the optimal values of Lagrange multipliers $\bm{v}_B = (\lambda_{{A}_{11}^\ast},\lambda_{{B}_{1}^\ast},\lambda_{{F}^\ast},\lambda_{{K}^{-1\ast}})$ are used.
The directional derivatives in the first line can be calculated as follows:
\begin{align}
&\left.\left<\frac{\partial {A}_{11}^\ast}{\partial \hat{{\textcolor{black}{\eta}}}},\delta \hat{{\textcolor{black}{\eta}}}\right>\right|_\mathrm{opt}
-\left.\left<\frac{\partial a_{\textcolor{black}{\eta}}}{\partial \hat{{\textcolor{black}{\eta}}}},\delta \hat{{\textcolor{black}{\eta}}}\right>\right|_\mathrm{opt}\nonumber\\
&=\sum_{m=i,e}\int_{\Omega_m}\frac{1}{\rho^{(m)}}\left(\bm{e_1}\cdot \nabla_y \delta \hat{{\textcolor{black}{\eta}}}^{(m)}   \right)d\Omega_y\nonumber\\
&-\sum_{m=i,e}\int_{\Omega_m}\nabla_y\cdot \left( \frac{1}{\rho^{(m)}}\nabla_y\delta \hat{{\textcolor{black}{\eta}}}^{(m)} \right)v_{\textcolor{black}{\eta}}^{(m)}d\Omega_y
+\int_{I_y^\pm}\bm{n}\cdot\left( \frac{1}{\rho^{(e)}} \nabla_y \delta \hat{{\textcolor{black}{\eta}}}^{(e)} \right)v_{\textcolor{black}{\eta}}^{(e)}d\Gamma_y\nonumber\\
&-\frac{1}{2}\int_{\Gamma_m}(\delta \hat{{\textcolor{black}{\eta}}}^{(i)} - \delta \hat{{\textcolor{black}{\eta}}}^{(e)})
\left[
\left\{
\bm{n}^{(i)}\cdot\left(\frac{1}{\rho^{(i)}}\nabla_y {v}_{\textcolor{black}{\eta}}^{(i)}\right) + \bm{n}^{(i)}\cdot\left(\frac{1}{\rho^{(i)}}\bm{e_1}\right)
\right\}\right.\nonumber\\
&\left.-\left\{
\bm{n}^{(e)}\cdot\left(\frac{1}{\rho^{(e)}}\nabla_y {v}_{\textcolor{black}{\eta}}^{(e)}\right) + \bm{n}^{(e)}\cdot\left(\frac{1}{\rho^{(e)}}\bm{e_1}\right)
\right\}
\right]d\Gamma_y\nonumber\\
&+\frac{1}{2}\int_{\Gamma_m}({v}_{\textcolor{black}{\eta}}^{(i)}+{v}_{\textcolor{black}{\eta}}^{(e)})\left[
\bm{n}^{(i)}\cdot \left(\frac{1}{\rho^{(i)}}\nabla_y \delta\hat{{\textcolor{black}{\eta}}}^{(i)}\right)
+\bm{n}^{(e)}\cdot \left(\frac{1}{\rho^{(e)}}\nabla_y \delta\hat{{\textcolor{black}{\eta}}}^{(e)}\right)
\right]d\Gamma_y\nonumber\\
&=-\sum_{m=i,e}\int_{\Omega_m}\delta \hat{{\textcolor{black}{\eta}}}^{(m)}\left\{
 \nabla_y\cdot\left(\frac{1}{\rho^{(m)}}\nabla_y v_{\textcolor{black}{\eta}}^{(m)}\right) + \nabla_y \cdot\left(\frac{1}{\rho^{(m)}}\bm{e_1}\right)
\right\}d\Omega_y\nonumber\\
&+\int_{I_y^\pm}\delta\hat{{\textcolor{black}{\eta}}}^{(e)}\left\{
\bm{n}\cdot \left(\frac{1}{\rho^{(e)}}\nabla_y v_{\textcolor{black}{\eta}}^{(e)}\right)
+\bm{n}\cdot \left(\frac{1}{\rho^{(e)}}\bm{e_1}\right)
\right\}d\Gamma_y\nonumber\\
&+\frac{1}{2}\int_{\Gamma_m}\delta\hat{{\textcolor{black}{\eta}}}^{(i)}\left[
\left\{
\bm{n}^{(i)}\cdot \left(\frac{1}{\rho^{(i)}}\nabla_y v_{\textcolor{black}{\eta}}^{(i)}\right)
+\bm{n}^{(i)}\cdot \left(\frac{1}{\rho^{(i)}}\bm{e_1}\right)
\right\}\right.\nonumber\\
&+\left.
\left\{
\bm{n}^{(e)}\cdot \left(\frac{1}{\rho^{(e)}}\nabla_y v_{\textcolor{black}{\eta}}^{(e)}\right)
+\bm{n}^{(e)}\cdot \left(\frac{1}{\rho^{(e)}}\bm{e_1}\right)
\right\}
\right]d\Gamma_y\nonumber\\
&+\frac{1}{2}\int_{\Gamma_m}\delta\hat{{\textcolor{black}{\eta}}}^{(e)}\left[
\left\{
\bm{n}^{(i)}\cdot \left(\frac{1}{\rho^{(i)}}\nabla_y v_{\textcolor{black}{\eta}}^{(i)}\right)
+\bm{n}^{(i)}\cdot \left(\frac{1}{\rho^{(i)}}\bm{e_1}\right)
\right\}\right.\nonumber\\
&+\left.
\left\{
\bm{n}^{(e)}\cdot \left(\frac{1}{\rho^{(e)}}\nabla_y v_{\textcolor{black}{\eta}}^{(e)}\right)
+\bm{n}^{(e)}\cdot \left(\frac{1}{\rho^{(e)}}\bm{e_1}\right)
\right\}
\right]d\Gamma_y\nonumber\\
&-\frac{1}{2}\int_{\Gamma_m}\bm{n}^{(i)}\cdot\left(\frac{1}{\rho^{(i)}}\nabla_y\delta\hat{{\textcolor{black}{\eta}}}^{(i)}
\right)(v_{\textcolor{black}{\eta}}^{(i)} - v_{\textcolor{black}{\eta}}^{(e)})d\Gamma_y\nonumber\\
&-\frac{1}{2}\int_{\Gamma_m}\bm{n}^{(e)}\cdot\left(\frac{1}{\rho^{(e)}}\nabla_y\delta\hat{{\textcolor{black}{\eta}}}^{(e)}
\right)(v_{\textcolor{black}{\eta}}^{(e)} - v_{\textcolor{black}{\eta}}^{(i)})d\Gamma_y
\end{align}
These directional derivatives are canceled if $v_{\textcolor{black}{\eta}} \in H^1_{\underline{\sharp}}(Y)$ satisfies the following adjoint equation:
\begin{align}
&\nabla_y\cdot \left(\frac{1}{\rho^{(m)}}\nabla_y v_{\textcolor{black}{\eta}}^{(m)}\right) + \nabla_y\cdot\left(\frac{1}{\rho^{(m)}}\bm{e_1}\right)
=0~~~\mathrm{in~}\Omega_m~(m=i,e),\nonumber\\
&v_{\textcolor{black}{\eta}}^{(i)}= v_{\textcolor{black}{\eta}}^{(e)}~~~\mathrm{on~}\Gamma_m,\nonumber\\
&\left\{\bm{n}^{(i)}\cdot\left(\frac{1}{\rho^{(i)}}\nabla_y v_{\textcolor{black}{\eta}}^{(i)}\right)
+\bm{n}^{(i)}\cdot\left(\frac{1}{\rho^{(i)}}\bm{e_1}\right)\right\}\nonumber\\
&+\left\{\bm{n}^{(e)}\cdot\left(\frac{1}{\rho^{(e)}}\nabla_y v_{\textcolor{black}{\eta}}^{(e)}\right)
+\bm{n}^{(e)}\cdot\left(\frac{1}{\rho^{(e)}}\bm{e_1}\right)\right\}=0~~~\mathrm{on~}\Gamma_m,\nonumber\\
&\bm{n}\cdot\left(\frac{1}{\rho^{(e)}}\nabla_y v_{\textcolor{black}{\eta}}^{(e)} \right)
+\bm{n}\cdot\left(\frac{1}{\rho^{(e)}}\bm{e_1} \right)=0~~~\mathrm{on~}I_y^{\pm}.
\end{align}
This is the same as that for the strong form of the cell problem for ${\textcolor{black}{\eta}}$, which indicates that $v_{\textcolor{black}{\eta}} = {\textcolor{black}{\eta}}$.
Similarly, considering the directional derivatives in the second and third lines in Eq.~(\ref{eq: dLdu_micro}), $v_{\xi 1}={\textcolor{black}{\eta}}$ and $v_{\xi 2}=\xi$. \textcolor{black}{In other words}, the microscale problem is considered as a self-adjoint problem.
By using variables $\bm{v}_{micro}=(v_{{\textcolor{black}{\eta}}},v_{\xi 1},v_{\xi 2})$, the optimality condition given in Eq.~(\ref{eq: opt_u_micro}) is satisfied.

\textcolor{black}{Furthermore, by} using the state and adjoint variables of $\bm{U}$ and $\bm{V}$, respectively, we can derive the shape derivative of $L$.
Here, we \textcolor{black}{employ the} formulas used in \cite{allaire2004structural} for deriving the shape derivative.
If functional $G$ is defined as a domain integral with its integrand $g$ expressed as 
\begin{align}
G=\int_{\Omega}g d\Omega,
\end{align}
then its shape derivative is derived as
\begin{align}
DG\cdot\bm{\theta}=\int_{\partial \Omega}(\bm{\theta}\cdot \bm{n})g d\Gamma,
\end{align}
where $\bm{n}$ is the outward-pointing normal-unit vector on $\partial \Omega$.
However, if \textcolor{black}{the} functional $G$ is defined as the following boundary integral:
\begin{align}
G=\int_{\partial \Omega}g d\Gamma,
\end{align}
then its shape derivative is obtained as 
\begin{align}
DG\cdot\bm{\theta}=\int_{\partial \Omega}(\bm{\theta} \cdot\bm{n})\left( \nabla g \cdot \bm{n}  +\kappa g     \right) d\Gamma,
\end{align}
where $\kappa\equiv\mathrm{div}\bm{n}$ is the mean curvature of $\partial \Omega$. 

By applying these formulas to \textcolor{black}{the} Lagrangian (\ref{eq: Lagrangian}), the shape derivative is obtained 
as follows:
\begin{align}
DL\cdot \bm{\theta} 
&=\lambda_{{A}_{11}^\ast}\left\{
DA_{11}^\ast(\Omega,{\textcolor{black}{\eta}})\cdot \bm{\theta}-Da_{\textcolor{black}{\eta}}(\Omega,{\textcolor{black}{\eta}},{\textcolor{black}{\eta}})\cdot \bm{\theta}
\right\}\nonumber\\
&+\lambda_{{B}_{1}^\ast}\left\{
DB_{1}^\ast(\Omega,\xi)\cdot \bm{\theta}-Da_{\xi 1}(\Omega,\xi,{\textcolor{black}{\eta}})\cdot \bm{\theta}
\right\}\nonumber\\
&+\lambda_{{K}^{-1 \ast}}DK^{-1\ast}(\Omega)\cdot \bm{\theta}
+\lambda_{{F}^\ast}Da_{\xi 2}(\Omega,\xi,\xi)\cdot \bm{\theta},
\label{eq: shape derivative befor limit}
\end{align} 
where each term is expressed as
\begin{align}
&DA_{11}^\ast(\Omega,{\textcolor{black}{\eta}})\cdot \bm{\theta}-Da_{\textcolor{black}{\eta}}(\Omega,{\textcolor{black}{\eta}},{\textcolor{black}{\eta}})\cdot \bm{\theta}\nonumber\\
&=\sum_{m=i,e}\int_{\Gamma_m}\frac{1}{\rho^{(m)}}\left(\bm{e_1}\cdot \nabla_y {\textcolor{black}{\eta}}^{(m)}+1\right)
(\bm{\theta}\cdot\bm{n}^{(m)})d\Gamma_y\nonumber\\
&+\sum_{m=i,e}\int_{\Gamma_m}\left\{
\frac{1}{\rho^{(m)}}\left(\nabla_y {\textcolor{black}{\eta}}^{(m)} \cdot \nabla_y {\textcolor{black}{\eta}}^{(m)}+\bm{e_1}\cdot\nabla_y{\textcolor{black}{\eta}}^{(m)}
\right)
\right\}(\bm{\theta}\cdot\bm{n}^{(m)})d\Gamma_y\nonumber\\
&-\int_{\Gamma_m}(\bm{\theta}\cdot\bm{n}^{(i)})\left[
\bm{n}^{(i)}\cdot\left\{\frac{1}{\rho^{(i)}}(\nabla_y{\textcolor{black}{\eta}}^{(i)}+\bm{e_1})\right\}\right.\nonumber\\
&\left.
-\bm{n}^{(e)}\cdot\left\{\frac{1}{\rho^{(e)}}(\nabla_y{\textcolor{black}{\eta}}^{(e)}+\bm{e_1})\right\}
\right](\nabla_y{\textcolor{black}{\eta}}^{(i)} - \nabla_y{\textcolor{black}{\eta}}^{(e)})\cdot \bm{n}^{(i)}d\Gamma_y,\\
&DB_{1}^\ast(\Omega,\xi)\cdot \bm{\theta}-Da_{\xi 1}(\Omega,\xi,{\textcolor{black}{\eta}})\cdot \bm{\theta}\nonumber\\
&=\sum_{m=i,e}\int_{\Gamma_m}\frac{1}{\rho^{(m)}}\frac{\partial \xi^{(m)}}{\partial y_1}
(\bm{\theta}\cdot\bm{n}^{(m)})d\Gamma_y
+\sum_{m=i,e}\int_{\Gamma_m}\frac{1}{\rho^{(m)}}\nabla_y\xi^{(m)}\cdot\nabla_y{\textcolor{black}{\eta}}^{(m)}(\bm{\theta}\cdot\bm{n}^{(m)})d\Gamma_y\nonumber\\
&-\frac{1}{2}\int_{\Gamma_m}(\bm{\theta}\cdot\bm{n}^{(i)})
\left[\bm{n}^{(i)}\cdot\left\{\frac{1}{\rho^{(i)}}(\nabla_y{\textcolor{black}{\eta}}^{(i)}+\bm{e_1})\right\}\right.\nonumber\\
&\left.-\bm{n}^{(e)}\cdot\left\{\frac{1}{\rho^{(e)}}(\nabla_y{\textcolor{black}{\eta}}^{(e)}+\bm{e_1})\right\}
\right](\nabla_y\xi^{(i)} - \nabla_y\xi^{(e)})\cdot\bm{n}^{(i)}d\Gamma_y\nonumber\\
&-\frac{1}{2}\int_{\Gamma_m}(\bm{\theta}\cdot\bm{n}^{(i)})
\left[\bm{n}^{(i)}\cdot\left(\frac{1}{\rho^{(i)}}\nabla_y\xi^{(i)}\right)-\bm{n}^{(e)}\cdot\left(\frac{1}{\rho^{(e)}}\nabla_y\xi^{(e)}\right)
\right](\nabla_y{\textcolor{black}{\eta}}^{(i)} - \nabla_y{\textcolor{black}{\eta}}^{(e)})\cdot\bm{n}^{(i)}d\Gamma_y,\\
&Da_{\xi 2}(\Omega,\xi,\xi)\cdot \bm{\theta}\nonumber\\
&=-\sum_{m=i,e}\int_{\Gamma_m}\frac{1}{\rho^{(m)}}\nabla_y \xi^{(m)}\cdot \nabla_y \xi^{(m)}
(\bm{\theta}\cdot\bm{n}^{(m)})d\Gamma_y\nonumber\\
&+\int_{\Gamma_m}(\bm{\theta}\cdot\bm{n}^{(i)})\left[
\bm{n}^{(i)}\cdot\left(\frac{1}{\rho^{(i)}}\nabla_y\xi^{(i)}\right)
-\bm{n}^{(e)}\cdot\left(\frac{1}{\rho^{(e)}}\nabla_y\xi^{(e)}\right)
\right](\nabla_y\xi^{(i)} - \nabla_y\xi^{(e)})\cdot \bm{n}^{(i)}d\Gamma_y,\\
&DK^{-1\ast}(\Omega)\cdot \bm{\theta}
=\sum_{m=i,e}\int_{\Gamma_m}\frac{1}{K^{(m)}}(\bm{\theta}\cdot \bm{n}^{(m)})d\Gamma_y.
\end{align}

~\\

\textbf{Step2:} Analysis of the asymptotic behaviors of state variables $({\textcolor{black}{\eta}}, \xi)$ as $\varepsilon\to 0$

~\\

To take the limit of the shape derivative defined in Eq.~(\ref{eq: shape derivative befor limit}), the asymptotic behavior of solutions in microscale $({\textcolor{black}{\eta}},\xi)$ should be evaluated when $\varepsilon\to 0$.

First, we consider the solution of cell problem ${\textcolor{black}{\eta}}$.
To simplify the boundary value problem for ${\textcolor{black}{\eta}}$,
we introduce $\overline{{\textcolor{black}{\eta}}}_\varepsilon(\bm{y})={\textcolor{black}{\eta}}_\varepsilon(\bm{y}) + y_1$, which satisfies the following boundary value problem:
\begin{align}
&\nabla_y\cdot \left(\frac{1}{\rho^{(m)}}\nabla_y \overline{{\textcolor{black}{\eta}}}_\varepsilon^{(m)}\right) 
=0~~~\mathrm{in~}\Omega_m~(m=i,e),\nonumber\\
&\overline{{\textcolor{black}{\eta}}}_\varepsilon^{(i)}= \overline{{\textcolor{black}{\eta}}}_\varepsilon^{(e)}~~~\mathrm{on~}\Gamma_m,\nonumber\\
&\bm{n}^{(i)}\cdot\left(\frac{1}{\rho^{(i)}}\nabla_y \overline{{\textcolor{black}{\eta}}}_\varepsilon^{(i)}\right)
+\bm{n}^{(e)}\cdot\left(\frac{1}{\rho^{(e)}}\nabla_y \overline{{\textcolor{black}{\eta}}}_\varepsilon^{(e)}\right)
=0~~~\mathrm{on~}\Gamma_m,\nonumber\\
&\bm{n}\cdot\left(\frac{1}{\rho^{(e)}}\nabla_y \overline{{\textcolor{black}{\eta}}}_\varepsilon^{(e)} \right)
=0~~~\mathrm{on~}I_y^{\pm},\nonumber\\
&\left.\overline{{\textcolor{black}{\eta}}}_\varepsilon^{(e)}\right|_{\Gamma_1}-\left.\overline{{\textcolor{black}{\eta}}}_\varepsilon^{(e)}\right|_{\Gamma_2}=\left. y_1\right|_{\Gamma_1} - \left. y_1\right|_{\Gamma_2},
\end{align}
where \textcolor{black}{the} subscript $\varepsilon$ represents quantities, as $\varepsilon$ approaches zero.
Then, we consider the expansion of $\overline{{\textcolor{black}{\eta}}}_\varepsilon(\bm{y}) = \overline{{\textcolor{black}{\eta}}}_0(\bm{y}) + h_\varepsilon(\bm{y})$, where $\overline{{\textcolor{black}{\eta}}}_0(\bm{y}) = {\textcolor{black}{\eta}}_0(\bm{y}) + y_1$, and subscript $0$ represents quantities when \textcolor{black}{the} domain $\Omega_i$ does not appear. The remainder of $h_\varepsilon(\bm{y})$ should satisfy the following boundary value problem:
\begin{align}
&\nabla_y\cdot \left(\frac{1}{\rho^{(m)}}\nabla_y {h}_\varepsilon^{(m)}\right) 
=-\nabla_y\cdot \left(\frac{1}{\rho^{(m)}}\nabla_y \overline{{\textcolor{black}{\eta}}}_0\right) ~~~\mathrm{in~}\Omega_m~(m=i,e),\nonumber\\
&{h}_\varepsilon^{(i)}= {h}_\varepsilon^{(e)}~~~\mathrm{on~}\Gamma_m,\nonumber\\
&\bm{n}^{(i)}\cdot\left(\frac{1}{\rho^{(i)}}\nabla_y {h}_\varepsilon^{(i)}\right)
+\bm{n}^{(e)}\cdot\left(\frac{1}{\rho^{(e)}}\nabla_y {h}_\varepsilon^{(e)}\right)\nonumber\\
&=-\left\{
\bm{n}^{(i)}\cdot\left(\frac{1}{\rho^{(i)}}\nabla_y \overline{{\textcolor{black}{\eta}}}_0 \right)
+\bm{n}^{(e)}\cdot\left(\frac{1}{\rho^{(e)}}\nabla_y \overline{{\textcolor{black}{\eta}}}_0 \right)
\right\}~~~\mathrm{on~}\Gamma_m,\nonumber\\
&\bm{n}\cdot\left(\frac{1}{\rho^{(e)}}\nabla_y {h}_\varepsilon^{(e)} \right)
=-\bm{n}\cdot\left(\frac{1}{\rho^{(e)}}\nabla_y \overline{{\textcolor{black}{\eta}}}_0 \right)~~~\mathrm{on~}I_y^{\pm},\nonumber\\
&\left.{h}_\varepsilon^{(e)}\right|_{\Gamma_1}-\left.{h}_\varepsilon^{(e)}\right|_{\Gamma_2}=0.
\end{align}
We approximate the remainder, $h_\varepsilon(\bm{y})$, by $\varepsilon H(\bm{\zeta})$, which is defined in the scaled coordinate, $\bm{\zeta} = (\bm{y}-\bm{y_0})/{\varepsilon}$, by using $\bm{y_0}$ that represents the center coordinate of $\Omega_i$.
$H(\bm{\zeta})$ satisfies the following approximated boundary value problem:
\begin{align}
&\nabla_\zeta \cdot \left(\frac{1}{\rho^{(m)}}\nabla_\zeta {H}^{(m)}\right) 
=0 ~~~\mathrm{in~}\Omega_m~(m=i,e),\nonumber\\
&{H}^{(i)}= {H}^{(e)}~~~\mathrm{on~}|\bm{\zeta}|=1,\nonumber\\
&\bm{n}^{(i)}\cdot\left(\frac{1}{\rho^{(i)}}\nabla_\zeta {H}^{(i)}\right)
+\bm{n}^{(e)}\cdot\left(\frac{1}{\rho^{(e)}}\nabla_\zeta {H}^{(e)}\right)\nonumber\\
&=-\left\{
\bm{n}^{(i)}\cdot \left.\left(\frac{1}{\rho^{(i)}}\nabla_y \overline{{\textcolor{black}{\eta}}}_0 \right)\right|_{\bm{y}=\bm{y_0}}
+\bm{n}^{(e)}\cdot \left.\left(\frac{1}{\rho^{(e)}}\nabla_y \overline{{\textcolor{black}{\eta}}}_0 \right)\right|_{\bm{y}=\bm{y_0}}
\right\}~~~\mathrm{on~}|\bm{\zeta}|=1,\nonumber\\
&{H}^{(e)}\to 0~~~\mathrm{when~}|\bm{\zeta}|\to \infty.
\end{align}
The solution of this problem can be elucidated in the scaled polar coordinate, $(\rho,\theta)$, with $\rho=|\bm{\zeta}|$ as follows:
\begin{align}
H^{(i)}(\rho,\theta)&=\rho(D_1 \sin \theta +D_2 \cos \theta),\nonumber\\
H^{(e)}(\rho,\theta)&=\frac{1}{\rho}(D_1 \sin \theta +D_2 \cos \theta),
\end{align}
where constants $D_1$ and $D_2$ are determined by the boundary conditions on $\rho=1$ as
\begin{align}
D_1 = -\left(\frac{1}{\rho^{(i)}}+\frac{1}{\rho^{(e)}}\right)^{-1}\left(\frac{1}{\rho^{(i)}}-\frac{1}{\rho^{(e)}}\right)\left.\frac{\partial \overline{{\textcolor{black}{\eta}}}_0}{\partial y_2}\right|_{\bm{y}=\bm{y_0}},\nonumber\\
D_2 = -\left(\frac{1}{\rho^{(i)}}+\frac{1}{\rho^{(e)}}\right)^{-1}\left(\frac{1}{\rho^{(i)}}-\frac{1}{\rho^{(e)}}\right)\left.\frac{\partial \overline{{\textcolor{black}{\eta}}}_0}{\partial y_1}\right|_{\bm{y}=\bm{y_0}}.\nonumber
\end{align}
This solution of $H(\bm{\zeta})$ expresses a leading part of the remainder, $h_\varepsilon$, when considering limit $\varepsilon \to 0$.
The error estimations for this approximation are possible \textcolor{black}{by} using the method in \cite{Amstutz2006}; however, we used these formulas without rigorous mathematical proofs.
Then, the asymptotic behavior of solution ${\textcolor{black}{\eta}}_\varepsilon$ when $\varepsilon \to 0$ can be expressed as follows:
\begin{align}
{\textcolor{black}{\eta}}_\varepsilon &\to {\textcolor{black}{\eta}}_0(\bm{y_0}),\nonumber\\
\nabla_y{\textcolor{black}{\eta}}_\varepsilon &\to \left.\nabla_y {\textcolor{black}{\eta}}_0\right|_{\bm{y}=\bm{y_0}}+\nabla_\zeta H(\bm{\zeta}),
\end{align}
where we used the smoothness of solution ${\textcolor{black}{\eta}}_0$ and the chain-rule for the derivative.

Similarly, the asymptotic behavior of solution $\xi_\varepsilon$ when $\varepsilon\to 0$ can be expressed as follows:
\begin{align}
\xi_\varepsilon &\to \xi_0(\bm{y_0}),\nonumber\\
\nabla_y\xi_\varepsilon &\to \left.\nabla_y \xi_0\right|_{\bm{y}=\bm{y_0}}+\nabla_\zeta K(\bm{\zeta}),
\end{align}
where $K(\bm{\zeta})$ is a solution to the following boundary value problem:
\begin{align}
&\nabla_\zeta \cdot \left(\frac{1}{\rho^{(m)}}\nabla_\zeta {K}^{(m)}\right) 
=0 ~~~\mathrm{in~}\Omega_m~(m=i,e),\nonumber\\
&{K}^{(i)}= {K}^{(e)}~~~\mathrm{on~}|\bm{\zeta}|=1,\nonumber\\
&\bm{n}^{(i)}\cdot\left(\frac{1}{\rho^{(i)}}\nabla_\zeta {K}^{(i)}\right)
+\bm{n}^{(e)}\cdot\left(\frac{1}{\rho^{(e)}}\nabla_\zeta {K}^{(e)}\right)\nonumber\\
&=-\left\{
\bm{n}^{(i)}\cdot \left.\left(\frac{1}{\rho^{(i)}}\nabla_y {\xi}_0 \right)\right|_{\bm{y}=\bm{y_0}}
+\bm{n}^{(e)}\cdot \left.\left(\frac{1}{\rho^{(e)}}\nabla_y {\xi}_0 \right)\right|_{\bm{y}=\bm{y_0}}
\right\}~~~\mathrm{on~}|\bm{\zeta}|=1,\nonumber\\
&{K}^{(e)}\to 0~~~\mathrm{when~}|\bm{\zeta}|\to \infty.
\end{align}
By using coordinate $(\rho,\theta)$, $K$ can be explicitly expressed as
\begin{align}
&K^{(i)}(\rho,\theta)=\rho(D_3\sin \theta + D_4 \cos \theta),\nonumber\\
&K^{(e)}(\rho,\theta)=\frac{1}{\rho}(D_3\sin \theta + D_4 \cos \theta),
\end{align}
according to the following coefficients:
\begin{align}
D_3 = -\left(\frac{1}{\rho^{(i)}}+\frac{1}{\rho^{(e)}}\right)^{-1}\left(\frac{1}{\rho^{(i)}}-\frac{1}{\rho^{(e)}}\right)\left.\frac{\partial {\xi}_0}{\partial y_2}\right|_{\bm{y}=\bm{y_0}},\nonumber\\
D_4 = -\left(\frac{1}{\rho^{(i)}}+\frac{1}{\rho^{(e)}}\right)^{-1}\left(\frac{1}{\rho^{(i)}}-\frac{1}{\rho^{(e)}}\right)\left.\frac{\partial {\xi}_0}{\partial y_1}\right|_{\bm{y}=\bm{y_0}}.\nonumber
\end{align}

\textbf{Step3:} The asymptotic solutions obtained in Step 2 are substituted into the shape derivative obtained in Step 1.
~\\

The asymptotic behavior of the shape derivative expressed in Eq.~(\ref{eq: shape derivative befor limit}) is \textcolor{black}{determined} using the asymptotic behavior of state variables $({\textcolor{black}{\eta}},\xi)$ obtained in Step 2. The limit form of the shape derivative is as follows:
\begin{align}
DJ\cdot \bm{\theta} =DL\cdot\bm{\theta} \to (-\theta_n)\varepsilon\left(
\lambda_{{A}_{11}^\ast}I_1 + \lambda_{{B}_{1}^\ast} I_2 + \lambda_{{K}^{-1 \ast}}I_3 + \lambda_{{F}^\ast} I_4
\right),
\end{align}
where $I_i~(i=1,...,4)$ are independent of $\varepsilon$ and expressed as
\begin{align}
I_1&=-\frac{4\pi(\rho^{(i)}-\rho^{(e)})}{\rho^{(e)}(\rho^{(i)} + \rho^{(e)})}\left\{
\left.\nabla_y {\textcolor{black}{\eta}}_0\right|_{\bm{y}=\bm{y_0}}\cdot\left.\nabla_y {\textcolor{black}{\eta}}_0\right|_{\bm{y}=\bm{y_0}}
+2\left.\frac{\partial {\textcolor{black}{\eta}}_0}{\partial y_1}\right|_{\bm{y}=\bm{y_0}}+1
\right\},\nonumber\\
I_2&= -\frac{4\pi(\rho^{(i)}-\rho^{(e)})}{\rho^{(e)}(\rho^{(i)} + \rho^{(e)})}\left\{
\left.\nabla_y \xi_0\right|_{\bm{y}=\bm{y_0}}\cdot\left.\nabla_y {\textcolor{black}{\eta}}_0\right|_{\bm{y}=\bm{y_0}}
+\left.\frac{\partial \xi_0}{\partial y_1}\right|_{\bm{y}=\bm{y_0}}
\right\},\nonumber\\
I_3&=2\pi\left(\frac{1}{K^{(i)}}-\frac{1}{K^{(e)}}\right),\nonumber\\
I_4&=\frac{4\pi(\rho^{(i)}-\rho^{(e)})}{\rho^{(e)}(\rho^{(i)} + \rho^{(e)})}
\left.\nabla_y \xi_0\right|_{\bm{y}=\bm{y_0}}\cdot\left.\nabla_y \xi_0\right|_{\bm{y}=\bm{y_0}}.
\end{align}
These are obtained considering that $\bm{\theta}\cdot \bm{n}^{(i)}=-\theta_n$ with negative constant $\theta_n$, as we focus on the shape change expressed as $\bm{\theta}=-\theta_n \bm{n}^{(i)}$.

~\\

\textbf{Step4:} The topological derivative is derived using ($\ref{eq:definition of TD}$).
~\\

By using the limit values of the shape derivative, the topological derivative is obtained based on the relationship between the shape and topological derivatives expressed in (\ref{eq:definition of TD}), as follows:
\begin{align}
D_T J &= \lim_{\varepsilon\to 0}\left\{
\frac{1}{V'(\varepsilon)|\theta_n|}DJ\cdot\bm{\theta}
\right\}\nonumber\\
&=\lim_{\varepsilon\to 0}\left[
\frac{1}{(-2\pi\varepsilon)(-\theta_n)}(-\theta_n)\varepsilon
\left\{
\lambda_{{A}_{11}^\ast}I_1 + \lambda_{{B}_{1}^\ast} I_2 + \lambda_{{K}^{-1 \ast}}I_3 + \lambda_{{F}^\ast} I_4
\right\}\right]\nonumber\\
&=-\frac{1}{2\pi}\left(
\lambda_{{A}_{11}^\ast}I_1 + \lambda_{{B}_{1}^\ast} I_2 + \lambda_{{K}^{-1 \ast}}I_3 + \lambda_{{F}^\ast} I_4
\right)\label{eq: complete form of topological derivative}
\end{align}
The obtained topological derivative, $D_T J$, contains the macroscopic contribution to objective function $J$ expressed by the optimal values of Lagrange multipliers $\bm{v}_B = (\lambda_{{A}_{11}^\ast},\lambda_{{B}_{1}^\ast},\lambda_{{F}^\ast},\lambda_{{K}^{-1\ast}})$. In addition, it comprises the microscopic contribution expressed by $I_i~(i=1,...,4)$, which are functions of the solutions of the cell problems.  

As the metasurface defined in design domain $D$ is composed of air and an elastic material, two \textcolor{black}{types} of topological derivatives are obtained: $D_T J^{\mathrm{air}\to \mathrm{elastic}}$ for air and $D_T J^{\mathrm{elastic}\to \mathrm{air}}$ for the elastic material.
The topological derivative for air, i.e., $D_T J^{\mathrm{air}\to \mathrm{elastic}}$ is obtained by substituting material parameters $(\rho^{(i)},K^{(i)})=(\rho^{\mathrm{elastic}},K^{\mathrm{elastic}})$ and $(\rho^{(e)},K^{(e)})=(\rho^{\mathrm{air}},K^{\mathrm{air}})$ \textcolor{black}{in} Eq.~(\ref{eq: complete form of topological derivative}).
Similarly, the topological derivative for the elastic material, i.e., $D_T J^{\mathrm{elastic}\to \mathrm{air}}$
is obtained by substituting material parameters $(\rho^{(i)},K^{(i)})=(\rho^{\mathrm{air}},K^{\mathrm{air}})$ and $(\rho^{(e)},K^{(e)})=(\rho^{\mathrm{elastic}},K^{\mathrm{elastic}})$ \textcolor{black}{in} Eq.~(\ref{eq: complete form of topological derivative}).

\bibliographystyle{elsarticle-num} 
\bibliography{reference}

\begin{thebibliography}{10}
\expandafter\ifx\csname url\endcsname\relax
  \def\url#1{\texttt{#1}}\fi
\expandafter\ifx\csname urlprefix\endcsname\relax\def\urlprefix{URL }\fi
\expandafter\ifx\csname href\endcsname\relax
  \def\href#1#2{#2} \def\path#1{#1}\fi

\bibitem{veselago1968electrodynamics}
V.~G. Veselago, The electrodynamics of substances with simultaneously negative
  values of $\epsilon$ and $\mu$, Soviet physics uspekhi 10~(4) (1968) 509.

\bibitem{liu2000locally}
Z.~Liu, X.~Zhang, Y.~Mao, Y.~Zhu, Z.~Yang, C.~T. Chan, P.~Sheng, Locally
  resonant sonic materials, Science 289~(5485) (2000) 1734--1736.

\bibitem{fang2006ultrasonic}
N.~Fang, D.~Xi, J.~Xu, M.~Ambati, W.~Srituravanich, C.~Sun, X.~Zhang,
  Ultrasonic metamaterials with negative modulus, Nature materials 5~(6) (2006)
  452.

\bibitem{huang2009negative}
H.~Huang, C.~Sun, G.~Huang, On the negative effective mass density in acoustic
  metamaterials, International Journal of Engineering Science 47~(4) (2009)
  610--617.

\bibitem{huang2009wave}
H.~Huang, C.~Sun, Wave attenuation mechanism in an acoustic metamaterial with
  negative effective mass density, New Journal of Physics 11~(1) (2009) 013003.

\bibitem{li2004double}
J.~Li, C.~Chan, Double-negative acoustic metamaterial, Physical Review E 70~(5)
  (2004) 055602.

\bibitem{ding2007metamaterial}
Y.~Ding, Z.~Liu, C.~Qiu, J.~Shi, Metamaterial with simultaneously negative bulk
  modulus and mass density, Physical review letters 99~(9) (2007) 093904.

\bibitem{zigoneanu2014three}
L.~Zigoneanu, B.-I. Popa, S.~A. Cummer, Three-dimensional broadband
  omnidirectional acoustic ground cloak, Nature materials 13~(4) (2014)
  352--355.

\bibitem{li2009experimental}
J.~Li, L.~Fok, X.~Yin, G.~Bartal, X.~Zhang, Experimental demonstration of an
  acoustic magnifying hyperlens., Nature materials 8~(12) (2009) 931--934.

\bibitem{ma2014acoustic}
G.~Ma, M.~Yang, S.~Xiao, Z.~Yang, P.~Sheng, Acoustic metasurface with hybrid
  resonances, Nature materials 13~(9) (2014) 873--878.

\bibitem{xie2014wavefront}
Y.~Xie, W.~Wang, H.~Chen, A.~Konneker, B.-I. Popa, S.~A. Cummer, Wavefront
  modulation and subwavelength diffractive acoustics with an acoustic
  metasurface, Nature communications 5 (2014).

\bibitem{bendsoe1988generating}
M.~P. Bends{\o}e, N.~Kikuchi, Generating optimal topologies in structural
  design using a homogenization method, Computer methods in applied mechanics
  and engineering 71~(2) (1988) 197--224.

\bibitem{sigmund2003systematic}
O.~Sigmund, J.~S. Jensen, Systematic design of phononic band--gap materials and
  structures by topology optimization, Philosophical Transactions of the Royal
  Society of London A: Mathematical, Physical and Engineering Sciences
  361~(1806) (2003) 1001--1019.

\bibitem{wadbro2006topology}
E.~Wadbro, M.~Berggren, Topology optimization of an acoustic horn, Computer
  methods in applied mechanics and engineering 196~(1) (2006) 420--436.

\bibitem{du2007minimization}
J.~Du, N.~Olhoff, Minimization of sound radiation from vibrating bi-material
  structures using topology optimization, Structural and Multidisciplinary
  Optimization 33~(4) (2007) 305--321.

\bibitem{duhring2008acoustic}
M.~B. D{\"u}hring, J.~S. Jensen, O.~Sigmund, Acoustic design by topology
  optimization, Journal of sound and vibration 317~(3) (2008) 557--575.

\bibitem{dilgen2019topology}
C.~B. Dilgen, S.~B. Dilgen, N.~Aage, J.~S. Jensen, Topology optimization of
  acoustic mechanical interaction problems: a comparative review, Structural
  and Multidisciplinary Optimization (2019) 1--23.

\bibitem{Diaz2010}
A.~R. Diaz, O.~Sigmund, A topology optimization method for design of negative
  permeability metamaterials, Structural and Multidisciplinary Optimization
  41~(2) (2010) 163--177.

\bibitem{lu2013topology}
L.~Lu, T.~Yamamoto, M.~Otomori, T.~Yamada, K.~Izui, S.~Nishiwaki, Topology
  optimization of an acoustic metamaterial with negative bulk modulus using
  local resonance, Finite Elements in Analysis and Design 72 (2013) 1--12.

\bibitem{noguchi2015acoustic}
Y.~Noguchi, T.~Yamada, M.~Otomori, K.~Izui, S.~Nishiwaki, An acoustic
  metasurface design for wave motion conversion of longitudinal waves to
  transverse waves using topology optimization, Applied Physics Letters
  107~(22) (2015) 221909.

\bibitem{christiansen2016designing}
R.~E. Christiansen, O.~Sigmund, Designing meta material slabs exhibiting
  negative refraction using topology optimization, Structural and
  Multidisciplinary Optimization 54~(3) (2016) 469--482.

\bibitem{roca2019computational}
D.~Roca, D.~Yago, J.~Cante, O.~Lloberas-Valls, J.~Oliver, Computational design
  of locally resonant acoustic metamaterials, Computer Methods in Applied
  Mechanics and Engineering 345 (2019) 161--182.

\bibitem{smith2005electromagnetic}
D.~Smith, D.~Vier, T.~Koschny, C.~Soukoulis, Electromagnetic parameter
  retrieval from inhomogeneous metamaterials, Physical review E 71~(3) (2005)
  036617.

\bibitem{fokin2007method}
V.~Fokin, M.~Ambati, C.~Sun, X.~Zhang, Method for retrieving effective
  properties of locally resonant acoustic metamaterials, Physical review B
  76~(14) (2007) 144302.

\bibitem{sanchez127non}
E.~Sanchez-Palencia, Non-homogeneous media and vibration theory. 1980, Lecture
  Notes in Physics 127.

\bibitem{bakhvalov2012homogenisation}
N.~S. Bakhvalov, G.~Panasenko, Homogenisation: averaging processes in periodic
  media: mathematical problems in the mechanics of composite materials,
  Vol.~36, Kluwer Academic Publishers, Dordrecht, 1989.

\bibitem{bensoussan1978asymptotic}
A.~Bensoussan, J.-L. Lions, G.~Papanicolaou, Asymptotic analysis for periodic
  structures, Vol.~5, North-Holland Publishing Company Amsterdam, 1978.

\bibitem{santosa1991dispersive}
F.~Santosa, W.~W. Symes, A dispersive effective medium for wave propagation in
  periodic composites, SIAM Journal on Applied Mathematics 51~(4) (1991)
  984--1005.

\bibitem{smyshlyaev2000rigorous}
V.~P. Smyshlyaev, K.~Cherednichenko, On rigorous derivation of strain gradient
  effects in the overall behaviour of periodic heterogeneous media, Journal of
  the Mechanics and Physics of Solids 48~(6) (2000) 1325--1357.

\bibitem{abdulle2014finite}
A.~Abdulle, M.~J. Grote, C.~Stohrer, Finite element heterogeneous multiscale
  method for the wave equation: long-time effects, Multiscale Modeling \&
  Simulation 12~(3) (2014) 1230--1257.

\bibitem{dohnal2014bloch}
T.~Dohnal, A.~Lamacz, B.~Schweizer, Bloch-wave homogenization on large time
  scales and dispersive effective wave equations, Multiscale Modeling \&
  Simulation 12~(2) (2014) 488--513.

\bibitem{allaire2016comparison}
G.~Allaire, M.~Briane, M.~Vanninathan, A comparison between two-scale
  asymptotic expansions and bloch wave expansions for the homogenization of
  periodic structures, SEMA journal 73~(3) (2016) 237--259.

\bibitem{craster2010high}
R.~V. Craster, J.~Kaplunov, A.~V. Pichugin, {High-frequency homogenization for
  periodic media}, Proceedings of the Royal Society A: Mathematical, Physical
  and Engineering Sciences 466~(2120) (2010) 2341--2362.

\bibitem{antonakakis2013asymptotics}
T.~Antonakakis, R.~V. Craster, S.~Guenneau, Asymptotics for metamaterials and
  photonic crystals, Proceedings of the Royal Society A: Mathematical, Physical
  and Engineering Sciences 469~(2152) (2013) 20120533.

\bibitem{noguchi2018topology}
Y.~Noguchi, T.~Yamada, K.~Izui, S.~Nishiwaki, Topology optimization for
  hyperbolic acoustic metamaterials using a high-frequency homogenization
  method, Computer Methods in Applied Mechanics and Engineering 335 (2018)
  419--471.

\bibitem{marigo2016homogenization}
J.-J. Marigo, A.~Maurel, Homogenization models for thin rigid structured
  surfaces and films, The Journal of the Acoustical Society of America 140~(1)
  (2016) 260--273.

\bibitem{marigo2016two}
J.-J. Marigo, A.~Maurel, Two-scale homogenization to determine effective
  parameters of thin metallic-structured films, Proceedings of the Royal
  Society A: Mathematical, Physical and Engineering Sciences 472~(2192) (2016)
  20160068.

\bibitem{rohan2010homogenization}
E.~Rohan, V.~Luke{\v{s}}, Homogenization of the acoustic transmission through a
  perforated layer, Journal of Computational and Applied Mathematics 234~(6)
  (2010) 1876--1885.

\bibitem{rohan2019homogenization}
E.~Rohan, V.~Luke{\v{s}}, Homogenization of the vibro--acoustic transmission on
  perforated plates, Applied Mathematics and Computation 361 (2019) 821--845.

\bibitem{rohan2010sensitivity}
E.~Rohan, V.~Luke{\v{s}}, Sensitivity analysis for acoustic waves propagating
  through homogenized thin perforated layer, in: Proceedings of ISMA, 2010.

\bibitem{rohan2013optimal}
E.~Rohan, V.~Luke{\v{s}}, Optimal design in vibro-acoustic problems involving
  perforated plates, in: Proc. 11th International Conference on Vibration
  Problems (ICOVP-2013), Z. Dimitrovov{\'a} et. al.(eds.) Lisbon, Portugal,
  AMPTAC, article, 2013, pp. 1--10.

\bibitem{cioranescu2008periodic}
D.~Cioranescu, A.~Damlamian, G.~Griso, The periodic unfolding method in
  homogenization, SIAM Journal on Mathematical Analysis 40~(4) (2008)
  1585--1620.

\bibitem{andkjaer2013topology}
J.~Andkj{\ae}r, O.~Sigmund, Topology optimized cloak for airborne sound,
  Journal of Vibration and Acoustics 135~(4) (2013) 041011.

\bibitem{carpio2008solving}
A.~Carpio, M.~Rapun, Solving inhomogeneous inverse problems by topological
  derivative methods, Inverse Problems 24~(4) (2008) 045014.

\bibitem{carpio2008topological}
A.~Carpio, M.~L. Rap{\'u}n, Topological derivatives for shape reconstruction,
  in: Inverse problems and imaging, Springer, 2008, pp. 85--133.

\bibitem{novotny2003topological}
A.~A. Novotny, R.~A. Feij{\'o}o, E.~Taroco, C.~Padra, Topological sensitivity
  analysis, Computer methods in applied mechanics and engineering 192~(7)
  (2003) 803--829.

\bibitem{feijoo2003topological}
R.~A. Feij{\'o}o, A.~A. Novotny, E.~Taroco, C.~Padra, The topological
  derivative for the poisson's problem, Mathematical Models and Methods in
  Applied Sciences 13~(12) (2003) 1825--1844.

\bibitem{yamada2010topology}
T.~Yamada, K.~Izui, S.~Nishiwaki, A.~Takezawa, A topology optimization method
  based on the level set method incorporating a fictitious interface energy,
  Computer Methods in Applied Mechanics and Engineering 199~(45) (2010)
  2876--2891.

\bibitem{allaire2004structural}
G.~Allaire, F.~Jouve, A.-M. Toader, Structural optimization using sensitivity
  analysis and a level-set method, Journal of computational physics 194~(1)
  (2004) 363--393.

\bibitem{MR3043640}
F.~Hecht, \href{https://freefem.org/}{New development in freefem++}, J. Numer.
  Math. 20~(3-4) (2012) 251--265.
\newline\urlprefix\url{https://freefem.org/}

\bibitem{dapogny2014three}
C.~Dapogny, C.~Dobrzynski, P.~Frey, Three-dimensional adaptive domain
  remeshing, implicit domain meshing, and applications to free and moving
  boundary problems, Journal of computational physics 262 (2014) 358--378.

\bibitem{semin2018homogenization}
A.~Semin, B.~Delourme, K.~Schmidt, On the homogenization of the helmholtz
  problem with thin perforated walls of finite length, ESAIM: Mathematical
  Modelling and Numerical Analysis 52~(1) (2018) 29--67.

\bibitem{Cea1986}
J.~Cea, Conception optimale ou identification de formes, calcul rapide de la
  d\'eriv\'ee directionnelle de la fonction co\^ut, ESAIM: Mathematical
  Modelling and Numerical Analysis - Mod\'elisation Math\'ematique et Analyse
  Num\'erique 20~(3) (1986) 371--402.

\bibitem{Amstutz2006}
S.~Amstutz, Sensitivity analysis with respect to a local perturbation of the
  material property, Asymptotic Analysis 49 (01 2006).

\end{thebibliography}


%
%
%
\end{document}